\documentclass[journal]{IEEEtran}
\usepackage[keeplastbox]{flushend}
\usepackage{amsfonts}
\usepackage{amsopn}
\usepackage{amsmath}
\usepackage{mathrsfs}
\usepackage{multirow}
\usepackage{setspace}
\usepackage{algorithm}
\usepackage{algorithmic}
\usepackage{booktabs} 
\usepackage[caption=false,font=footnotesize]{subfig}
\usepackage{graphicx}
\usepackage{epstopdf}
\usepackage{booktabs}
\usepackage{boldline,multirow}
\usepackage{cite}
\usepackage{threeparttable}
\usepackage{makecell}
\usepackage{multirow}
\usepackage{boldline}
\usepackage{colortbl}
\usepackage{xcolor}

\usepackage{lineno}

\definecolor{Stem beans}{RGB}{255 0 0}
\definecolor{Peas}{RGB}{90 11 255}
\definecolor{Forest}{RGB}{0 131 74}
\definecolor{Lucerne}{RGB}{0 252 255}
\definecolor{Wheat}{RGB}{255 182 229}
\definecolor{Beet}{RGB}{183 0 255}
\definecolor{Potatoes}{RGB}{255 255 0}
\definecolor{Bare soil}{RGB}{171 138 80}
\definecolor{Grass}{RGB}{0 255 0}
\definecolor{Rapeseed}{RGB}{255 128 0}
\definecolor{Barley}{RGB}{148 0 0}
\definecolor{Wheat2}{RGB}{191 191 255}
\definecolor{Wheat3}{RGB}{191 255 191}
\definecolor{Water}{RGB}{0 0 255}
\definecolor{Building}{RGB}{255 217 157}

\definecolor{Road}{RGB}{255 255 0}
\definecolor{Building}{RGB}{255 0 0}
\definecolor{Bare land}{RGB}{140 90 0}
\definecolor{Farmland1}{RGB}{0 255 255}
\definecolor{Farmland2}{RGB}{255 120 0}

\definecolor{Water}{RGB}{0 0 255}
\definecolor{Vegetation}{RGB}{0 255 0}
\definecolor{High-density urban}{RGB}{255 255 0}
\definecolor{Low-density urban}{RGB}{255 0 0}
\definecolor{Developed}{RGB}{255 0 255}

\begin{document}
\title{Multi-frequency PolSAR Image Fusion Classification Based on Semantic Interactive Information and Topological Structure
}
\author{\IEEEauthorblockN{Yice Cao,
Yan Wu,~\IEEEmembership{Member,~IEEE,}
Ming Li,~\IEEEmembership{Member,~IEEE,}
Mingjie Zheng,
Peng Zhang, 
and Jili Wang}
\thanks{This work was supported in part by the Natural Science Foundation of China under Grant 62172321, Grant 61871312, by the Civil Space Thirteen Five Years Pre-Research Project under Grant D040114. \emph{(Corresponding author: Yan Wu.)}

Yice Cao and Yan Wu are with the Remote Sensing Image Processing and Fusion Group, School of Electronic Engineering, Xidian University, Xi'an 710071, China. (e-mail: ywu@mail.xidian.edu.cn)

Ming Li and Peng Zhang are with the National Key Laboratory of Radar Signal Processing, Xidian University, Xi'an 710071, China.

Mingjie Zheng and Jili Wang are with the Aerospace Information Research Institute, Chinese Academy of Sciences, Beijing, China.

}}

\markboth{}%
{Shell \MakeLowercase{\textit{et al.}}: Bare Demo of IEEEtran.cls for IEEE Transactions on Magnetics Journals}
\IEEEtitleabstractindextext{%
\begin{abstract}
Compared with the rapid development of single-frequency multi-polarization SAR image classification technology, there is less research on the land cover classification of multi-frequency polarimetric SAR (MF-PolSAR) images. In addition, the current deep learning methods for MF-PolSAR classification are mainly based on convolutional neural networks (CNNs), only local spatiality is considered but the nonlocal relationship is ignored. Therefore, based on semantic interaction and nonlocal topological structure, this paper proposes the MF semantics and topology fusion network (MF-STFnet) to improve MF-PolSAR classification performance. In MF-STFnet, two kinds of classification are implemented for each band, semantic information-based (SIC) and topological property-based (TPC). They work collaboratively during MF-STFnet training, which can not only fully leverage the complementarity of bands, but also combine local and nonlocal spatial information to improve the discrimination between different categories. For SIC, the designed cross-band interactive feature extraction module (CIFEM) is embedded to explicitly model the deep semantic correlation among bands, thereby leveraging the complementarity of bands to make ground objects more separable. For TPC, the graph sample and aggregate network (GraphSAGE) is employed to dynamically capture the representation of nonlocal topological relations between land cover categories. In this way, the robustness of classification can be further improved by combining nonlocal spatial information. Finally, an adaptive weighting fusion (AWF) strategy is proposed to merge inference from different bands, so as to make the MF joint classification decisions of SIC and TPC. The effectiveness of the proposed modules is proved by ablation experiments on three measured MF-PolSAR datasets. In addition, the comparative experiments show that MF-STFnet can achieve more competitive classification performance than some state-of-the-art methods.

\end{abstract}

}
\maketitle
\IEEEdisplaynontitleabstractindextext
%
\IEEEpeerreviewmaketitle

\section{Introduction}
\IEEEPARstart{T}{he} polarimetric synthetic aperture radar (PolSAR), as a well-developed active microwave imaging system, has day-night and all-weather capabilities for earth monitoring \cite{Lee-2011-book}. Compared with the traditional SAR systems, PolSAR systems have the mechanism of alternately receiving and transmitting signals of different polarization modes, which can obtain more abundant ground target information \cite{Xiao-2019-CRPMnet}. In the radar remote sensing field, PolSAR data has found successful applications in land cover and use assessment \cite{Xiao-2019-CRPMnet, Gadhiya-2018-OWN, Gao-2014-WMM, Chen-1996-DNN, Wever-1995-identify, Yang-2015-SSRC, Famil-2001-alpha}, change detection \cite{Nielsen-2015-dection}, and so on \cite{Yi-2021-arctic}.

Among PolSAR data interpretation technology, the land cover classification of PolSAR images is an important link. To date, a variety of effective PolSAR image classification methods have been developed from different perspectives. The traditional PolSAR image classification methods mainly include polarimetric measurement data analysis-based \cite{Liu-2014-subspace, Yang-2018-kernel}, statistical characteristics-based \cite{Wu-2019-wishart}, and polarimetric target decomposition-based \cite{Feng-2014-sparse}. However, their classification accuracy is limited due to the weak discriminative ability of hand-crafted features \cite{Xiao-2019-CRPMnet}. With the deep learning (DL) algorithms erupting, deep feature learning-based methods have been proposed for PolSAR image classification, and have shown promising competence because of the ability to flexibly learn discriminative feature representation. In the early stage, some researchers tried to use simple neural networks \cite{Liu-2016-DBN, Geng-2018-MLP}. Recently, with the rise of CNN, CNN-based models and their derivatives have been widely studied for PolSAR image classification \cite{Zhang-2019-CNN, Hua-2022-CNN}. Typical methods among them include 3D-CNN \cite{Dong-2020-3D}, fully convolutional network (FCN) \cite{F-2019-FCN}, residual network (ResNet) \cite{Ding-2021-ResNet}, complex-valued network \cite{Cao-2019-FCN}, transformer \cite{Dong-2022-transformer}, and so on.

The above classification methods are all for single-frequency (SF) PolSAR images. With the development and application of multi-frequency and multi-polarization SAR systems, MF-PolSAR remote sensing techniques offer more efficient and reliable means of information collection for the earth surface observation \cite{ Gadhiya-2018-OWN}. Due to different frequency band attributes, the ground objects presented in PolSAR images of different bands will be different. This difference in imaging effect implies the existence of complementarity between bands \cite{Gao-2014-WMM}. For example, the P-band microwave has longer wavelength, lower resolution, and strong penetrating ability. It can penetrate canopies and even reach the ground, and has an obvious effect on dense ground objects such as forests \cite{Chen-1996-DNN}. Therefore, P-band PolSAR is suitable for the observation of forestry and soil types. The C-band microwave has shorter wavelength, higher resolution, and weak penetration ability. The physical details and edge features of targets are obvious in C-band PolSAR images. Therefore, C-band PolSAR is more suitable for observing geological structures, land erosion, and so on. Compared with C and P bands, the L-band microwave has moderate wavelength and certain penetration ability. It can detect not only surface scatterings, but also other scattering forms through the canopy to a certain extent. Therefore, L-band PolSAR can better distinguish similar crops, and is suitable for agricultural land-use cover \cite{Wever-1995-identify}. Based on the imaging difference analysis of PolSAR in different bands, combining multiple bands contains more abundant and comprehensive ground object information, which is theoretically beneficial to PolSAR image interpretation \cite{Yang-2015-SSRC}. In addition, utilizing the information complementarity of bands is expected to eliminate the inaccuracy of object recognition in a single band. 

The aim of the land cover classification of MF-PolSAR data is to determine the ground object coverage according to the characteristics of band data and the complementarity of bands \cite{Chen-1996-DNN}. Accurate MF-PolSAR image classification contributes to further promoting the application of PolSAR in civil and military fields. However, compared with the long research focus on SF-PolSAR image classification, the land cover classification of MF-PolSAR image has not been widely and deeply studied due to its particularity and challenges. There are two main reasons. On the one hand, the complementary information of bands has not been fully exploited. The extracted features are not enough to capture and reflect the complementarity, making it difficult to distinguish similar objects in MF classification results, and failing to eliminate the classification inaccuracy under single bands. On the other hand, the MF information fusion methods with good generalization ability remain to be proposed. There are attribute differences among MF features, and a simple fusion method may not be able to effectively resist and eliminate such differences. Thus, it may be impossible to obtain robust classification results. In addition, the research of adaptive MF information fusion methods, which are not specific to processing data and the number of bands, is a necessary technical support for the practical application of MF-PolSAR systems in the future. Therefore, how to leverage the complementarity and effectively fusing MF information is the key to achieving more satisfactory classification results under MF conditions than under SF conditions.

In recent years, some MF-PolSAR image classification methods have been proposed. The early traditional methods, such as statistical modeling \cite {Gao-2014-WMM, Famil-2001-alpha}, support vector machine (SVM) \cite{Lardeux-2009-SVM}, Stein-sparse representation \cite {Yang-2015-SSRC}, and tensor representation \cite{Liu-2014-TRMLS}, try to merge MF complementary information to permit better classification performance than SF cases. Besides, there are some research on DL-based methods for the MF-PolSAR image classification task recently. For example, in \cite{Ratha-2018-ANN}, features extracted from the Kronecker product matrix were used as the input of an artificial neural network (ANN) for MF-PolSAR classification. In addition, in \cite{Xiao-2019-CRPMnet}, by concatenating all bands as one band, MF-PolSAR classification was carried out through the dilated convolution and pixel-refining parallel mapping network (CRPM-Net).

Notably, the above neural networks for MF-PolSAR image classification are all based on the CNN structure. However, CNN performs convolution on regular regions with a fixed-size convolution kernel, so it can only model short-range local spatial relationships \cite{Ding-2021-aggregate}, which will limit the improvement of classification performance. To address this problem, based on the graph structure to model the topological relationship between samples, graph convolutional networks (GCNs) \cite{Kipf-2016-graph} have been proposed and have shown great advantages in many fields. The topological relationship describes the overall characteristics between samples (nodes), which belong to the medium and long-range context relationship, and are not limited by the coordinates in images \cite{Hong-2021-graph}. Therefore, the use of topological structure can effectively capture the nonlocal spatial information, which better describes the overall characteristics between categories, and in turn, helps to improve the classification performance \cite{Zhang-2021-topological}. At present, some methods based on GCNs have been used to classify PolSAR images \cite{Cheng-2021-graph, Liu-2021-adaptive, Ren-2021-graph}, and have achieved good classification results. Worthly, the spatial-based GCNs \cite{Wu-2021-comprehensive, Hamilton-2017-inductive, Ding-2021-graph} have attracted much attention due to their better flexibility and efficiency. This kind of GCN defines graph convolution based on the spatial relationship of nodes, which can flexibly aggregate new nodes and reduce computational complexity. Among them, the graph sample and aggregate network (GraphSAGE) \cite{Hamilton-2017-inductive} with inductive learning ability has good generalization. It makes the distributed training of large-scale graph data possible and expands the application scope of GCN. So far, spatial-based GCNs have not been used in MF-PolSAR image classification.

\begin{figure*}[htp]
\centering
\subfloat{\includegraphics[width=18cm,height=6cm]{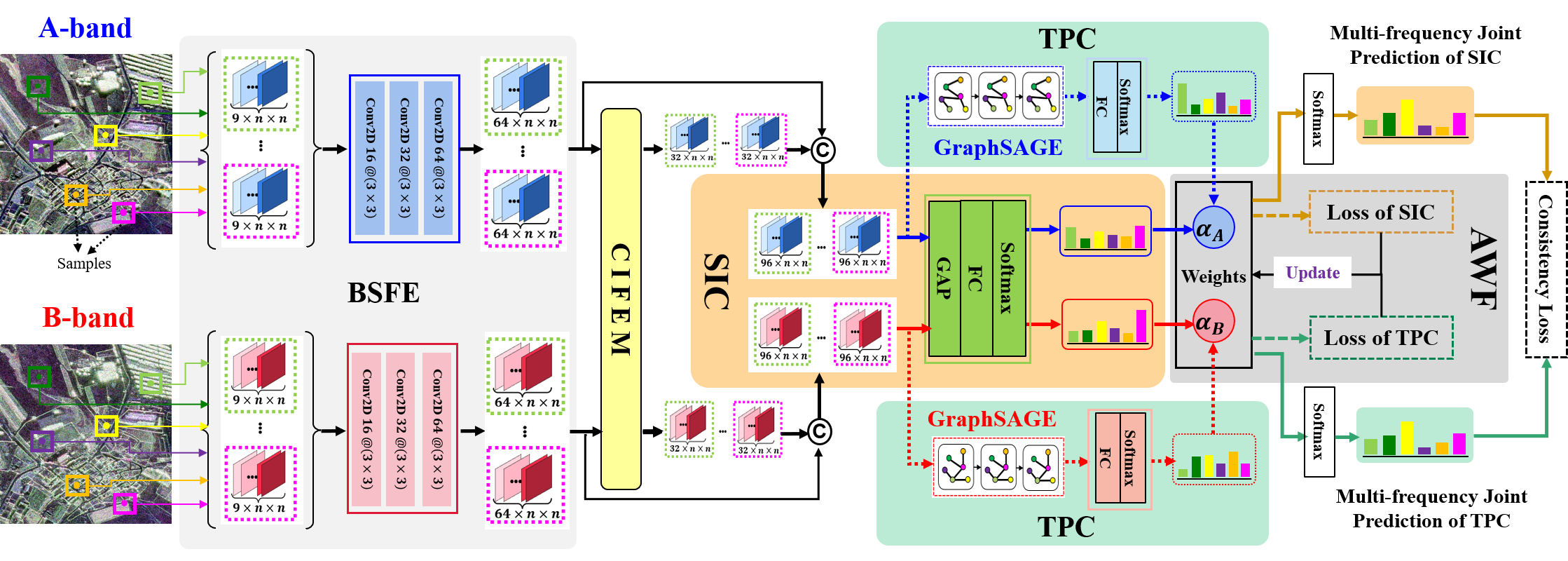}}

\caption{Architecutre of the proposed MF-STFnet. The C icon represents the feature concatenation operation.}
\label{fig:MF-STFnet}
\end{figure*}

Based on the above analysis, to further enhance MF-PolSAR classification performance, a novel MF-PolSAR classification model named multi-frequency semantics and topology fusion network (MF-STFnet) is proposed in this paper. MF-STFnet is mainly based on semantic interactive information and topological structure, and can fully mine and leverage the complementarity of bands as well as combine local and nonlocal spatial information to improve the accuracy and robustness of classification. In MF-STFnet, for each band, the semantic information-based (SIC) and topological property-based (TPC) classifications are adopted. For SIC, on the one hand, the band-specific CNNs with different parameters are designed to extract band-specific features, which preserve the inter-band discriminative representation and ensure the diversity of band attributes. On the other hand, to further mine and leverage the complementarity of bands, a shared part named cross-band interactive feature extraction module (CIFEM) is adopted and embedded in MF-STFnet. CIFEM is designed to use the correlations between attributes of band-specific representations to explicitly model the deep semantic correlation among bands. It can extract more discriminative cross-band interactive features, and enhance the interactive fusion of information between bands, thereby improving the accuracy of classifying ground objects. For TPC, the subgraphs for each band are firstly constructed based on semantic features. After that, two layers of GraphSAGE are used to dynamically capture the representation of nonlocal topological relations between samples, so as to further improve the robustness of classification by combining nonlocal spatial information. Finally, an adaptive weighting fusion (AWF) strategy is adopted to merge inference from different bands in an adaptive weighted manner, thereby making the multi-frequency joint classification decisions of SIC and TPC. It should be emphasized that the weights related to different bands are adaptively updated based on the total loss during network optimization.

The main contributions can be summarized as follows:
\begin{enumerate}
\item A MF-PolSAR image classification framework named MF-STFnet is proposed, including two kinds of classifications, SIC and TPC. MF-STFnet can not only make full use of the complementary and interaction information among bands, but also combine local and nonlocal spatial information, thereby obtaining more accurate and robust classification results.

\item The designed CIFEM is used in SIC to explicitly model semantic interaction between different bands. The extracted interactive features are concatenated with band-specific features to realize the semantic information-based classification. In addition, the GraphSAGE model is adopted in TPC to dynamically capture the representation of nonlocal topological relations, so as to perform the topological property-based classification.

\item The AWF strategy with good generalization is proposed to flexibly fuse the results of different bands, thereby adaptively obtaining the final multi-frequency joint classification decision. Notably, its weights are updated adaptively based on the total loss.
\end{enumerate}

The rest of this paper is organized as follows. Section II describes the proposed MF-STFnet for MF-PolSAR image classification. In Section III, experimental results and analysis on three measured MF-PolSAR datasets are presented. The conclusion is in Section IV.

\section{Proposed MF-STFnet for Multi-frequency PolSAR Image Classification}
For brevity and clarity, taking the case of two frequency bands (A-band and B-band) as an example, Fig. \ref{fig:MF-STFnet} intuitively shows the architecture of the proposed MF-STFnet for MF-PolSAR image classification. On this basis, the architecture of more than two bands can be extended. As shown in Fig.1, the proposed method mainly concludes band-specific semantic feature extraction (BSFE), CIFEM, SIC, TPC, and the final multi-frequency joint classification decision based on the AWF strategy. In addition, Table \ref{label-MF-STFnet} reports the detailed structure of MF-STFnet.

\begin{table}[htp]
\centering
\caption{Detailed Structure of the Proposed MF-STFnet}
\fontsize{6.5}{12}\selectfont 
\begin{tabular}{|c|c|c|c|} \hline
\multicolumn{3}{|c|}{\bf Structure} & \bf{Output Size} \\ \hline
\multicolumn{2}{|c|}{\multirow{2}{*}{Input layer}} & A-band & 9 $\times$ 13 $\times$ 13 \\ \cline{3-4}
\multicolumn{2}{|c|}{} & B-band & 9 $\times$ 13 $\times$ 13 \\ \hline

\multirow{6}{*}{BSFE} & \multirow{3}{*}{A-band} & Conv(16 $\times$ 3 $\times$ 3) / BN / ReLU & 16 $\times$ 13 $\times$ 13 \\ \cline{3-4}
\multirow{6}{*}{} & \multirow{3}{*}{} & Conv(32 $\times$ 3 $\times$ 3) / BN / ReLU & 32 $\times$ 13 $\times$ 13 \\ \cline{3-4}
\multirow{6}{*}{} & \multirow{3}{*}{} & Conv(64 $\times$ 3 $\times$ 3) / BN / ReLU & 64 $\times$ 13 $\times$ 13 \\ \cline{2-4}

\multirow{6}{*}{} & \multirow{3}{*}{B-band} & Conv(16 $\times$ 3 $\times$ 3) / BN / ReLU & 16 $\times$ 13 $\times$ 13 \\ \cline{3-4}
\multirow{6}{*}{} & \multirow{3}{*}{} & Conv(32 $\times$ 3 $\times$ 3) / BN / ReLU & 32 $\times$ 13 $\times$ 13 \\ \cline{3-4}
\multirow{6}{*}{} & \multirow{3}{*}{} & Conv(64 $\times$ 3 $\times$ 3) / BN / ReLU & 64 $\times$ 13 $\times$ 13 \\ \hline

\multicolumn{2}{|c|}{\multirow{2}{*}{CIFEM}} & A-band & 32 $\times$ 13 $\times$ 13 \\ \cline{3-4}
\multicolumn{2}{|c|}{} & B-band & 32 $\times$ 13 $\times$ 13 \\ \hline

\multicolumn{2}{|c|}{\multirow{2}{*}{Concatenation}} & A-band & 96 $\times$ 13 $\times$ 13 \\ \cline{3-4}
\multicolumn{2}{|c|}{} & B-band & 96 $\times$ 13 $\times$ 13 \\ \hline

\multirow{2}{*}{SIC} & A-band &{Global Pool / FC / Softmax} & $C$ $\times$ 1 \\ \cline{2-4}
\multirow{2}{*}{} & B-band &{Global Pool / FC / Softmax} & $C$ $\times$ 1 \\ \hline

\multirow{2}{*}{TPC} & A-band &{GraphSAGE(64-32) / FC / Softmax} & $C$ $\times$ 1 \\ \cline{2-4}
\multirow{2}{*}{} & B-band &{GraphSAGE(64-32) / FC / Softmax} & $C$ $\times$ 1 \\ \hline

\multirow{2}{*}{\makecell[c]{Multi-frequency \\ Joint Prediction}} & SIC &\multirow{2}{*}{AWF / Softmax} & $C$ $\times$ 1 \\ \cline{2-2} \cline{4-4}
\multirow{2}{*}{} & TPC &\multirow{2}{*}{} & $C$ $\times$ 1 \\ \hline

\end{tabular}
\label{label-MF-STFnet}
\end{table}

\subsection{Band-specific Semantic Feature Extraction (BSFE)}
At the beginning of MF-STFnet, for each band, the band-specific CNN is utilized for band-specific semantic feature extraction (BSFE). The BSFE part aims to capture various band-specific information for multi-frequency learning enhancement, so as to preserve the inter-band discriminative representation and ensure the diversity of band attributes.

In the monostatic backscattering case, the scattering characteristics of each resolution cell in a PolSAR image can be described by the 3 $\times$ 3 polarimetric coherency matrix $\mathrm{T}$ \cite{Lee-2011-book}:

\begin{equation}
\label{coherency-T}
\begin{aligned}
\mathrm{T}  =\left\langle\mathrm{u}_{L} \cdot \mathrm{u}_{L}^{\mathrm{H}}\right\rangle =\left[\begin{array}{ccc}
\mathrm{T}_{11} & \mathrm{T}_{12} & \mathrm{T}_{13} \\
\mathrm{T}_{21} & \mathrm{T}_{22} & \mathrm{T}_{23} \\
\mathrm{T}_{31} & \mathrm{T}_{32} & \mathrm{T}_{33}
\end{array}\right], 
\end{aligned}
\end{equation}
where $\langle.\rangle$ indicates temporal or spatial ensemble averaging. $\mathrm{u}_{L}$ is the polarimetric target vector, and the superscript $\mathrm{H}$ denotes the conjugate transpose. In this paper, a 9-dimensional vector, which is expended by the upper triangular of $\mathrm{T}$, is used as the input feature of each pixel cell. The input vector can be represented by

\begin{scriptsize}
\begin{equation} 
\left[\mathrm{T}_{11}, \mathrm{~T}_{22}, \mathrm{~T}_{33}, \Re\left(\mathrm{T}_{12}\right), \Re\left(\mathrm{T}_{13}\right), \Re\left(\mathrm{T}_{23}\right), \Im\left(\mathrm{T}_{12}\right), \Im\left(\mathrm{T}_{13}\right), \Im\left(\mathrm{T}_{23}\right)\right]
\end{equation}
\end{scriptsize} 
where $\Re\left(.\right)$ and $\Im\left(.\right)$ denote the real and imaginary parts of complex elements, respectively.

Suppose there are $K$ bands. First, $K$ small data patches of size $9 \times n \times n$ centered on the same pixel in all band images are generated as input to the BSFE part. $n \times n$ is the spatial dimension of small data patch, which is set to 13 $\times$ 13 in this paper. For each band, the band-specific CNN in BSFE consists of three cascaded convolution blocks, and its corresponding structure is shown in Fig. \ref{fig:MF-STFnet}. Each convolution block is stacked by a convolutional layer with kernel size 3 $\times$ 3, a batch normalization (BN) layer, and a rectified linear unit (ReLU) layer. Let $X_{k} \in \mathfrak{R}^{9 \times n \times n}$ be an input data patch of the $k$-th band, the output features $X_{k}^{b1}$ obtained by the first convolution block $F_{b1}(\cdot)$ can be formulated as

\begin{equation}
X_{k}^{b 1}=F_{b 1}\left(X_{k}\right)=f_{b 1}^{\text{ReLU}}\left(f_{b 1}^{\text{BN}}\left(f_{b 1}^{\text{Conv}}\left(X_{k}\right)\right)\right)
\end{equation}
where ${k}=\{1,...,K\}$, and $K$ is the total number of frequency bands. $f_{b 1}^{\text{Conv}}(\cdot)$, $f_{b 1}^{\text{BN}}(\cdot)$, and $f_{b 1}^{\text{ReLU}}(\cdot)$ denote the convolution, BN, and ReLU in the first convolution block of CNN, respectively. The other two convolution blocks have the same processing as $F_{b 1}(\cdot)$. The only difference between these three blocks lies in the output dimension. Therefore, for each band, the whole procedure of BSFE can be formulated as
\begin{equation}
X_{k}^{\text{BSFE}}=F_{b 3}\left(F_{b 2}\left(F_{b 1}\left(X_{k}\right)\right)\right)
\end{equation}
where $X_{k}^{\text{BSFE}}  \in \mathfrak{R}^{64 \times n \times n}$ denotes the corresponding BSFE output of the $k$-th band. $F_{b 2}(\cdot)$ and $F_{b 3}(\cdot)$ respectively represent the second and third convolution blocks. The output features $\left\{X_{k}^{\text{BSFE}}\right\}_{{k} \in \{1,...,K\}}$ extracted by the BSFE part can not only preserve the discriminative representation between bands, but also ensure the inter-band diversity, which is a key aspect of multi-frequency learning enhancement.

\subsection{Cross-band Interactive Feature Extraction}
The features extracted by BSFE are only specific to bands, and there is no information interaction between bands. This may not fully utilize the complementary information of bands, thereby limiting classification performance. To focus on the correlation of MF-PolSAR data, based on the band-specific semantic representation of BSFE, the CIFEM is utilized to explicitly model the deep semantic correlation between bands, so as to extract discriminative enough features to improve category differentiation.

Formally, with respect to the $k$-th band, a set $S^{k}$ containing different band pairs is denoted as
\begin{equation}
S^{k}=\{(k, \bar{k})\}_{\bar{k}=\{1,...,K\} \backslash k}
\end{equation}

Based on the semantic representation extracted by BSFE, the correlative feature maps $\text{X}_{k, \bar{k}}$ between $X_{k}^{\text{BSFE}}$ and $X_{\bar{k}}^{\text{BSFE}}$ can be formulated as
\begin{equation}
\text{X}_{k, \bar{k}}=F_{\text{Cor}}\left(X_{k}^{\text{BSFE}}, X_{\bar{k}}^{\text{BSFE}}\right)
\end{equation}
where $F_{\text{Cor}}(\cdot)$ corresponds the correlation operation. In detail, assume $X_{k_i}^{\text{BSFE}}$ represents any feature map in $X_{k}^{\text{BSFE}}$, where ${i}=\{1,...,m\}$. $m$ is the number of feature maps. We first do a dot products between $X_{k_i}^{\text{BSFE}}$ and each feature map of $X_{\bar{k}}^{\text{BSFE}}$ to obtain a set of feature maps $\text{x}_{k_i, \bar{k}}$, which can be denoted as 
\begin{equation}
\text{x}_{k_i, \bar{k}}=X_{k_{i}}^{\text{BSFE}} * X_{\bar{k}}^{\text{BSFE}}
\end{equation}
where $*$ denotes the dot product operator. Thus, for all $m$ feature maps in $X_{k}^{\text{BSFE}}$, the correlative feature maps $\text{X}_{k, \bar{k}}$ can be described as
\begin{equation}
\text{X}_{k, \bar{k}}=\left\{\text{x}_{k_{i}, \bar{k}}\right\}_{i \in\{1, \ldots, m\}}
\end{equation}

Based on the above process, for the $k$-th band, all correlative feature maps with respect to other bands can form a set $D_{S^{k}}=\left\{\text{X}_{k, \bar{k}}\right\}_{(k, \bar{k}) \in S^{k}}$. Finally, to reduce computational complexity, a shared convolution block $F_{b}(\cdot)$ with 1 $\times$ 1 convolutional kernel and 32 output dimension is adopted to project each $\text{X}_{k, \bar{k}}$ in $D_{S^{k}}$. The projected output $\text{X}_{k, \bar{k}}^{\text{Cor}}$ is
\begin{equation}
\text{X}_{k, \bar{k}}^{\text{Cor}}=F_{b}\left(\text{X}_{k, \bar{k}}\right)=f_{b}^{\text{ReLU}}\left(f_{b}^{\text{BN}}\left(f_{b}^{\text{Conv}}\left(\text{X}_{k, \bar{k}}\right))\right)\right)
\end{equation}
where $f_{b}^{\text{Conv}}(\cdot)$, $f_{b}^{\text{BN}}(\cdot)$, and $f_{b}^{\text{ReLU}}(\cdot)$ denote the convolution, BN, and ReLU in $F_{b}(\cdot)$, respectively.

Thus, for the $k$-th band, the correlative feature maps with respect to other bands are collected in a set $\left\{\text{X}_{k, \bar{k}}^{\text {Cor }}\right\}_{(k, \bar{k}) \in S^{k}}$. We concatenate all elements in this set as the final cross-band interactive feature maps $\text{X}_{k}^{\text{Cor}}$ with respect to the $k$-th band. 

\begin{figure}[htp]%
\centering
\subfloat{\includegraphics[width=9cm,height=5cm]{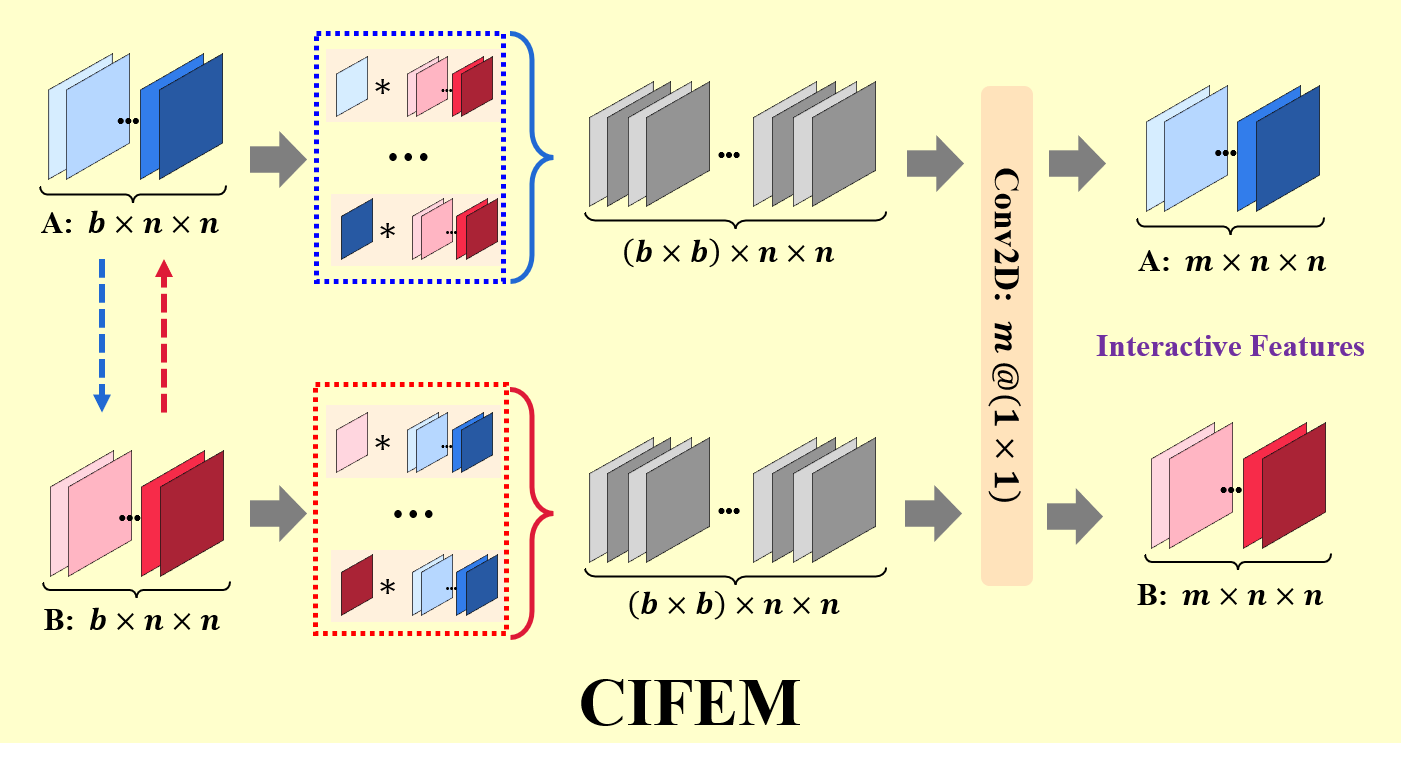}}

\caption{Illustration of the CIFEM of two bands.}
\label{fig:CIFEM}
\end{figure}%

We illustrate the CIFEM of two bands as shown in Fig. \ref{fig:CIFEM}. Respecting to A-band, the correlative feature maps between A- and B-bands are calculated by the traversing multiplication between each feature map of A and all features maps of B. In Fig. \ref{fig:CIFEM}, this calculation process is highlighted by the blue dashed rectangle. While for the B-band, the correlative feature maps are calculated by the traversing multiplication between each feature map of B and all features maps of A, and the calculation process is highlighted by the red dashed rectangle. Finally, these correlative features are projected by the shared convolution block to obtain the final cross-band interactive features. Notably, when the number of frequency bands is greater than 2, for any band, the final cross-band interactive features are the concatenation of the calculation results of the current band and all other bands.

CIFEM can utilize the correlation between the attributes of semantic representation to explicitly model the deep semantic interaction between bands. The more discriminative cross-band interactive features extracted by CIFEM realize the interactive fusion and enhancement of information among bands, thereby improving the discrimination between different categories and leading to better classification performance.
  
\subsection{Classification of Each Band Based on Semantic Information and Topological Structure}
BSFE and CIFEM based on CNN frameworks only model the local spatial relationship of samples. As a result, they are unable to model the nonlocal topological relationship between land cover categories that describes the overall characteristics, which limits the further improvement of classification performance. For this reason, this paper uses GraphSAGE to dynamically capture the representation of topological relations between samples, so as to further improve the classification results with the assistance of nonlocal spatial information \cite{Zhang-2021-topological}. In this way, two classifications are implemented for each band, namely semantic information-based classification (SIC) and topological property-based classification (TPC). SIC and TPC achieve classification prediction from the local and nonlocal spatial perspectives, respectively. During the MF-STFnet training, the two classifications work cooperatively to capture more discriminative information, which is conducive to achieving more accurate model learning.

\subsubsection{SIC}
Based on the two subsections above, for each band, the band-specific and interactive feature maps are concatenated as SIC input. SIC consists of a global average pooling (GAP) layer, a fully connected (FC) layer with $\text{C}$ output, and a softmax layer. Where $\text{C}$ is the total number of categories. The processing of SIC is illustrated in green rectangles in Fig. \ref{fig:MF-STFnet}. 

Represent the concatenated features of $k$-th band as $\text{X}_{k}^{\text{Con}}$, the output $Z_{k}^{\text{SIC}}$ of the semantic information-based classifier can be denoted as
\begin{equation}
Z_{k}^{\text{SIC}}=f_{\text{SIC}}^{\text{Softmax}}\left(f_{\text{SIC}}^{\text{FC}}\left(f_{\text{SIC}}^{\text{GAP}}\left(\text{X}_{k}^{\text{Con}}\right)\right)\right)
\end{equation}
where $f_{\text{SIC}}^{\text{GAP}}(\cdot)$, $f_{\text{SIC}}^{\text{FC}}(\cdot)$, and $f_{\text{SIC}}^{\text{Softmax}}(\cdot)$ respectively the GAP, FC, and Softmax operations in the SIC part. 

Based on the combination of band-specific and interactive information, SIC obtains the category prediction from the semantic and local spatial perspectives, which can fully utilize the complementarity of bands to improve the distinction of ground object categories.

\subsubsection{TPC}
To consider the nonlocal spatial properties reflected by topological structure, for each band, GraphSAGE is adopted in TPC to dynamically update the nonlocal graph for more accurate feature embedding. GraphSAGE, as a classical aggregation function-based GCN model, can flexibly aggregate the sample information in data with arbitrary structure, thereby improving the ability to capture nonlocal spatial relationships.

Specifically, this paper utilizes two layers of GraphSAGE model in TPC to automatically learn nonlocal topology information, which collaborates with SIC during MF-STFnet training to help improve the accuracy and robustness of class discrimination. For each band, the nonlocal subgraph of node relationship is first constructed and initialized based on the training samples. It should be emphasized that nodes in the constructed subgraph are the samples. Suppose that the concatenated semantic features of sample $v$ in any band are denoted by $\text{X}_{(v)}^{\text{Con}}$, $u$ is the neighborhood nodes of $v$ in the constructed nonlocal graph, and $\mathcal{N}(\cdot)$ represents the neighborhood function. After aggregating the information of neighborhood nodes in the first layer of GraphSAGE, the features of node $v$ are updated as  

\begin{scriptsize}
\begin{equation} 
\mathrm{G}_{(v)}^{\text{TPC}_{1}} \leftarrow \sigma \left(\mathrm{~W}^{\text{TPC}_{1}} \cdot F_{\text{Mean}}\left(\mathrm{X}_{(v)}^{\text{Con}} \cup\left\{\mathrm{X}_{(u)}^{\text{Con}}, \forall u \in \mathcal{N}(v)\right\}\right)\right)
\end{equation}
\end{scriptsize}
where $\mathrm{~W}^{\text{TPC}_{1}}$ represents the weights of the first layer, $\sigma$ is the nonlinear activation function, $F_{\text{Mean}}(\cdot)$ represents the mean aggregation function, and $\mathrm{G}_{(v)}^{\text{TPC}_{1}}$ is the updated features of node $v$. Then, after the information aggregation in the second layer of GraphSAGE, the updated features $\mathrm{G}_{(v)}^{\text{TPC}_{2}}$ are
 
\begin{scriptsize}
\begin{equation} 
\mathrm{G}_{(v)}^{\text{TPC}_{2}} \leftarrow \sigma \left(\mathrm{~W}^{\text{TPC}_{2}} \cdot F_{\text{Mean}}\left(\mathrm{G}_{(v)}^{\text{TPC}_{1}} \cup\left\{\mathrm{G}_{(u)}^{\text{TPC}_{1}}, \forall u \in \mathcal{N}(v)\right\}\right)\right)
\end{equation}
\end{scriptsize}

Finally, the features $\mathrm{G}_{(v)}^{\text{TPC}_{2}}$ extracted by two layers of GraphSAGE will pass through a FC layer and a Softmax layer to obtain the final classification result of TPC
\begin{equation}
Z_{(v)}^{\text{TPC}}=f_{\text {TPC}}^{\text{Softmax}}\left(f_{\text{TPC}}^{\text{FC}}\left(\mathrm{G}_{(v)}^{\text{TPC}_{2}}\right)\right)
\end{equation}
where $f_{\text{TPC}}^{\text{FC}}(\cdot)$ and $f_{\text{TPC}}^{\text{Softmax}}(\cdot)$ respectively the FC and Softmax operations in the TPC part. 

\begin{figure}[htp]%
\centering
\subfloat{\includegraphics[width=7cm,height=5cm]{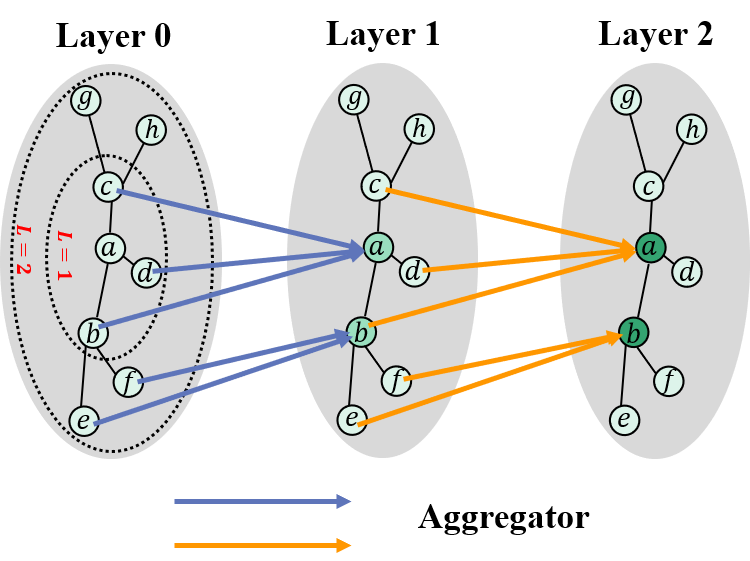}}

\caption{Illustration of the information aggregation and propagation in a simple two layers of GraphSAGE. $L$ denotes the neighborhood order.}
\label{fig:GraphSAGE}
\end{figure}%

Fig. \ref{fig:GraphSAGE} illustrates the information aggregation and propagation in a simple two layers of GraphSAGE, with a total of eight samples $a$-$g$. For the $k$-th band, based on the concatenated features $\text{X}_{k}^{\text{Con}}$, the dynamic subgraph is first constructed and initialized, which is shown as Layer 0 in Fig. \ref{fig:GraphSAGE}. Then, based on this subgraph, GraphSAGE aggregates neighborhood nodes to update the feature representation of each central node. For example, in Fig. \ref{fig:GraphSAGE}, the central node $a$ in the first layer aggregates the information of nodes $c$ and $d$ in its first-order neighborhood ($L=1$). Through the repeated aggregation of two layers of GraphSAGE model, the node information is extended to the second-order neighborhood. As shown in Fig. \ref{fig:GraphSAGE}, the central node $a$ in the second layer also aggregates the information of nodes $e$ and $f$ in its second-order neighborhood ($L=2$). Accordingly, the feature representation of nodes is aggregated and updated through information propagation to capture nonlocal spatial information, thereby improving the robustness of classification. 

TPC obtains the category prediction by mining nonlocal spatial information. It works collaboratively with SIC during MF-STFnet training, so the proposed network can not only exploit semantic information, but also simultaneously capture the local and nonlocal spatial context information. In this way, more discriminative feature representation can be effectively obtained, which is beneficial to improve the accuracy and robustness of classification.

\subsection{Final Multi-frequency Adaptive Fusion Classification Decision and Parameter Optimization}
For each band, there are two probability outputs of SIC and TPC. To obtain the final multi-frequency joint classification decisions of SIC and TPC, the adaptive weighting fusion (AWF) strategy is adopted to adaptively and flexibly combine the results of all bands. In addition, the final losses of SIC and TPC are also calculated based on AWF. These losses are not only used to guide MF-STFnet training, but also to update the weights of AWF.

\subsubsection{Multi-frequency Joint Decision Based on AWF Stragtegy}
For the $k$-th band, two probability outputs $Z_{k}^{\text{SIC}}$ and $Z_{k}^{\text{TPC}}$ are obtained. Assume that $\alpha \in \mathfrak{R}^{K}$ denotes the adaptive weight, the multi-frequency joint classification results of SIC and TPC calculated by AWF are respectively defined as
\begin{equation}
Y^{\text{SIC}}=\sum_{k=1}^{K} \alpha_{k}^{\gamma} Z_{k}^{\text{SIC}}
\end{equation}
\begin{equation}
Y^{\text{TPC}}=\sum_{k=1}^{K} \alpha_{k}^{\gamma} Z_{k}^{\text{TPC}}
\end{equation}
where $\sum_{k=1}^{K} \alpha_{k}=1, \alpha_{k} \geq 0$. $\gamma>1$ denotes the power exponent parameter of weights, which is used to avoid the trivial solution of $\alpha_{k}$ \cite{Xu-2020-embed}.

Then, the Softmax function $F_{\text{Softmax}}(\cdot)$ is utilized to calculate the probability distribution of each category, which can be described as
\begin{equation}
F_{\text{Softmax}}\left(Y^{\text{SIC}}\right)_{c}=e^{Y_{c}^{\text{SIC}}} / \sum_{j=1}^{C} e^{Y_{j}^{\text{SIC}}}, c=1, \ldots, C 
\end{equation}
\begin{equation}
F_{\text{Softmax}}\left(Y^{\text{TPC}}\right)_{c}=e^{Y_{c}^{\text{TPC}}} / \sum_{j=1}^{C} e^{Y_{j}^{\text{TPC}}}, c=1, \ldots, C
\end{equation}

\subsubsection{Objective Function for MF-STFnet Optimization}
The above-mentioned modules describe the forward propagation process of the whole MF-STFnet. To optimize this network, the cross-entropy loss \cite{Golik-2013-entropy} is first used in both SIC and TPC to calculate the semantic and topological classification losses, respectively. Based on the AWF strategy, the two losses $\ell_{\text{SIC}}$ and $\ell_{\text{TPC}}$ are represented as

\begin{equation}
\ell_{\text{SIC}}=\sum_{k=1}^{K} \alpha_{k}^{\gamma} L_{k}\left(\mathrm{Z}_{k}^{\text{SIC}}, { Label}\right) 
\end{equation}
\begin{equation}
\ell_{\text{TPC}}=\sum_{k=1}^{K} \alpha_{k}^{\gamma} L_{k}\left(\mathrm{Z}_{k}^{\text{TPC}}, { Label }\right)
\end{equation}
where $L_{k}$ represents the cross-entropy loss function, ${ Label }$ denotes the true label of samples.

In addition, to make SIC and TPC work together better during the training process, a consistency loss is used \cite{Gao-2021-coastal} to constrain the multi-frequency joint probability predictions of SIC and TPC. It calculates the difference between the SIC probability prediction distribution and the TPC probability prediction distribution of samples, and is described as
\begin{equation}
\ell_{\text {\text{consistency}}}=\frac{1}{N}\left\|Y^{\text{SIC}}-Y^{\text{TPC}}\right\|_{2}
\end{equation}
where $N$ is the number of samples. For any sample, the smaller the $\ell_{\text {\text{consistency}}}$ value, the more consistent the prediction results of SIC and TPC, which reflects the better cooperation between SIC and TPC. 

Thus, the final objective function, that is, the total loss is
\begin{equation}
\ell_{\text {total}}=\ell_{\text{SIC}}+\ell_{\text{TPC}}+\lambda . \ell_{\text {consistency}}
\end{equation}
where $\lambda$ is the regularization parameters. In this paper, $\lambda$ is set to 0.1. Based on the total loss, the proposed MF-STFnet can be trained for network optimization by the back-propagation algorithm.

\subsubsection{Update weights $\alpha$ of AWF}
In addition to updating MF-STFnet parameters, the adaptive weight $\alpha$ of AWF also need to be updated at the same time. According to the total loss $\ell_{\text {total}}$, only $\ell_{\text{SIC}}$ and $\ell_{\text{TPC}}$ are related to $\alpha$. Therefore, by fixing the MF-STFnet parameters, the optimization problem for updating $\alpha$ is

\begin{scriptsize}
\begin{equation}
\min _{\alpha}\left(\ell_{\text{SIC}}+\ell_{\text{TPC}}\right)=\min _{\alpha} \sum_{k=1}^{K} \alpha_{k}^{\gamma}\left[L_{k}\left(\mathrm{Z}_{k}^{\text{SIC}}, \text {label}\right)+L_{k}\left(\mathrm{Z}_{k}^{\text{TPC}}, \text {label}\right)\right]
\end{equation}
\end{scriptsize}

Here, we use $L_k$ to represent $L_{k}\left(\mathrm{Z}_{k}^{\text{SIC}}, \text {label}\right)+L_{k}\left(\mathrm{Z}_{k}^{\text{TPC}}, \text {label}\right)$. According to $\sum_{k=1}^{K} \alpha_{k}=1, \alpha_{k} \geq 0$, the corresponding Lagrangian function is
\begin{equation}
\omega(\alpha, \zeta)=\sum_{k=1}^{K} \alpha_{k}^{\gamma} L_{k}-\zeta\left(\sum_{k=1}^{K} \alpha_{k}^{\gamma}-1\right)
\label{alpha}
\end{equation}
where $\zeta$ is the Lagrange multiplier. 

According to (\ref{alpha}), taking the derivatives respect to $\alpha_{k}$ and  $\zeta$. The updating equation of weight $\alpha_{k}$ can be obtained by setting derivatives to zero
\begin{equation}
\alpha_{k}=\frac{L_{k}^{1 / 1-\gamma}}{\sum_{m=1}^{K} L_{m}^{1 / 1-\gamma}}
\end{equation}

\subsection{MF-STFnet Training and Inference}
During MF-STFnet training, MF-PolSAR training samples are first input to the BSFE part to capture band-specific semantic features. Then, based on these semantic representations, CIFEM is used to extract cross-band interactive features of each band. For any band, two classification results are respectively obtained from SIC and TPC based on the concatenation of band-specific semantic features and cross-band interactive features. After that, AWF is adopted to combine the results of all bands for the final multi-frequency joint classification decisions of SIC and TPC. Finally, the parameters of MF-STFnet and the adaptive weights of AWF are updated according to the calculated total loss. The above process is repeated continuously until the termination condition of network learning is reached, and the whole MF-STFnet training is completed.  

In the prediction of the whole MF-PolSAR image, the number of prediction samples is much larger than the number of training samples. The TPC part requires a fixed number of samples to construct subgraphs of the same scale as in model training to achieve prediction, thus greatly increasing the time of classification prediction. In contrast, the prediction of SIC is not subject to this restriction, and is simpler and more flexible than TPC. Therefore, the prediction time of SIC is shorter than that of TPC. Moreover, the multi-frequency joint prediction results of SIC and TPC are closer due to the constraint of consistency loss in model training. Therefore, to achieve efficient MF-PolSAR image classification, only the SIC results of each band are combined as the final multi-frequency joint classification prediction. That is, any pixel in the MF-PolSAR image can get the corresponding label prediction only through BSFE, CIFEM, and SIC parts.

\begin{figure}[th] 
\centering
\subfloat[C-band Pauli]{\includegraphics[width=0.3\linewidth,height=2.5cm]{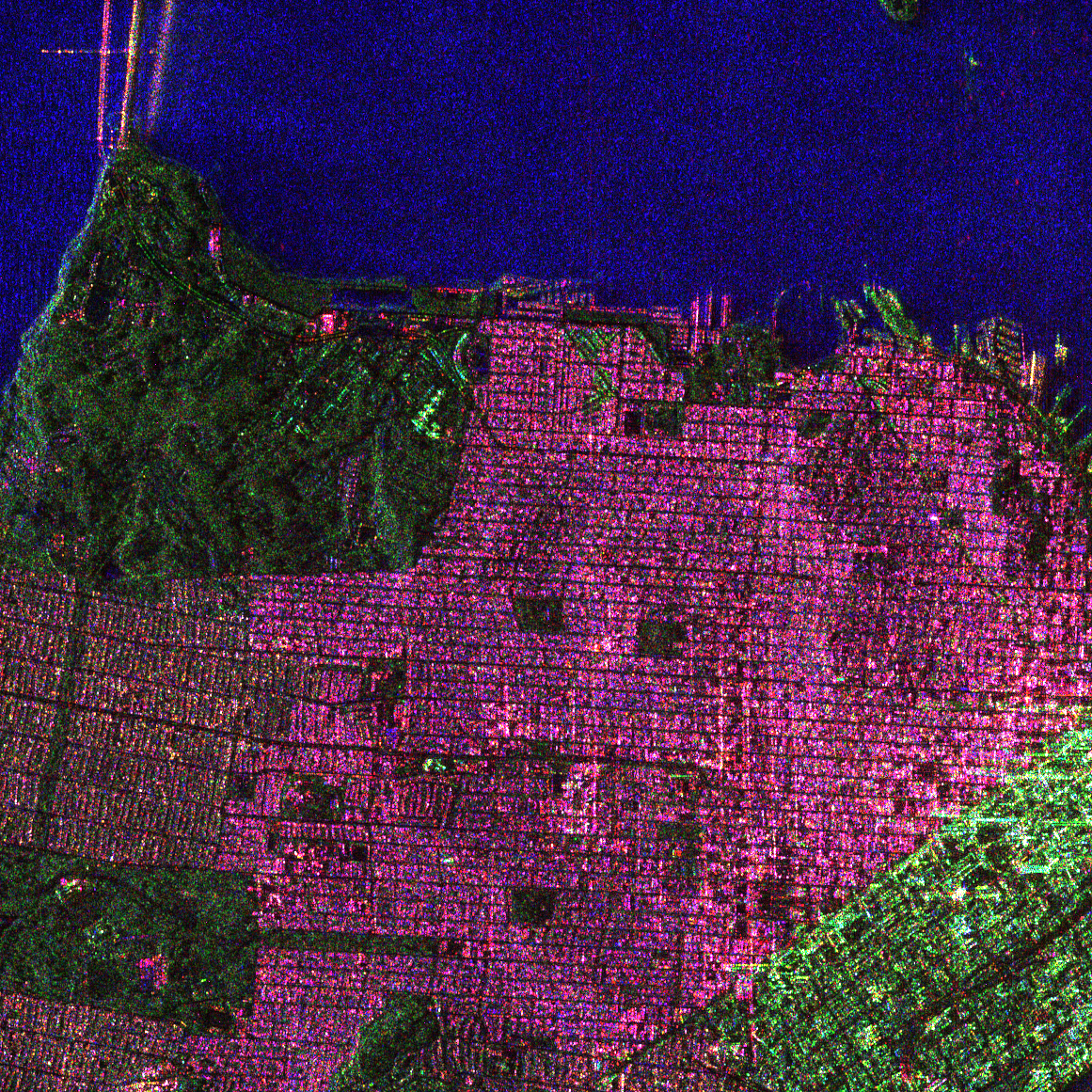}}
\label{}
\subfloat[L-band Pauli]{\includegraphics[width=0.3\linewidth,height=2.5cm]{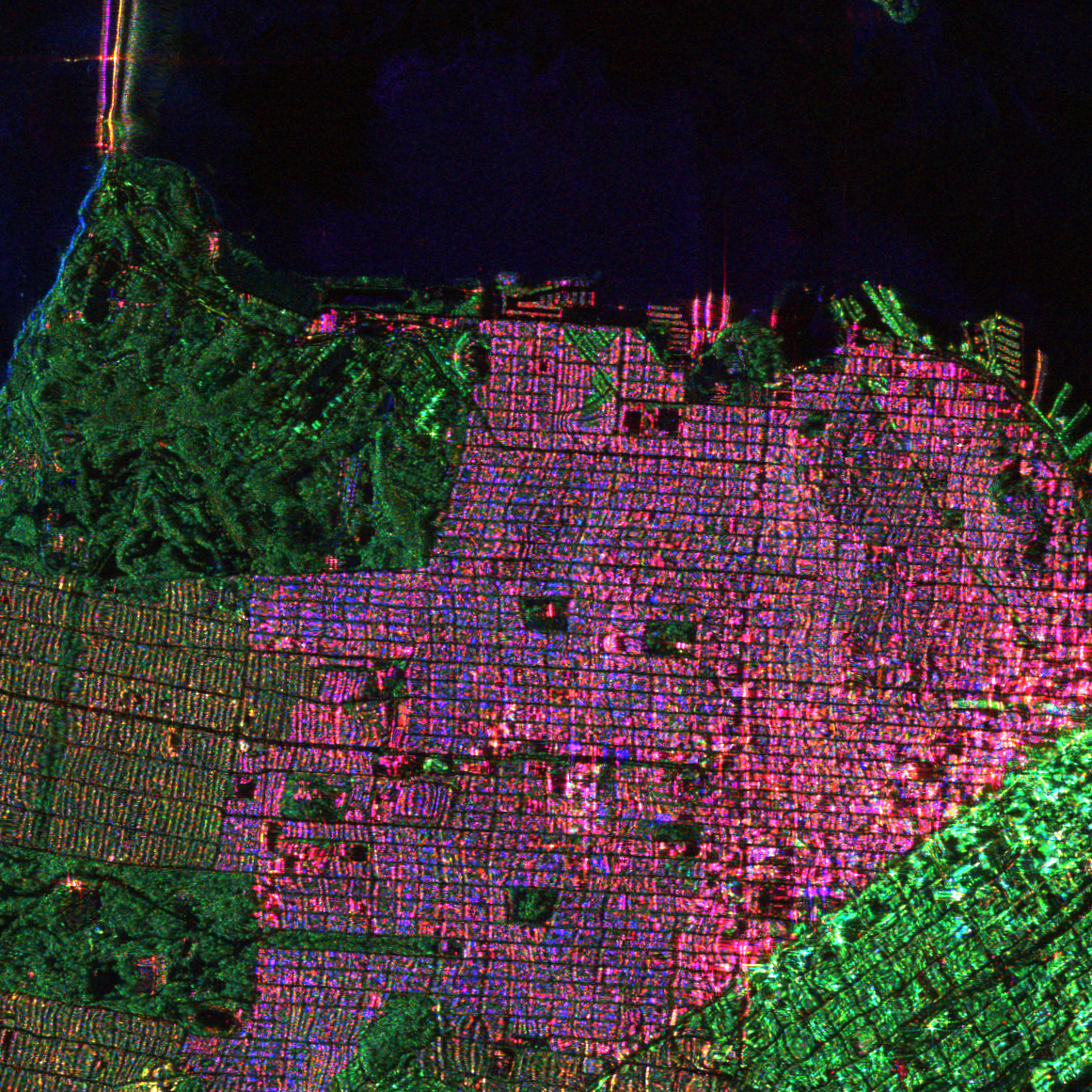}}
\label{}
\subfloat[Ground truth]{\includegraphics[width=0.3\linewidth,height=2.5cm]{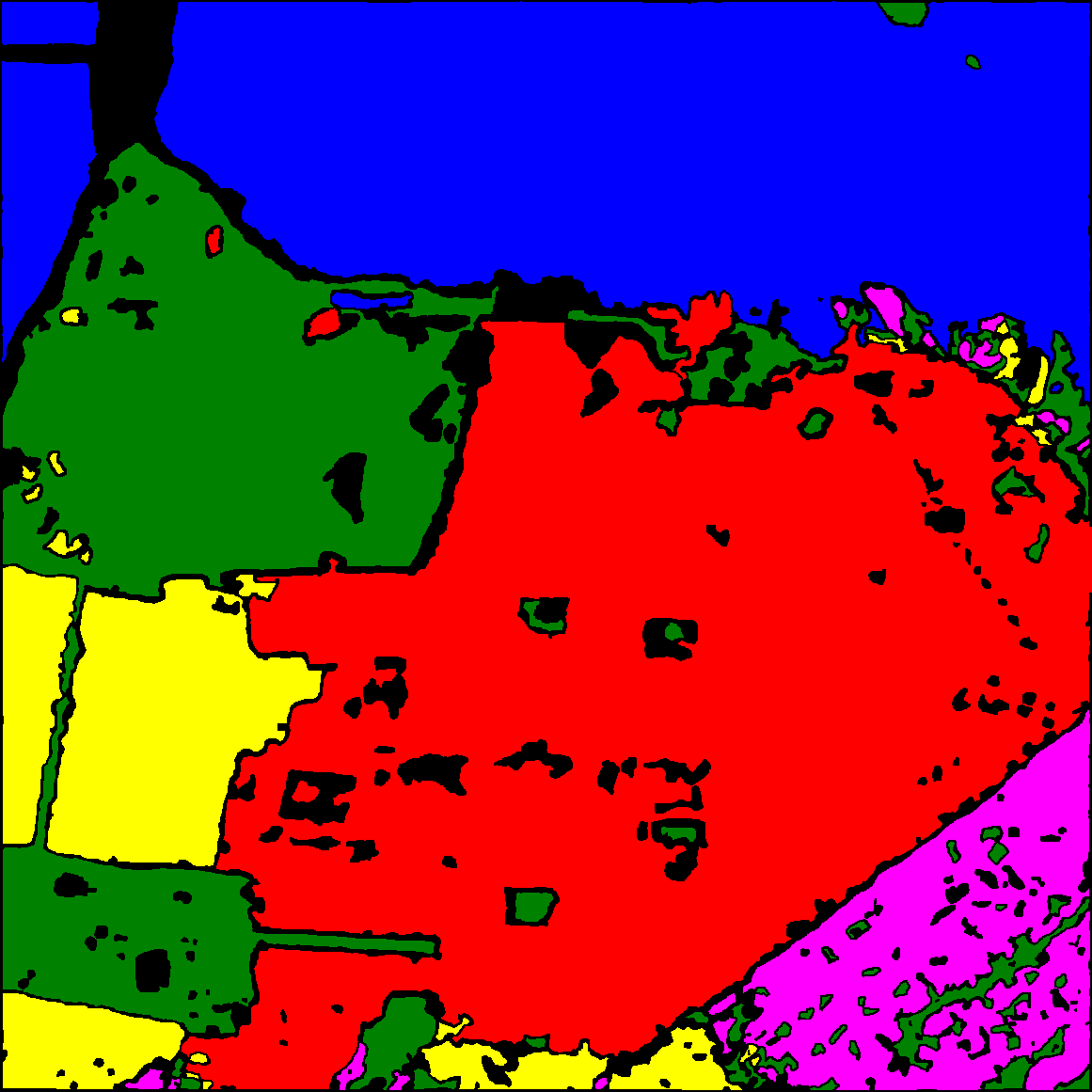}}
\label{}
\subfloat{\includegraphics[width=0.95\linewidth,height=0.5cm]{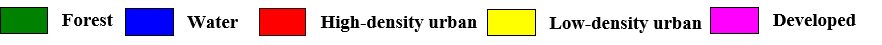}}

\caption{ Dual-frequency full PolSAR image over the San Francisco region.}
\label{figure-SanFran}
\end{figure} 

\begin{figure}[th] 
\centering
\subfloat[S-band Pauli]{\includegraphics[width=0.3\linewidth,height=2.5cm]{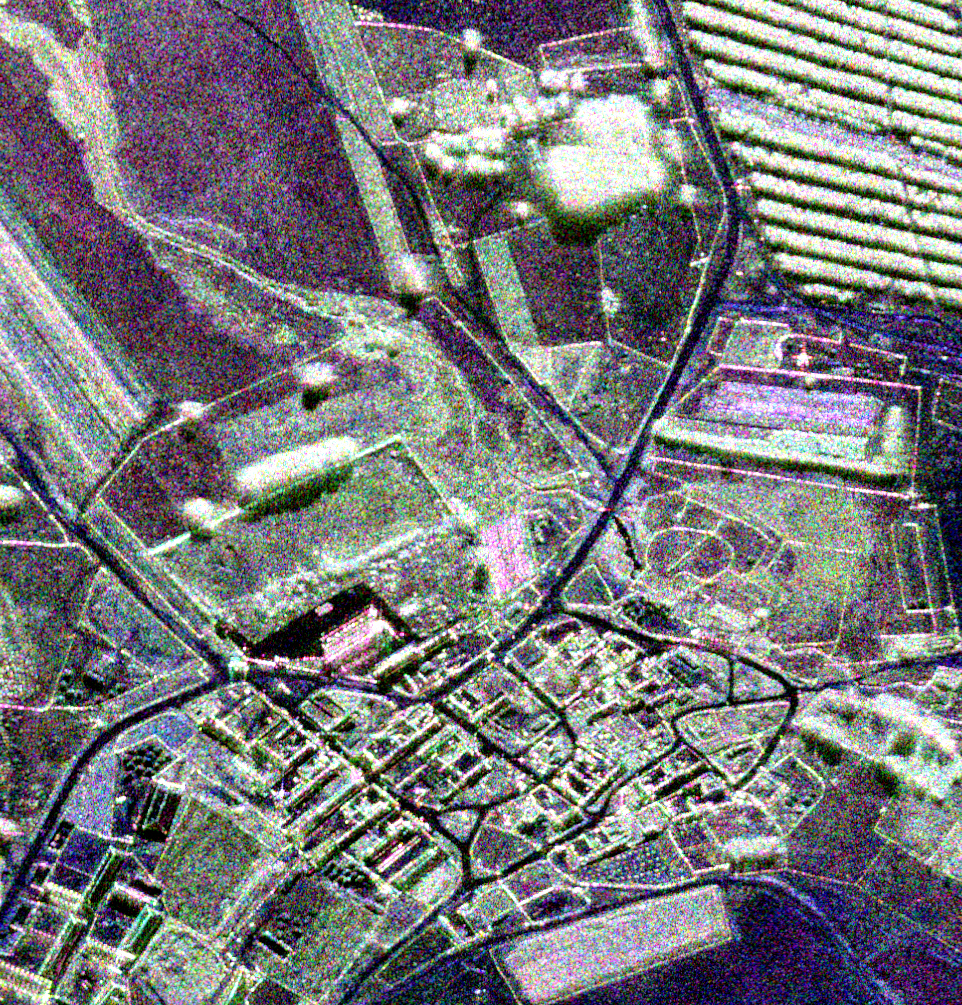}}
\label{}
\subfloat[L-band Pauli]{\includegraphics[width=0.3\linewidth,height=2.5cm]{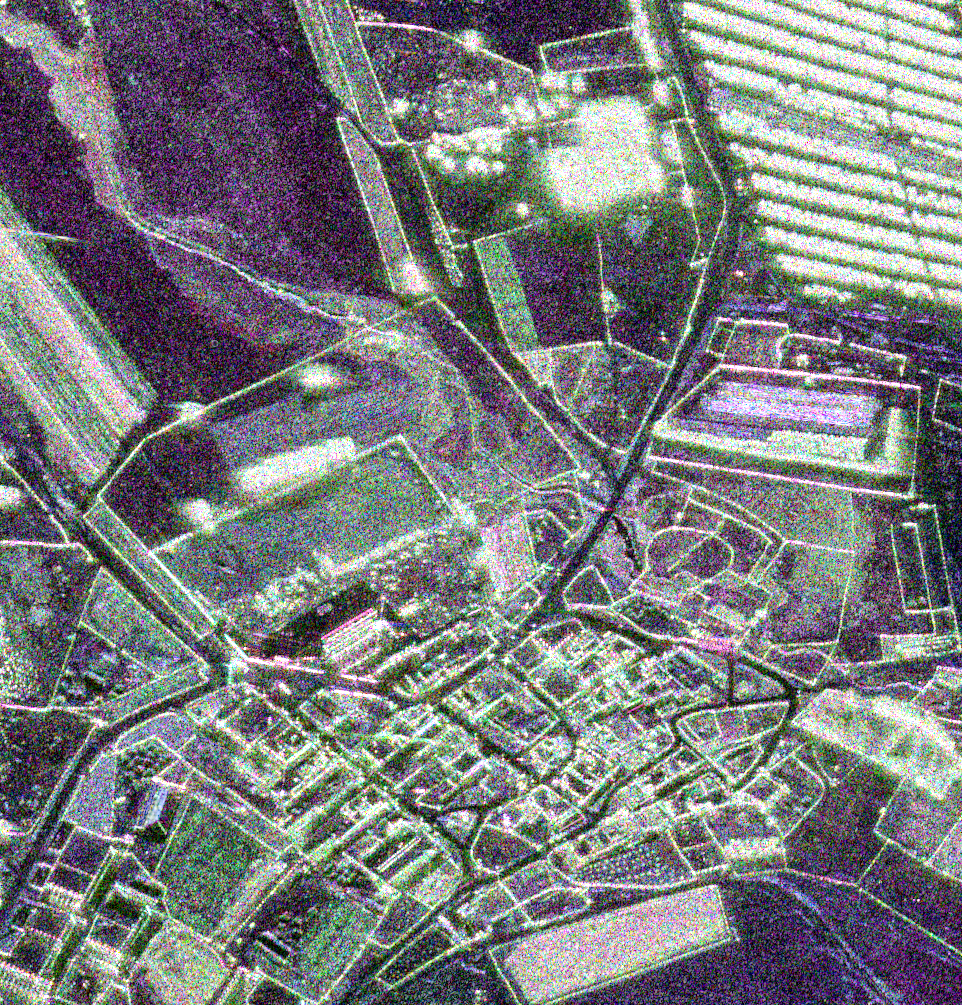}}
\label{}
\subfloat[Ground truth]{\includegraphics[width=0.3\linewidth,height=2.5cm]{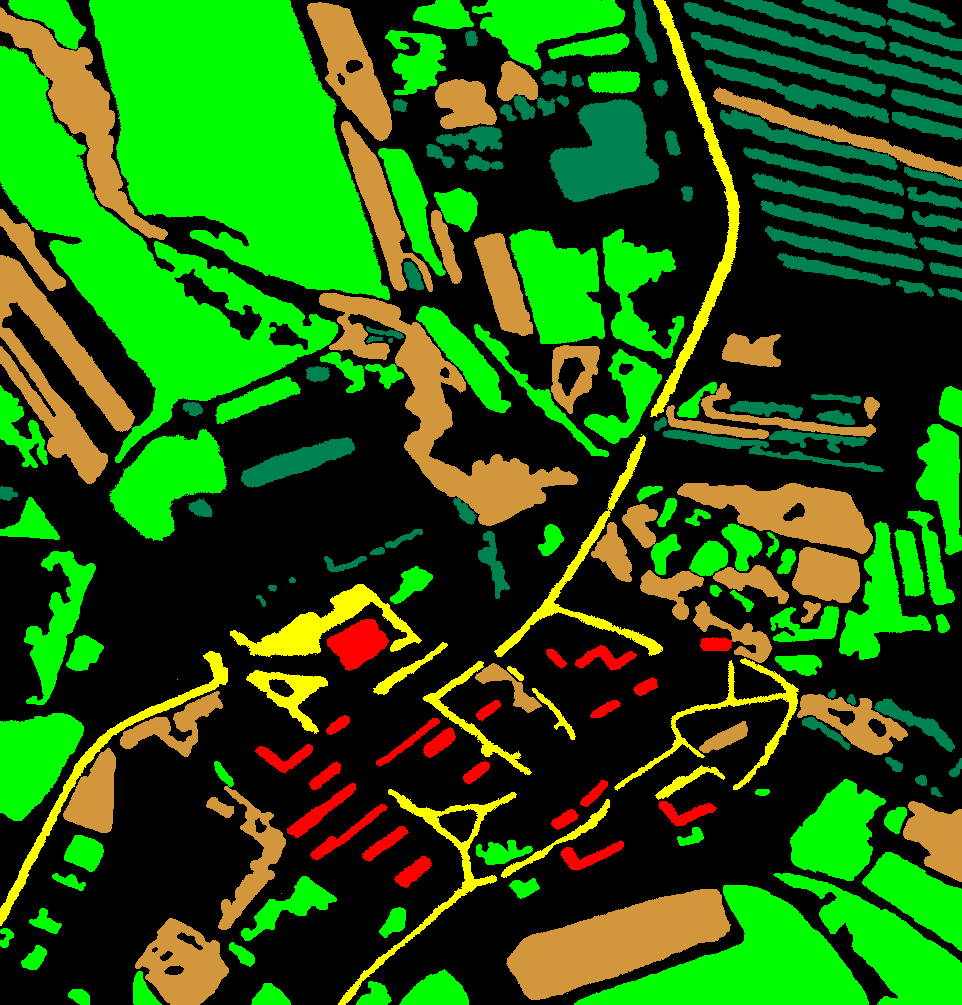}}
\label{}
\subfloat{\includegraphics[width=0.95\linewidth,height=0.5cm]{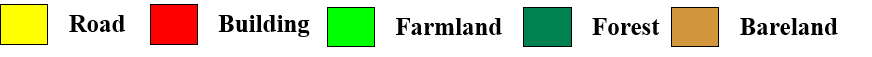}}

\caption{ Dual-frequency full PolSAR image over the Woniupan region.}
\label{figure-Woniupan}
\end{figure} 

\begin{figure}[th] 
\centering
\subfloat[C-band Pauli]{\includegraphics[width=0.4\linewidth,height=3.0cm]{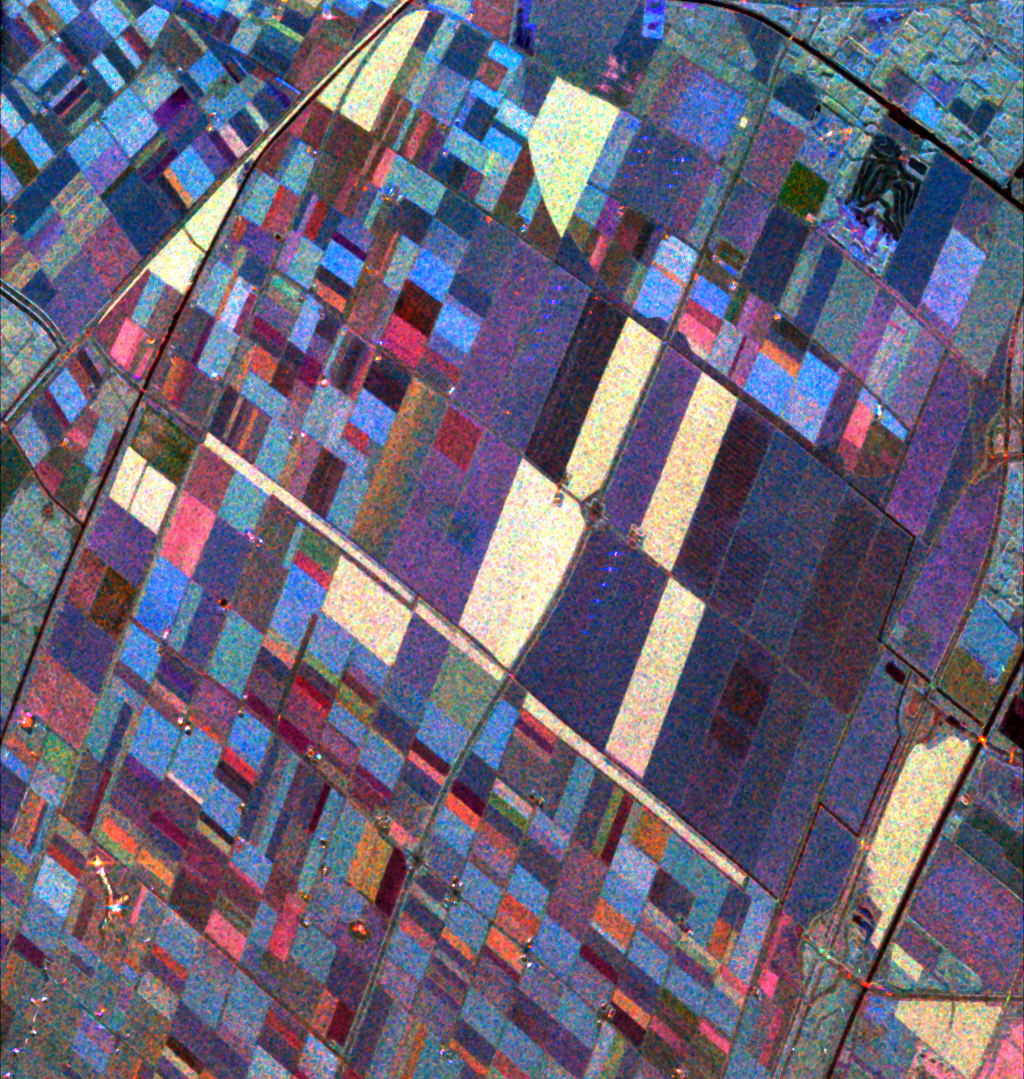}}
\label{}
\subfloat[L-band Pauli]{\includegraphics[width=0.4\linewidth,height=3.0cm]{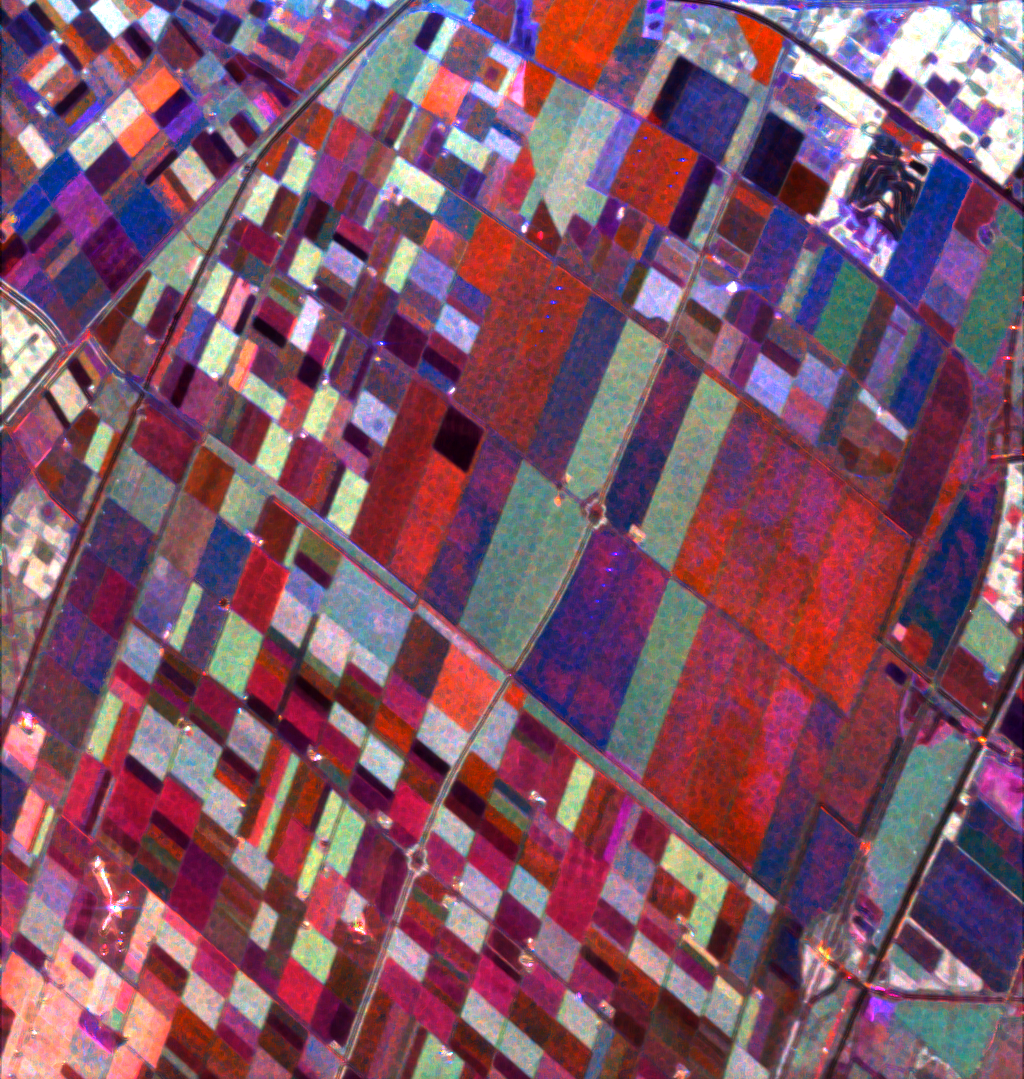}}
\label{}
\subfloat[P-band Pauli]{\includegraphics[width=0.4\linewidth,height=3.0cm]{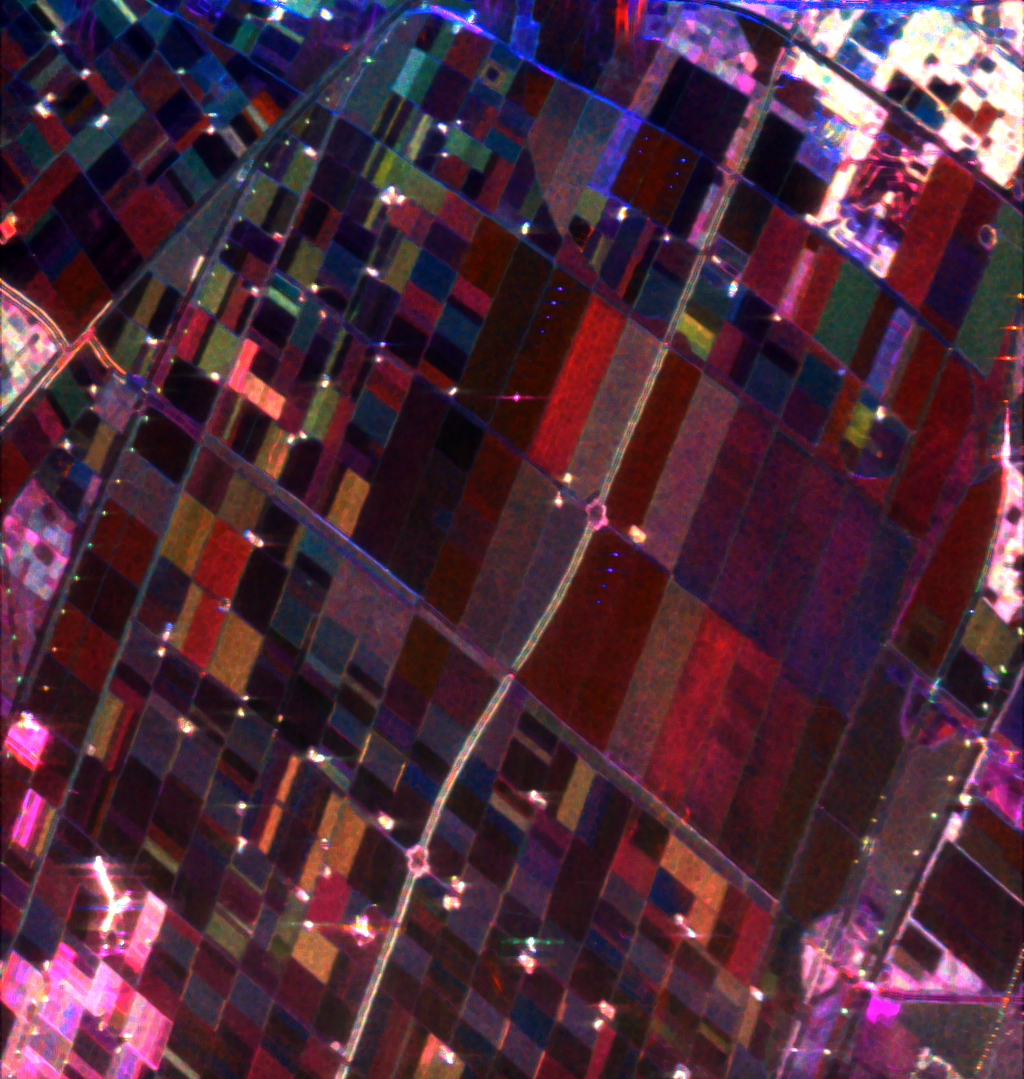}}
\label{}
\subfloat[Ground truth]{\includegraphics[width=0.4\linewidth,height=3.0cm]{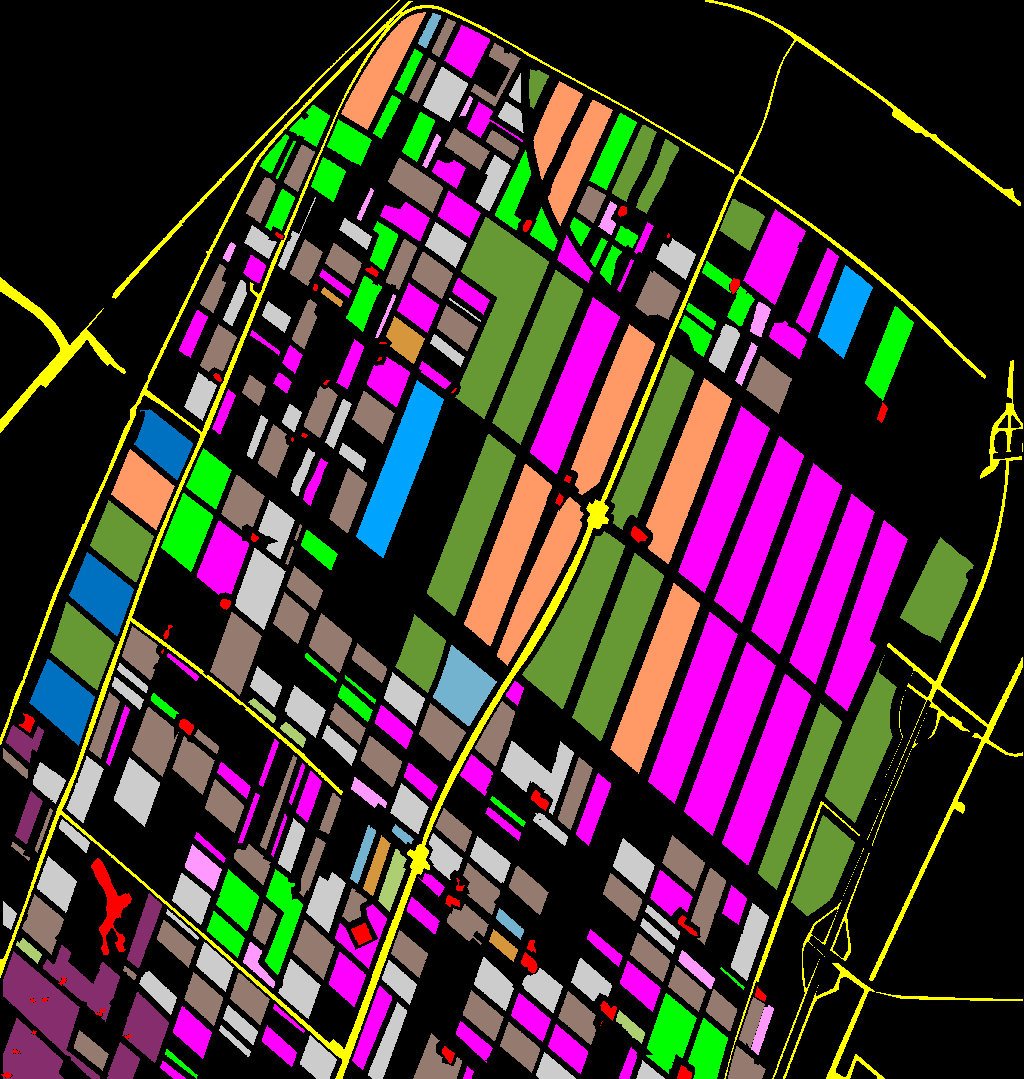}}
\label{}
\subfloat{\includegraphics[width=0.9\linewidth,height=0.7cm]{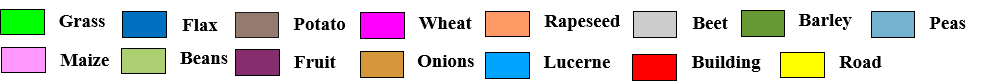}}

\caption{Multi-frequency full PolSAR image over the Flevoland region.}
\label{figure-Flevoland}
\end{figure} 

\section{Experimental Results and Analysis}
In this section, we conduct experiments to evaluate the performance of the proposed MF-STFnet on three measured MF-PolSAR datasets. 

\subsection{Experimental Datasets Description}
The three experimental MF-PolSAR datasets include the dual-band SanFrancisco, the dual-band Woniupan, and the three-band Flevoland datasets.

\subsubsection{SanFrancisco}
The SanFrancisco dataset is composed of the full polarimetric L-band ALOS and C-band GaoFen-3 data. Its image size is 1161 $\times$ 1161. According to the corresponding optical image, the dataset contains 5 categories: Forest, Water, High-density urban, Low-density urban, and Developed. Fig. \ref{figure-SanFran} shows the Pauli RGB images and the ground truth image.

\subsubsection{Woniupan} 
The Woniupan dataset was acquired by an airborne system in 2021, which concludes S- and L-bands. Its image size is 1005 $\times$ 962, and the Pauli RGB images are shown in Fig. \ref{figure-Woniupan}(a) and (b). There are five land cover categories identified in this dataset: Road, Building, Farmland, Forest, and Bareland. Fig. \ref{figure-Woniupan}(c) shows the ground truth image.

\subsubsection{Flevoland}
The Flevoland dataset is acquired by the NASA/JPL AIRSAR system in the C-, L-, and P-bands. It contains 15 categories and has a size of 1079 $\times$ 1024. The Pauli RGB and the ground truth images are shown in Fig. \ref{figure-Flevoland}.

\subsection{Experimental Setup and Evaluation Criteria}
\begin{figure}[htp]%
\centering
\subfloat{\includegraphics[width=8.8cm,height=3.5cm]{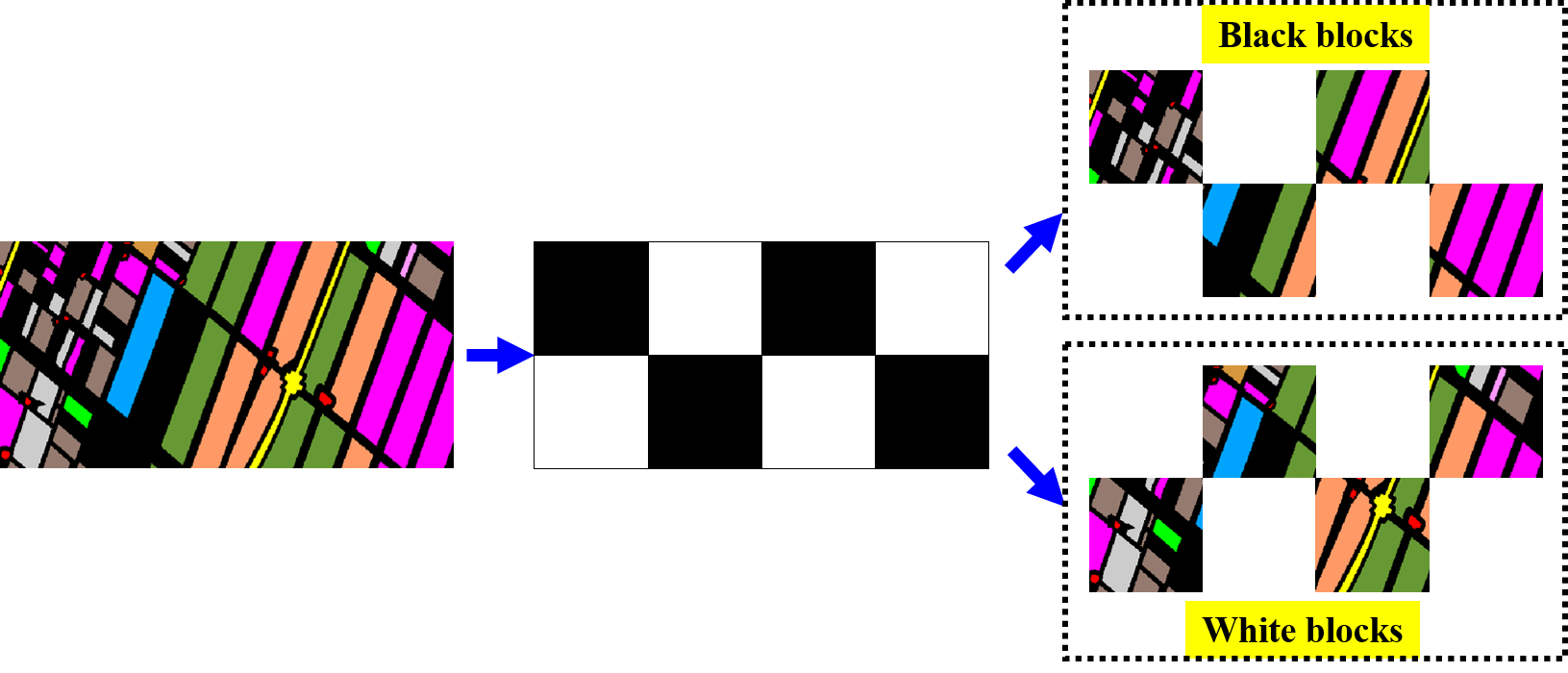}}

\caption{ A simple process of non-overlapping data partition.}
\label{figure-chessboard}
\end{figure}

For the non-overlapping collection of the training and test samples, we adopt the chessboard-like sampling strategy \cite{Liang-2020-aware} rather than the random sampling strategy. The chessboard-like sampling strategy can greatly reduce the spatial correlation between training and test data, thereby making the classification results not overly optimistic and making algorithms more suitable for practical applications. Its key step is the non-overlapping data partition, whose simple demonstration is shown in Fig. \ref{figure-chessboard}. In Fig. \ref{figure-chessboard}, the large image is segmented into non-overlapping chessboard-like sub-images. Among them, the sub-images corresponding to all-black blocks are one part, and the sub-images corresponding to all-white blocks are the other part. The two non-overlapping parts are alternately used as training data and test data. The results obtained when each part is used as test data are stitched as the final classification of the entire image. Specifically, for each experimental image, 400 nonoverlapping small images of the same size are divided. We first take 200 of them as training images and another 200 for the test. Then, do the same after swapping. We randomly choose 200 labeled pixels in training images to learn networks, and stitch the two test results as the final classification.

In addition, to avoid overfitting, five data augmentation strategies are utilized to expand the training dataset, including horizontal flip, vertical flip, random $90^{\circ}$, $180^{\circ}$, and $270^{\circ}$ rotations. The center pixels with 13 $\times$ 13 neighborhoods are used as the input patches. Adam with momentum 0.9 is used to update network parameters. The training epoch is set to 150 for all datasets. The learning rate and the batch size are set to 0.001 and 100, respectively. The proposed MF-STFnet is implemented on the PyTorch framework. All the experiments were run on a Lenovo Y720 cube gaming PC with an Intel Core i7-7700 CPU, an Nvidia GeForce GTX 1080 GPU, and 16GB RAM under Ubuntu 20.04 LTS operating system. 

To quantify the classification performance, the accuracy of each category, overall accuracy (OA), average accuracy (AA), and kappa coefficient ($\kappa$) are employed as evaluation metrics. In addition, to ensure fairness and reliability, all experiments are conducted ten times, and the mean values of evaluation metrics are reported.

\subsection{Performance Analysis of the Proposed MF-STFnet}

\subsubsection{Effect of the Regularization Parameter $\lambda$}
\begin{figure}[htp]%
\centering
\subfloat{\includegraphics[width=6cm,height=5cm]{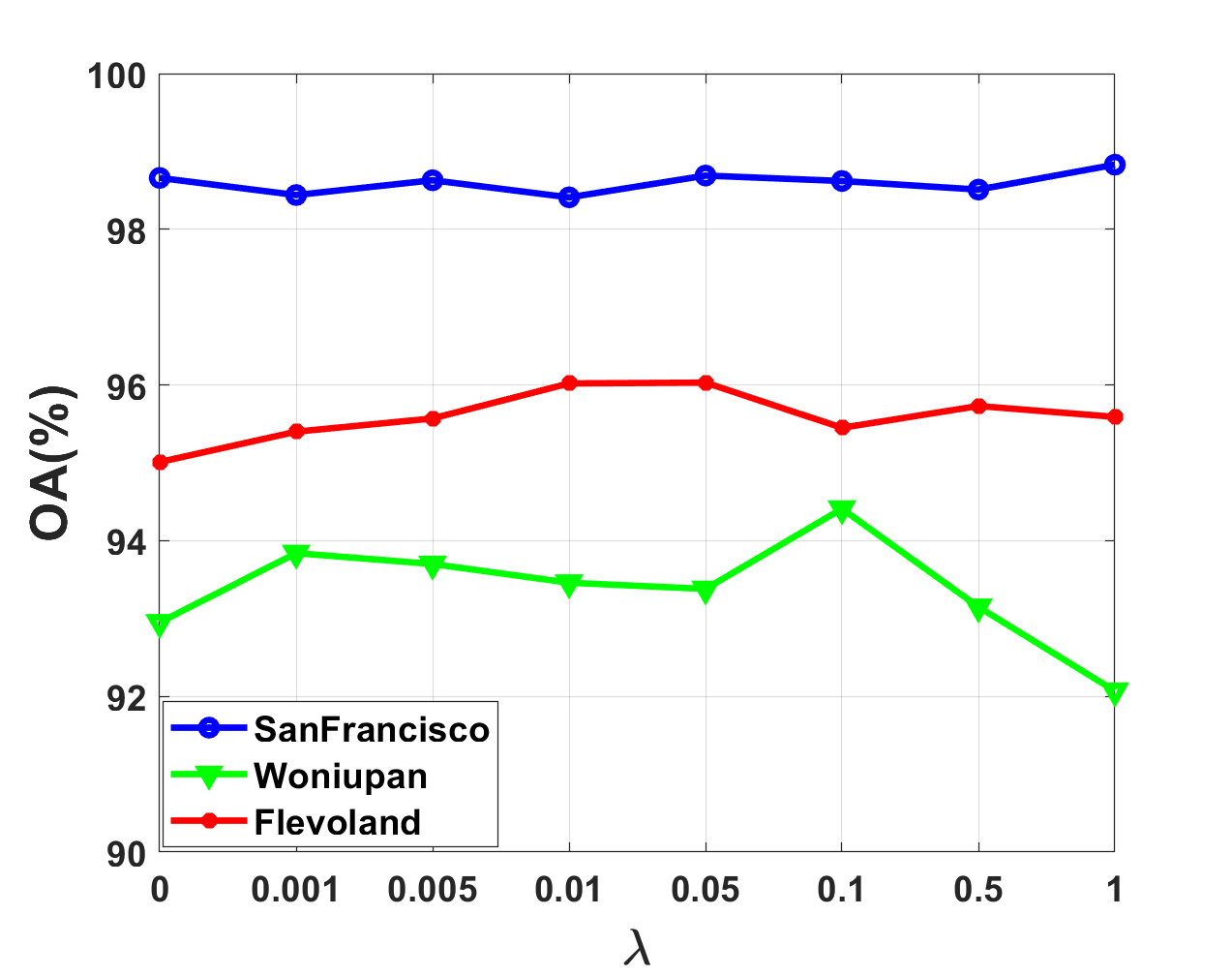}}

\caption{Classification accuracy of the proposed MF-STFnet with different values of the regularization parameter $\lambda$ on the three experimental datasets.}
\label{figure-lambda}
\end{figure}

The regularization parameter $\lambda$ is used to adjust the weight of the consistency loss in the total loss. We vary $\lambda$ to 0, 0.001, 0.005, 0.01, 0.05, 0.1, 0.5, and 1 to explore its influence on classification results. Fig. \ref{figure-lambda} exhibits the relationship between OA and different $\lambda$. Here, the power exponent parameter $\gamma$ is fixed to 3 for all experimental datasets.  

As shown in Fig. \ref{figure-lambda}, for the SanFrancisco and Flevoland datasets, the value of $\lambda$ has few effects on the classification performance. However, for the Woniupan dataset, when the value of $\lambda$ is too large, the classification performance will decline significantly. Therefore, to ensure the generalization, the regularization parameter $\lambda$ for the three experimental datasets is all selected as 0.1.

\begin{table*}[htp]
\centering
\caption{Overall Accuracy (\%) of the Proposed MF-STFnet with Different $\gamma$}
\label{table-gamma} 
\fontsize{8}{12}\selectfont 
\begin{tabular}{|c|c|c|c|c|c|c|c|c|c|c|c|c|c|c|} \hline
\textbf{{$\gamma$}} & \textbf{1.5} & \textbf{2} & \textbf{2.5} & \textbf{3} & \textbf{3.5} & \textbf{4} & \textbf{4.5} & \textbf{5} & \textbf{5.5} & \textbf{6}  & \textbf{6.5} & \textbf{7} & \textbf{7.5} & \textbf{8}\\ \hline

\textbf{SanFrancisco} & 97.77 & 98.26 & 98.38 & 98.62 & \textbf{98.76} & 98.51 & 98.68  & 98.09  & 98.69  & 98.46  & 98.66  & 85.36 & 76.39  & 44.01 \\ \hline

\textbf{Woniupan} &91.16 & 92.82 &93.93  & 94.41 &93.92 & \textbf{95.41} &94.62  & 92.44 &91.46  & 89.75 &87.70 & 70.02 & 64.84 & 61.64 \\ \hline

\textbf{Flevoland} & 94.65 & 95.44 & 95.39 & \textbf{95.45} & 95.41 & 93.97 & 56.01 & 43.92 & 21.03 & 8.76 & 8.02 & 3.30 & 7.98 & 3.61 \\ \hline

\end{tabular}
\end{table*}

\begin{table*}[htp] 
\centering  
\fontsize{9}{11}\selectfont  
\caption{Classification Performance of Different Networks on Three MF-PolSAR Datasets}  
\label{table-ablation}  
\begin{tabular}{c|c|c|c|c|c} \hline
\toprule[0.3pt]

\multicolumn{1}{c|}{\multirow{2}{*}{ }}
& \multicolumn{1}{c|}{\multirow{2}{*}{\bf CIFEM}}
& \multicolumn{1}{c|}{\multirow{2}{*}{\bf TPC}}
& \multicolumn{2}{c}{\bf OA(\%) / AA(\%) / \textbf{$\kappa$} }  \\ \cline{4-6}

\multicolumn{1}{c|}{}
& \multicolumn{1}{c|}{}
& \multicolumn{1}{c|}{}
& \multicolumn{1}{c|}{\bf SanFrancisco}&{\bf Woniupan}&{\bf Flevoland} \\ \hline 
{Net1} &{$\times$} &{$\times$}&{97.50 / 97.38 / 0.9656} &{90.21 / 87.61 / 0.8509} &{92.65 / 87.21 / 0.9156}  \cr \hline 
{Net2} &{\checkmark} &{$\times$} &{98.29 / 98.28 / 0.9765} &{93.17 / 91.86 / 0.8960} &{94.16 / 90.65 / 0.9329} \cr \hline 
{Net3} &{$\times$} &{\checkmark} &{98.64 / 98.23 / 0.9817} &{92.62 / 91.37 / 0.8871} &{93.13 / 88.09 / 0.9211} \cr \hline 
{MF-STFnet} &{\checkmark} &{\checkmark} &{\textbf{98.76} / \textbf{98.69} / \textbf{0.9830}} &{\textbf{95.41} / \textbf{93.02} / \textbf{0.9290}} &{\textbf{95.45} / \textbf{92.36} / \textbf{0.9477}} \cr
\toprule[0.3pt]
\end{tabular}
\end{table*}

\begin{figure*}[htp] 
\centering
\subfloat[Net1]{\includegraphics[width=0.18\linewidth,height=2.4cm]{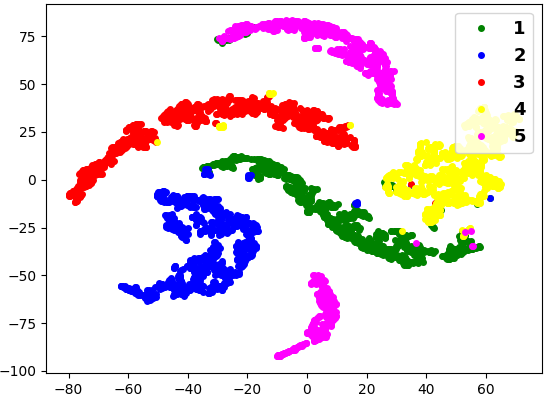}}
\label{Woniu-Net1}
\subfloat[Net\_common]{\includegraphics[width=0.18\linewidth,height=2.4cm]{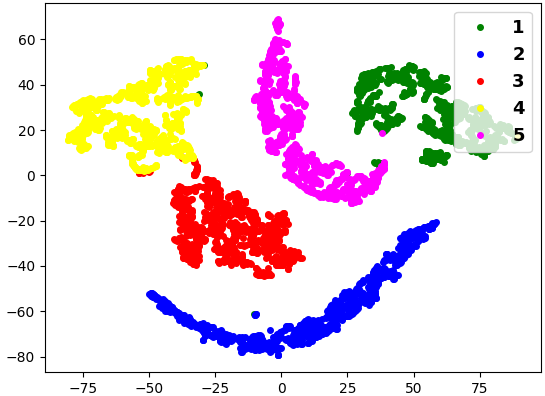}}
\label{Woniu-Netcommon}
\subfloat[Net2]{\includegraphics[width=0.18\linewidth,height=2.4cm]{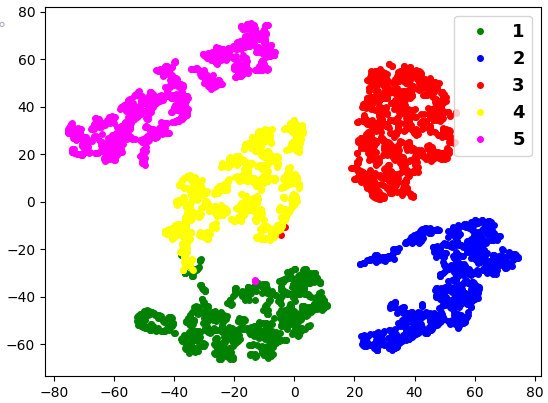}}
\label{Woniu-Net2}
\subfloat[Net3]{\includegraphics[width=0.18\linewidth,height=2.4cm]{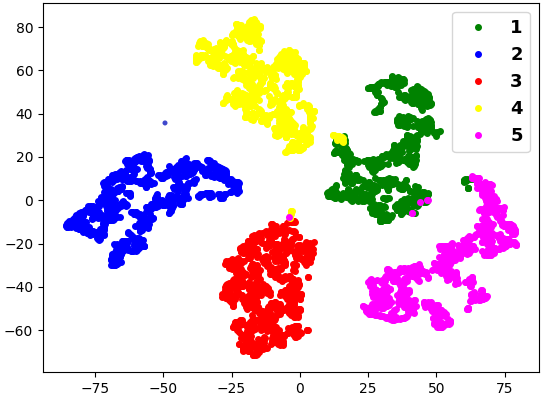}}
\label{Woniu-Net3}
\subfloat[MF-STFnet]{\includegraphics[width=0.18\linewidth,height=2.4cm]{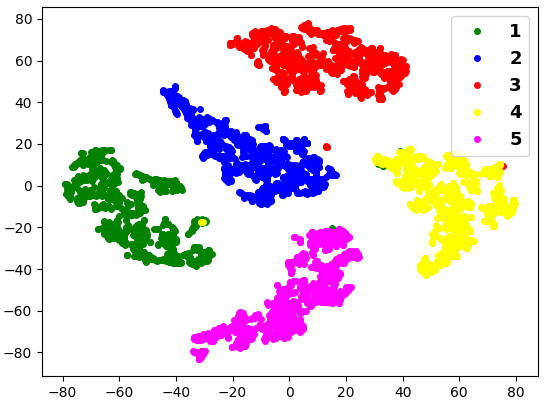}}
\label{Woniu-MF-STFnet}
\subfloat[Net1]{\includegraphics[width=0.18\linewidth,height=2.4cm]{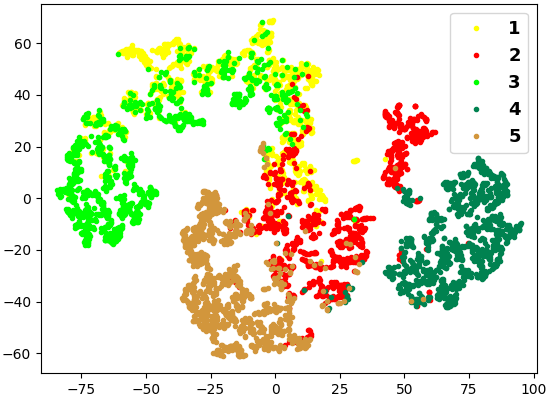}}
\label{Woniu-Net1}
\subfloat[Net\_common]{\includegraphics[width=0.18\linewidth,height=2.4cm]{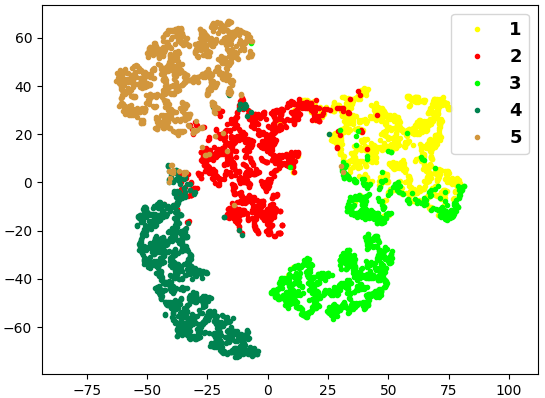}}
\label{Woniu-Netcommon}
\subfloat[Net2]{\includegraphics[width=0.18\linewidth,height=2.4cm]{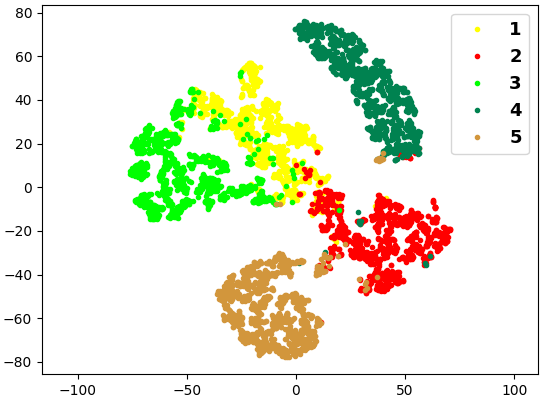}}
\label{Woniu-Net2}
\subfloat[Net3]{\includegraphics[width=0.18\linewidth,height=2.4cm]{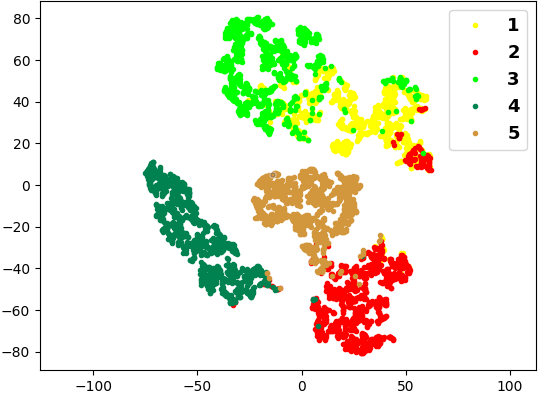}}
\label{Woniu-Net3}
\subfloat[MF-STFnet]{\includegraphics[width=0.18\linewidth,height=2.4cm]{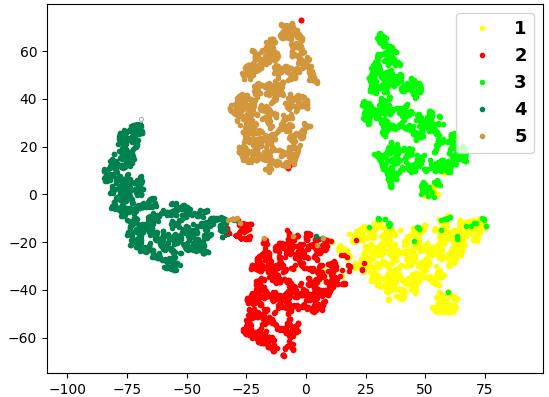}}
\label{Woniu-MF-STFnet}
\subfloat[Net1]{\includegraphics[width=0.18\linewidth,height=2.5cm]{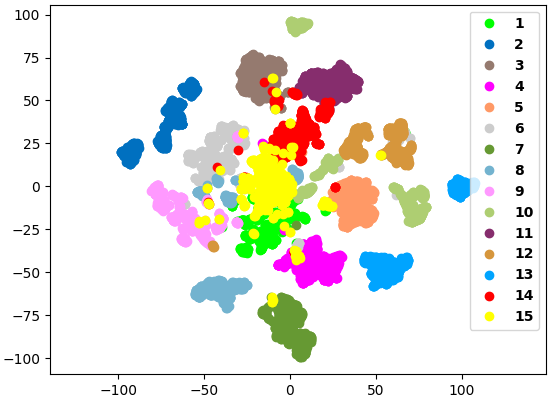}}
\label{Woniu-Net1}
\subfloat[Net\_common]{\includegraphics[width=0.18\linewidth,height=2.5cm]{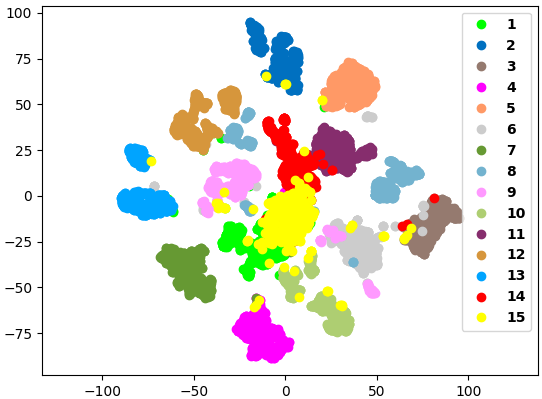}}
\label{Woniu-Netcommon}
\subfloat[Net2]{\includegraphics[width=0.18\linewidth,height=2.5cm]{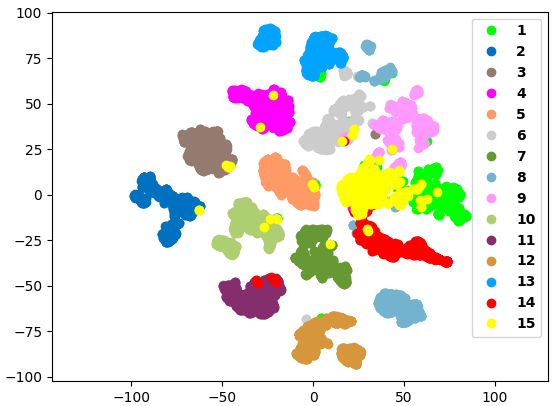}}
\label{Woniu-Net2}
\subfloat[Net3]{\includegraphics[width=0.18\linewidth,height=2.5cm]{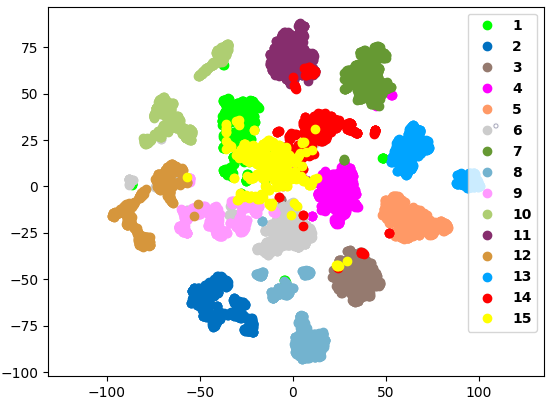}}
\label{Woniu-Net3}
\subfloat[MF-STFnet]{\includegraphics[width=0.18\linewidth,height=2.5cm]{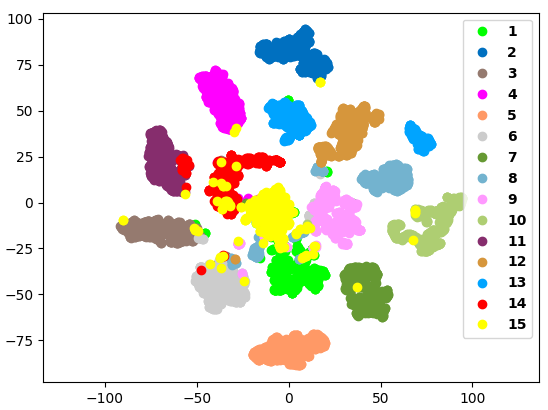}}
\label{Woniu-MF-STFnet}

\caption{The t-SNE visualization of features extracted by different classification networks on three MF-PolSAR datasets. (a)-(d) show the results on the SanFrancisco dataset, (e)-(h) show the results on the Woniupan dataset, and (i)-(l) show the results on the Flevoland dataset.}
\label{figure-t-SNE}
\end{figure*} 

\subsubsection{Effect of the Power Exponent Parameter $\gamma$}
The power exponent parameter $\gamma$ is one of the key parameters affecting the proposed MF-STFnet performance. As mentioned earlier, the value of $\gamma$ must be greater than 1. Therefore, for the three experimental datasets, the optimal $\gamma$ is selected from \{1.5, 2, 2.5, 3, 3.5, 4, 4.5, 5, 5.5, 6, 6.5, 7, 7.5, 8\}. Table \ref{table-gamma} shows the classification accuracy under the varying $\gamma$ on these datasets. Notably, here, the regularization parameter $\lambda$ is fixed to 0.1 for all experimental datasets.  
  
As shown in Table \ref{table-gamma}, for the SanFrancisco dataset, the results show that MF-STFnet achieves the highest results when $\gamma$ is 3.5. For the Woniupan dataset, the best classification index is obtained when $\gamma$ is 4. In addition, for the Flevoland dataset, the best result is realized when $\gamma$ is 3. Therefore, $\gamma=3.5$, $\gamma=4$, and $\gamma=3$ are the best choice for the three datasets, respectively.

\subsubsection{Ablation Experiment}
As analyzed earlier, the proposed MF-STFnet mainly relies on CIFEM and TPC to enhance the classification performance. Therefore, we conduct the related ablation experiment to verify the effectiveness of the two key components. Table \ref{table-ablation} shows the ablation results, where $\sqrt{ }$ and $\times$ respectively denote with and without this part. 

As shown in Table \ref{table-ablation}, it is obvious that the complete MF-STFnet model with all key parts achieves the highest results. Without CIFEM and TPC, Net1 achieves the lowest classification results on all datasets. When considering CIFEM, for the three datasets, Net2 outperforms Net1 by 0.79\%, 2.95\%, and 1.51\% on OA, respectively. This result safely demonstrates the superiority of capturing interactive information between bands. In addition, compared with Net1, Net3 with TPC achieves 1.14\%, 2.4\%, and 0.48\% improvements on OA for the three datasets, which demonstrates the effectiveness of nonlocal topological information in improving classification performance. It should be noted that MF-STFnet outperforms Net2 and Net3. This suggests that the simultaneous extraction of interactive and topological information can provide more comprehensive information for classification interpretation. To sum up, the experimental results shown in Table \ref{table-ablation} demonstrate the effectiveness of CIFEM and TPC. In addition, they also demonstrate the proposed MF-STFnet can obtain a more accurate classification due to simultaneously considering interactive and topological information.

\subsubsection{Validation of CIFEM}
To further verify the effectiveness of the CIFEM part, we construct a comparison module composed of two common convolution blocks, whose input is the concatenated features of different bands output by the BSFE part. The convolution output dimensions are set to 64 and 32. The comparison network using this module is called Net\_common. Table \ref{table-CIFEM} shows the classification results of Net\_common and Net2. In addition, Fig. \ref{figure-t-SNE} shows the t-SNE visualization of features extracted by Net\_common and Net2. Notably, these features are the multi-frequency joint output features of networks in the prediction stage. 

\begin{table}[thp]
\centering
\caption{Classification Results of Different Networks on Three MF-PolSAR Datasets}
\label{table-CIFEM}
\fontsize{7}{9}\selectfont
\begin{tabular}{|c|c|c|c|c|} \hline
\bf{Dataset} & \bf{Network} & \bf{OA(\%)} & \bf{AA(\%)} & \bf{Kappa}\\\hline
\multirow{2}{*}{SanFrancisco} & Net\_common & 97.99 & 97.69 & 0.9725 \\ \cline{2-5}
\multirow{2}{*}{} & Net 2 & 98.29 & 98.28 & 0.9765 \\ \cline{2-5} \hline

\multirow{2}{*}{Woniupan} & Net\_common & 91.19 & 89.21 & 0.8726 \\ \cline{2-5}
\multirow{2}{*}{} & Net 2 & 93.17 & 91.86 & 0.8960 \\ \cline{2-5} \hline

\multirow{2}{*}{Flevoland} & Net\_common & 93.19 & 90.02 & 0.9218 \\ \cline{2-5}
\multirow{2}{*}{} & Net 2 & 94.16 & 90.65 & 0.9329 \\ \cline{2-5} \hline
\end{tabular}
\end{table}  

As shown in Table \ref{table-CIFEM}, Net2 outperforms Net\_common on all datasets. Compared with Net\_common, Net2 could increase OA by 0.3\%-1.98\%, AA by 0.59\%-2.65\%, and Kappa by 0.004-0.0234. In addition, as illustrated in Fig. \ref{figure-t-SNE}, the feature distribution between categories in Net2 is more separable than in Net\_common. Specifically, for the SanFrancisco dataset, the feature distribution clusters of Categories 1 and 2 in Fig. \ref{figure-t-SNE}(c) are more compact than those in Fig. \ref{figure-t-SNE}(b). For the Woniupan dataset, Categories 4 and 5 in Fig. \ref{figure-t-SNE}(h) can be better distinguished from other categories than in Fig. \ref{figure-t-SNE}(g). For the Flevoland dataset, compared with Fig. \ref{figure-t-SNE}(l), the distribution overlap of Category 15 with other categories in Fig. \ref{figure-t-SNE}(m) is reduced. Furthermore, the feature distribution within categories in Fig. \ref{figure-t-SNE}(m) is more compact than in Fig. \ref{figure-t-SNE}(l). Therefore, summarizing the above analysis, we can conclude that the proposed CIFEM has advantages over common convolution and is more effective for improving the classification performance.

\subsubsection{Validation of TPC}
Fig. \ref{figure-t-SNE} also visualizes the feature extracted by Net1, Net3, and the proposed MF-STFnet to validate the importance of TPC. 

As shown in Fig. \ref{figure-t-SNE}, for the three datasets, the feature discrimination between categories in Net3 is improved than in Net1. Comparing Fig. \ref{figure-t-SNE}(a) with Fig. \ref{figure-t-SNE}(d), it can be seen that the compactness within categories in Fig. \ref{figure-t-SNE}(d) is better than in Fig. \ref{figure-t-SNE}(a). For the Woniupan dataset, compared with Fig. \ref{figure-t-SNE}(f), there are clear distribution boundaries between categories in Fig. \ref{figure-t-SNE}(i), and Categories 2 and 4 in Fig. \ref{figure-t-SNE}(i) can be better distinguished from other categories. In addition, the distances between category distribution clusters in Fig. \ref{figure-t-SNE}(n) are relatively farther than in Fig. \ref{figure-t-SNE}(k). These results illustrate that the features extracted by Net3 are more discriminative than by Net1, which can prove that the use of TPC is effective for better category distinction and improvement of classification performance. Notably, for each dataset, the feature distribution obtained by MF-STFnet is better than other networks, which indicates that the network structure we constructed is reasonable and effective, and the proposed modules make the learned features of MF-STFnet more discriminative.

\begin{figure}[th] 
\centering

\subfloat[SanFrancisco]{\includegraphics[width=0.33\linewidth,height=2.5cm]{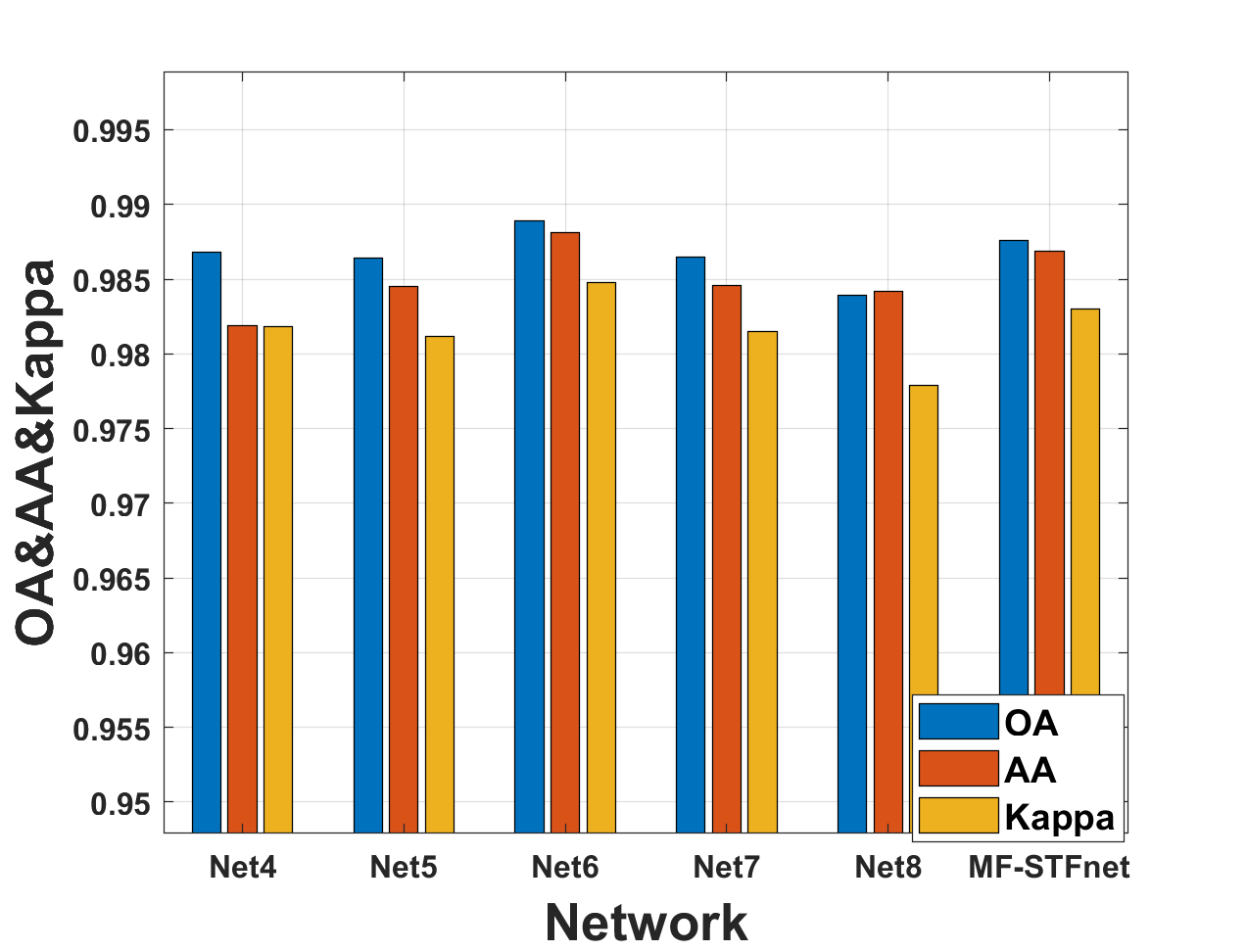}}
\subfloat[Woniupan]{\includegraphics[width=0.33\linewidth,height=2.5cm]{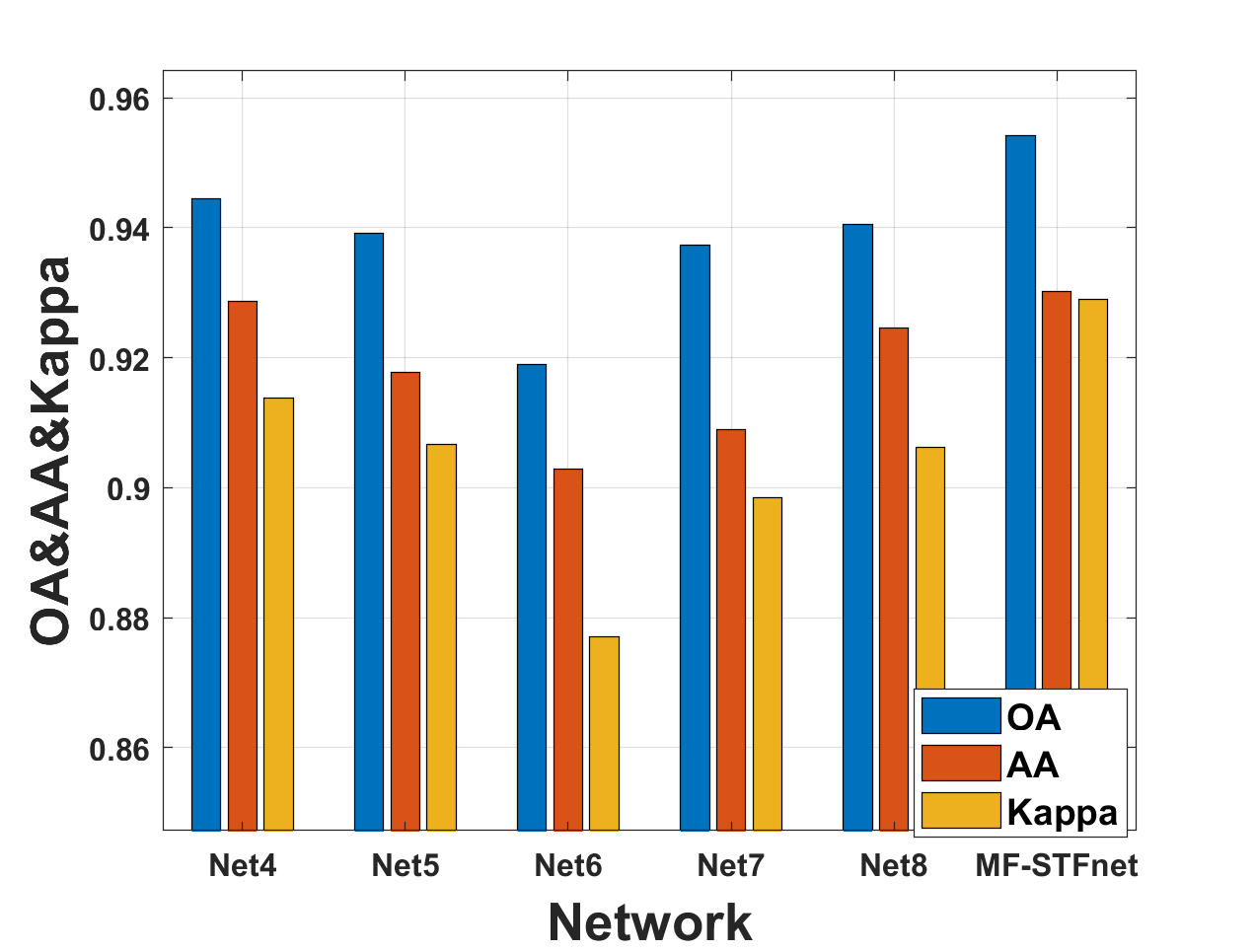}}
\subfloat[Flevoland]{\includegraphics[width=0.33\linewidth,height=2.5cm]{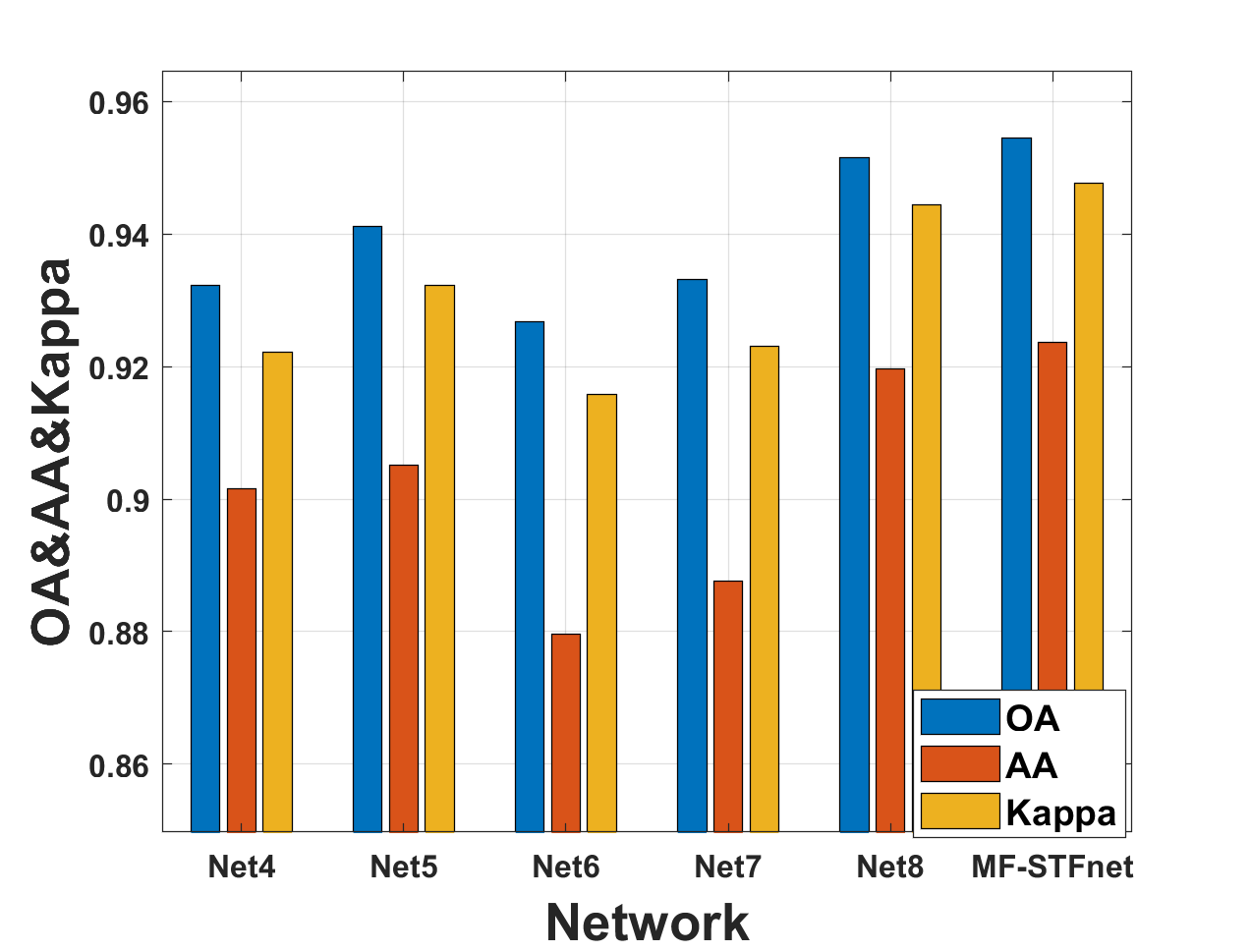}}

\caption{Comparison of classification results between the AWF strategy and other fusion strategies on the three experimental datasets.}
\label{figure-fusion}
\end{figure} 

\subsubsection{Comparison of Fusion Strategies}
Several classical feature fusion strategies are selected to replace the AWF strategy in MF-STFnet for comparison, including concatenation fusion, maximization fusion, multiplication fusion, and summation fusion. The corresponding networks are represented by Net4, Net5, Net6, and Net7, respectively. In addition, based on AWF, the equal weight summation fusion is also compared, which is denoted by Net8. Fig. \ref{figure-fusion} summarizes the classification results on three experimental datasets.

Comparing Net4, Net5, Net6, and Net7, it is found in Fig. \ref{figure-fusion} that the classical fusion strategies for different datasets to achieve the best classification results are different. Specifically, for the three datasets, the optimal classical fusions are multiplication, concatenation, and maximization, respectively. Notably, for all datasets, the proposed AWF strategy can achieve comparable or even better classification results than the optimal classical fusions. For example, for the SanFrancisco dataset, the multiplication fusion holds the highest-level classification performance compared to other fusions. Meanwhile, it is found that the classification results of the proposed AWF strategy differ the least from the best results. For the Woniupan and Flevoland datasets, the proposed AWF strategy has slight classification advantages over other fusion strategies. Furthermore, with the adaptive adjustment of weights, MF-STFnet is better than Net8. These prove that our AWF strategy is effective and rather stable, which is conducive to adaptively obtaining more robust classification results.

\begin{figure}[th] 
\centering
\subfloat[SanFrancisco]{\includegraphics[width=0.33\linewidth,height=2.5cm]{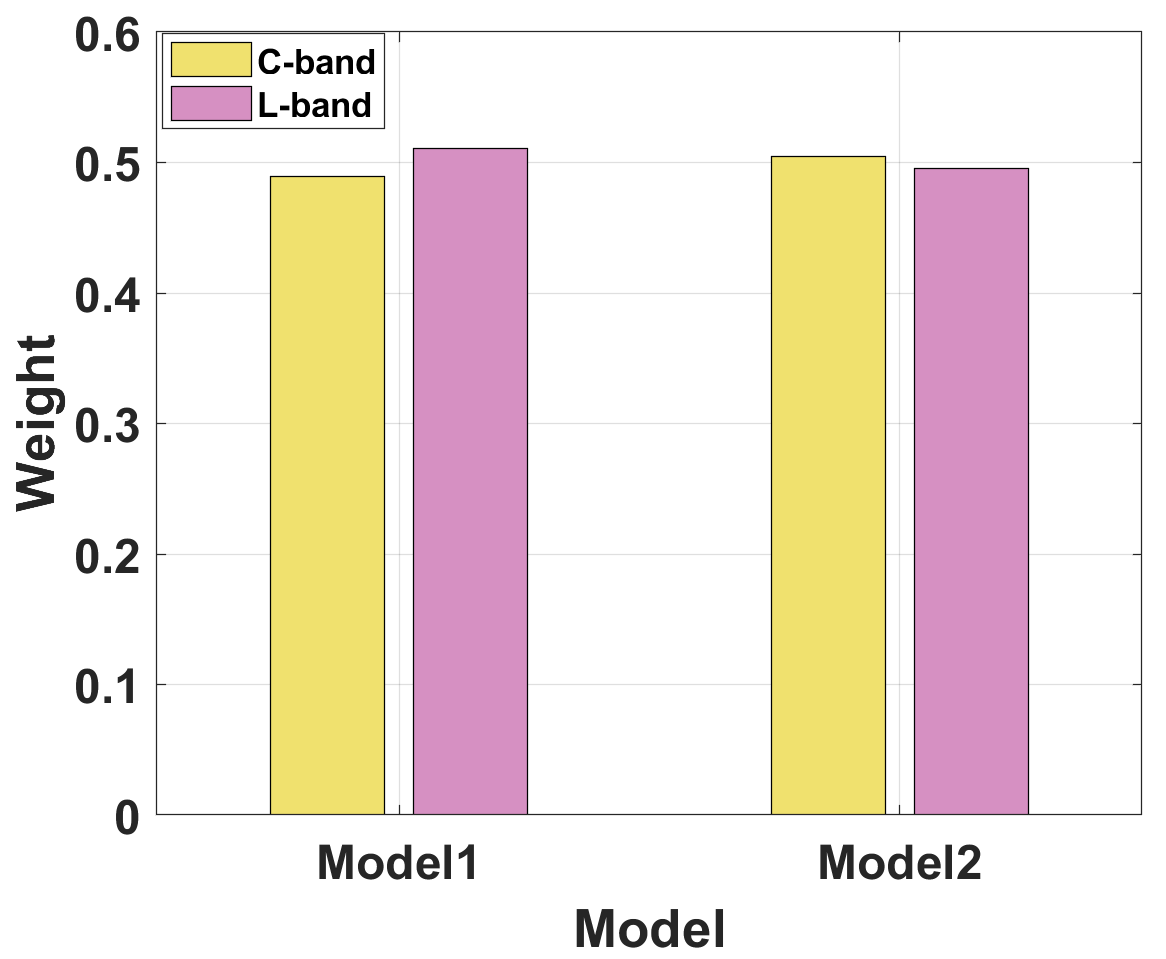}}
\subfloat[Woniupan]{\includegraphics[width=0.33\linewidth,height=2.5cm]{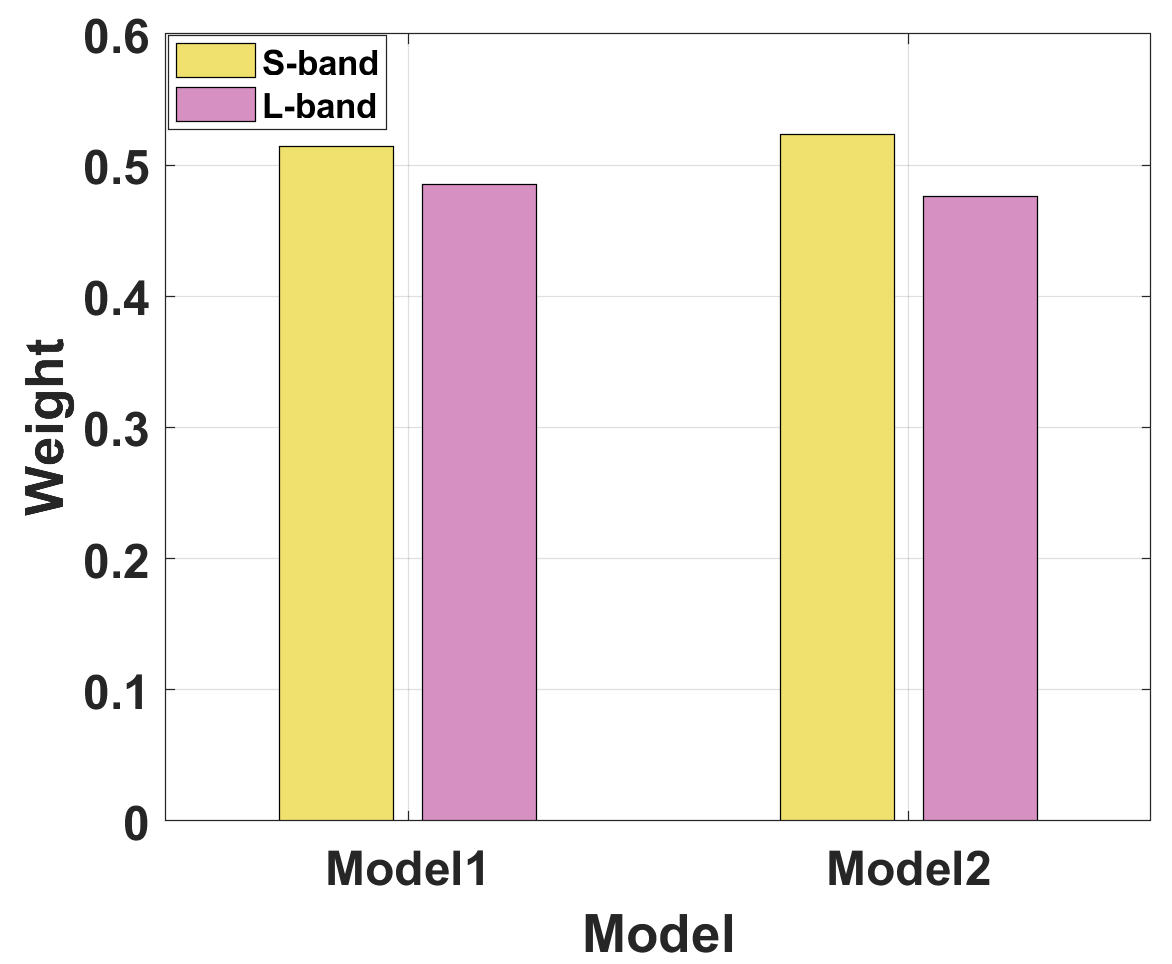}}
\subfloat[Flevoland]{\includegraphics[width=0.33\linewidth,height=2.5cm]{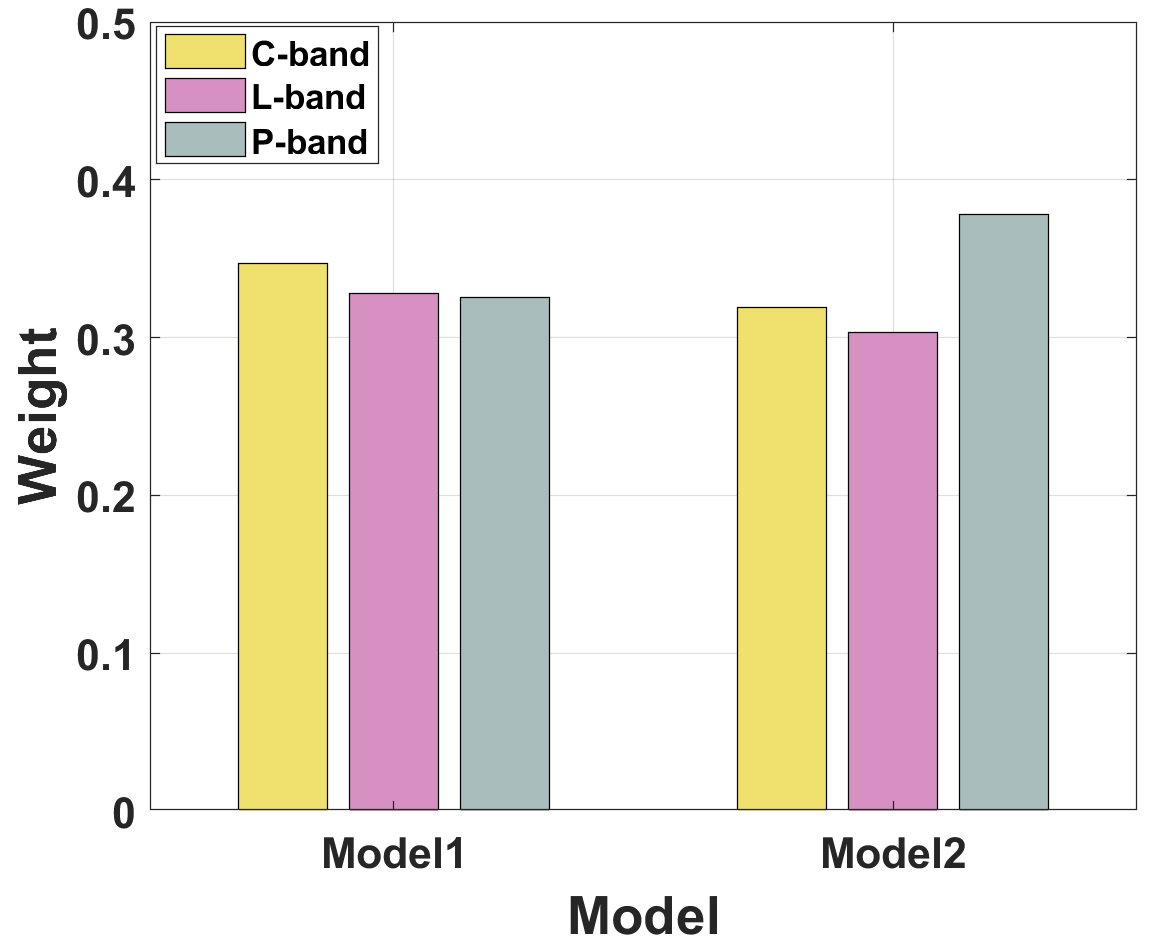}}

\caption{Learning weights of different frequency bands by MF-STFnet on the three datasets.}
\label{figure-weight}
\end{figure} 

Furthermore, we also show the learning weights of AWF in Fig. \ref{figure-weight}. As mentioned earlier, because the chessboard-like sampling strategy is used, there are two training results, denoted by model1 and model2, respectively. As shown in Fig. \ref{figure-weight}, The contribution of different frequency bands to the final classification decision is different. 

\subsection{Classification Results and Comparision}
According to the proposed MF-STFnet, the corresponding networks under all single bands and various combinations of different bands are constructed for classification performance comparison. For the two-band SanFrancisco and Woniupan datasets, the band combinations are C\&L and S\&L, respectively. While for the three-band Flevoland dataset, the different band combinations conclude C\&L, C\&P, L\&P, and C\&L\&P. Additionally, some competing works on multi-source or PolSAR image classification are employed as comparison algorithms, including wishart mixture model (WMM) \cite{Gao-2014-WMM}, object-based SVM (O-SVM) \cite{Lardeux-2009-SVM}, stein-sparse representation-based classification (S-SRC) \cite{Yang-2015-SSRC}, tensor feature-based ANN (TF-ANN) \cite{Ratha-2018-ANN}, CRPM-Net \cite{Xiao-2019-CRPMnet}, two-branch CNN (Tb-CNN) \cite{Xu-2018-TbCNN}, depthwise separable convolution based multi-task CNN (DMCNN) \cite{Zhang-2019-CNN}, and vision transformer (ViT) \cite{Dong-2022-transformer}. To use DMCNN and ViT for MF-PolSAR image classification, the proposed AWF strategy in this paper is adopted to combine bands for the final multi-frequency labeling decision.

For a fair comparison, the above DL-based methods all adopt the same sampling strategy as MF-STFnet. It is noticeable that we only report the results of these comparison methods under the full-band combination.

\begin{table*}[htp] 
\centering  
\fontsize{7.5}{10}\selectfont  
\caption{Classification Performance of Different Methods on the SanFrancisco Dataset}  
\label{table-SanFran-CL}  
\begin{tabular}{c|c|c|c|c|c|c|c|c|c|c|c} \hline
\toprule[0.3pt]

\multicolumn{1}{c|}{\multirow{2}{*}{\bf Category}}
& \multicolumn{3}{c|}{\bf MF-STFnet}
& \multicolumn{1}{c|}{\multirow{2}{*}{\bf WMM}}
& \multicolumn{1}{c|}{\multirow{2}{*}{\bf O-SVM}}
& \multicolumn{1}{c|}{\multirow{2}{*}{\bf S-SRC}}
& \multicolumn{1}{c|}{\multirow{2}{*}{\bf TF-ANN}}
& \multicolumn{1}{c|}{\multirow{2}{*}{\bf CRPM-Net}}
& \multicolumn{1}{c|}{\multirow{2}{*}{\bf Tb-CNN}}
& \multicolumn{1}{c|}{\multirow{2}{*}{\bf DMCNN}}
& \multicolumn{1}{c}{\multirow{2}{*}{\bf ViT}} \\ \cline{2-4}

\multicolumn{1}{c|}{}
& \multicolumn{1}{c|}{\bf C}& {\bf L} & {\bf C\&L}
& \multicolumn{1}{c|}{}
& \multicolumn{1}{c}{} \\ \hline 
{Forest} &{96.41} &{89.66} &{97.33} &\textbf{{98.27}} &{97.85} &{94.18} &{94.45} &{96.20} &{97.02} &{96.57} &{97.13} \cr 
{Water} &{99.97} &{99.80} &\textbf{{99.99}}  &{99.87} &{99.93} &{99.33} &{99.81} &{99.48} &{99.98}  &{99.99}  &{98.62}\cr 
{High-density urban} &{94.38} &{98.47} &\textbf{{98.56}}  &{95.82} &{96.30} &{96.57}  &{96.63} &{94.89} &{97.39} &{97.56} &{98.28}\cr 
{Low-density urban} &{94.50} &{96.78} &{98.82}  &{96.03} &{97.08} &\textbf{{99.04}}  &{91.59} &{95.27}&{96.71}  &{97.82} &{97.04}\cr 
{Developed} &{94.78} &{96.64} &\textbf{{98.78}}  &{86.06} &{94.92} &{96.19} &{91.67} &{95.84} &{95.76}  &{97.38} &{96.39}\cr \hline 
{OA(\%)} &{96.32} &{96.96} &\textbf{{98.76}} &{96.78} &{97.56} &{97.08} &{96.36} &{96.48}  &{97.85}  &{98.05} &{97.93}\cr
{AA(\%)} &{96.01} &{96.27} &\textbf{{98.69}} &{95.56} &{97.22} &{97.06} &{94.83} &{96.34} &{97.36}  &{97.96} &{97.49}\cr 
{$\kappa$} & {0.9497} &{0.9581} &\textbf{{0.9830}} &{0.9558} &{0.9665} &{0.9599}  &{0.9500} &{0.9518} &{0.9705}  &{0.9733} &{0.9716}\cr \hline

\toprule[0.3pt]
\end{tabular}
\end{table*}

\begin{figure}[htp] 
\centering
\subfloat[Ground Truth]{\includegraphics[width=0.32\linewidth,height=2.6cm]{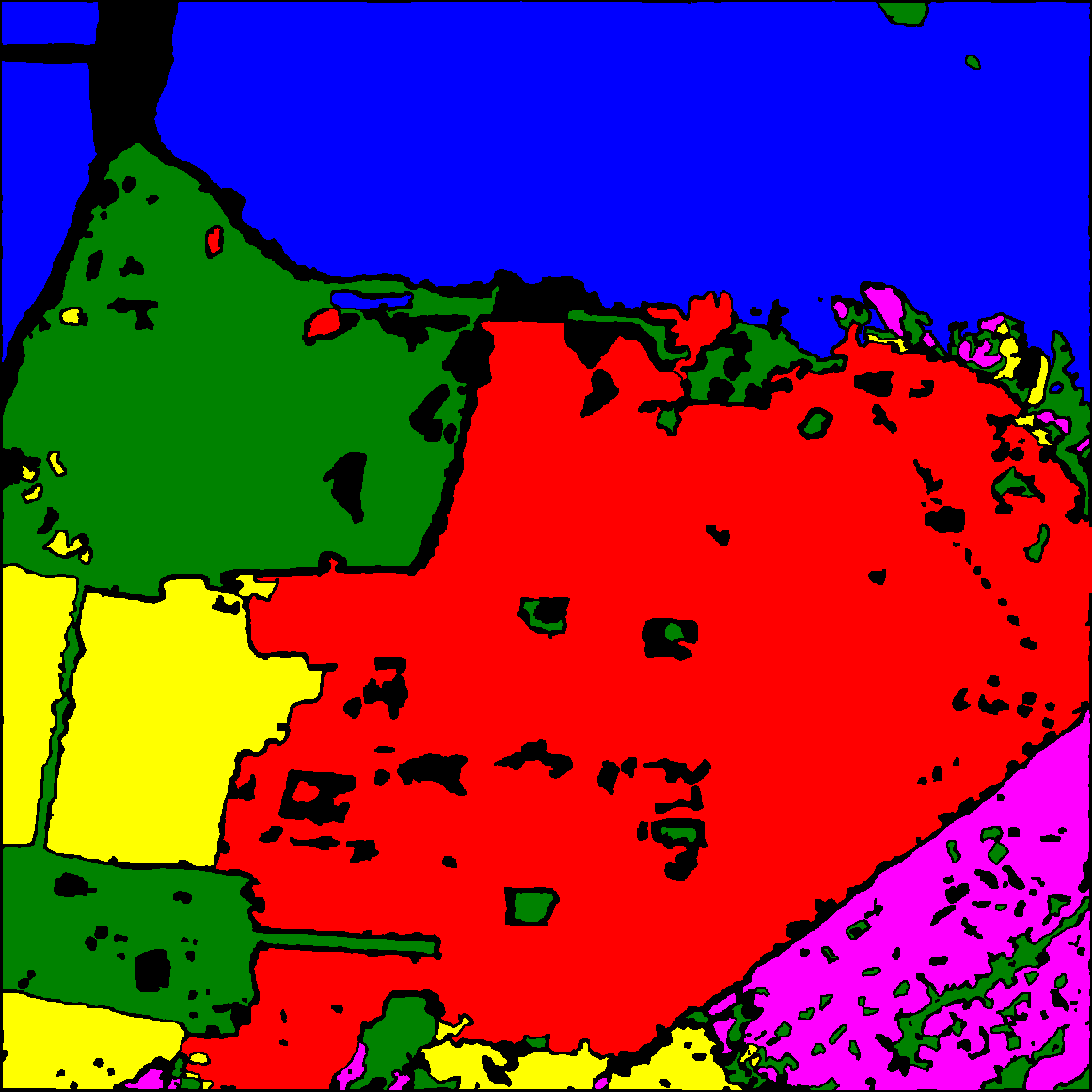}}
\label{Flev-GT}
\subfloat[MF-STFnet(C)]{\includegraphics[width=0.32\linewidth,height=2.6cm]{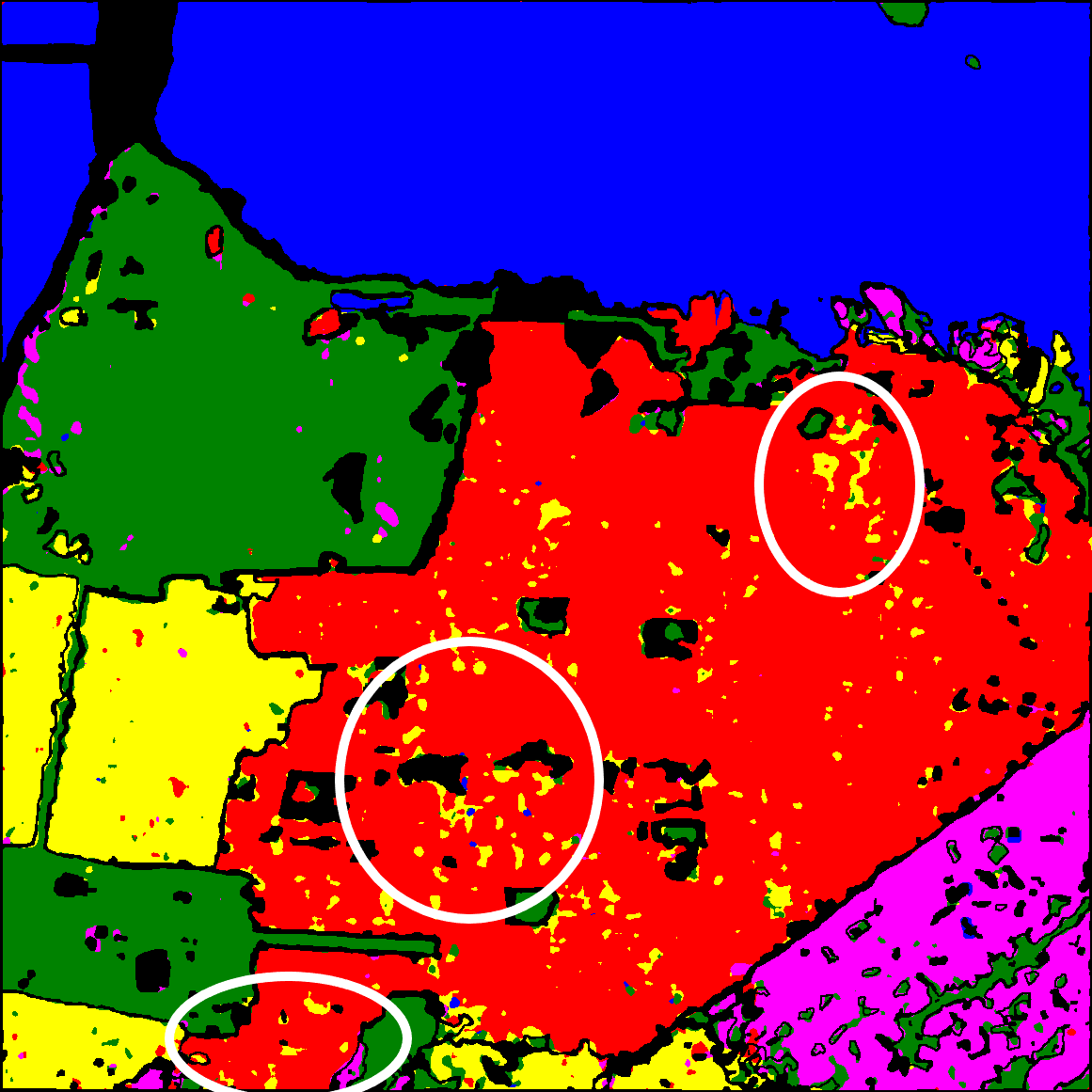}}
\label{STFnet-C}
\subfloat[MF-STFnet(L)]{\includegraphics[width=0.32\linewidth,height=2.6cm]{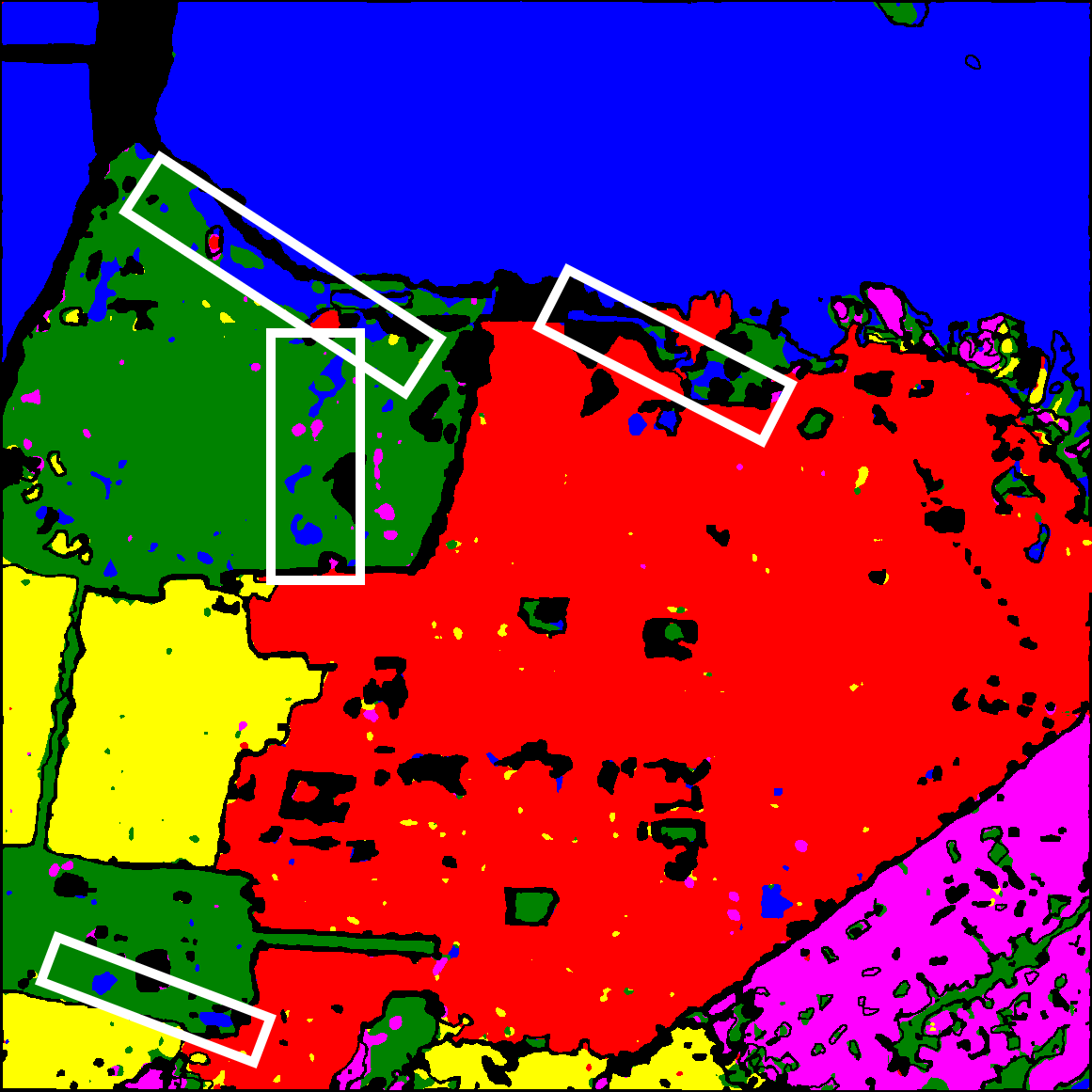}}
\label{STFnet-L}
\subfloat[MF-STFnet(C\&L)]{\includegraphics[width=0.32\linewidth,height=2.6cm]{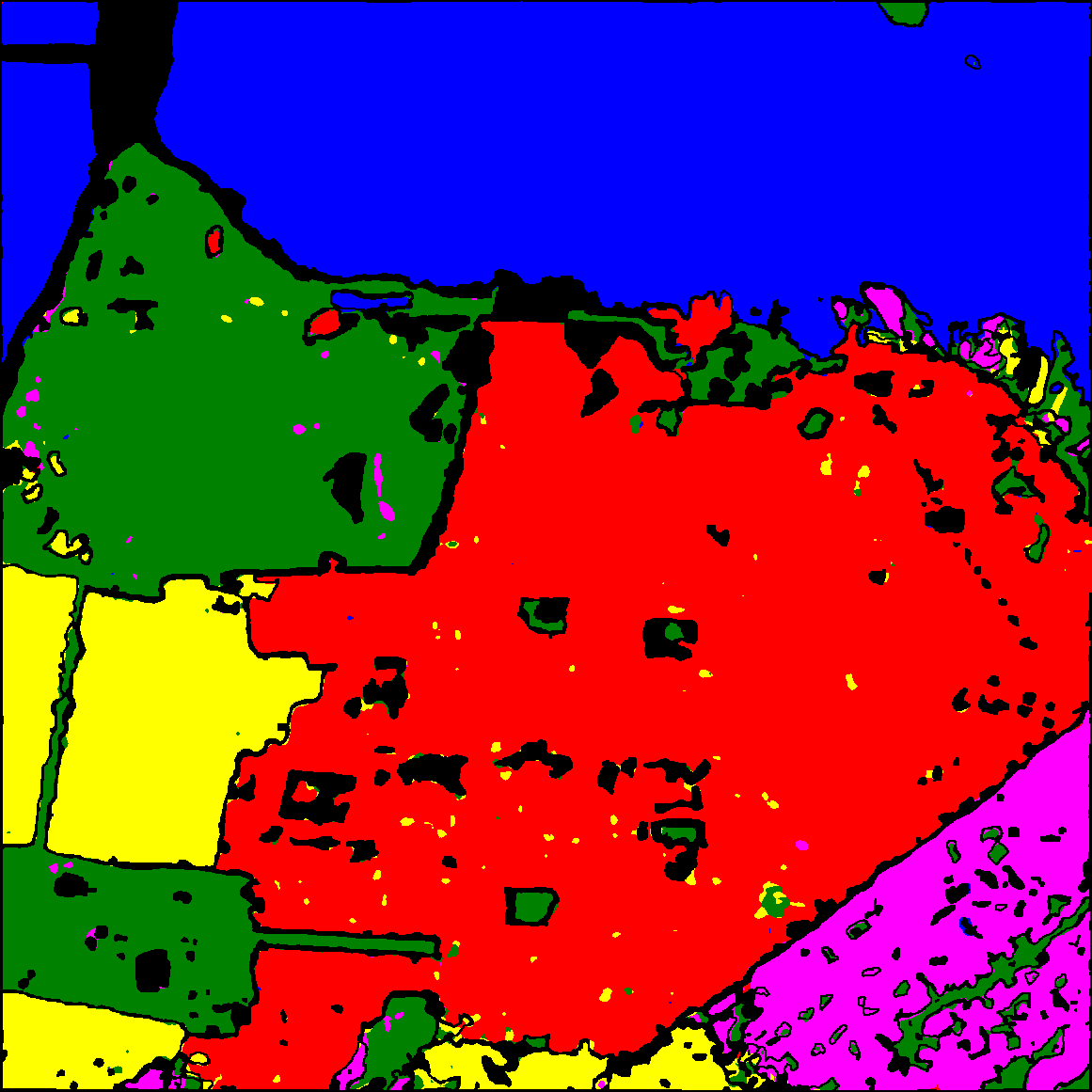}}
\label{STFnet-CL}
\subfloat[WMM]{\includegraphics[width=0.32\linewidth,height=2.6cm]{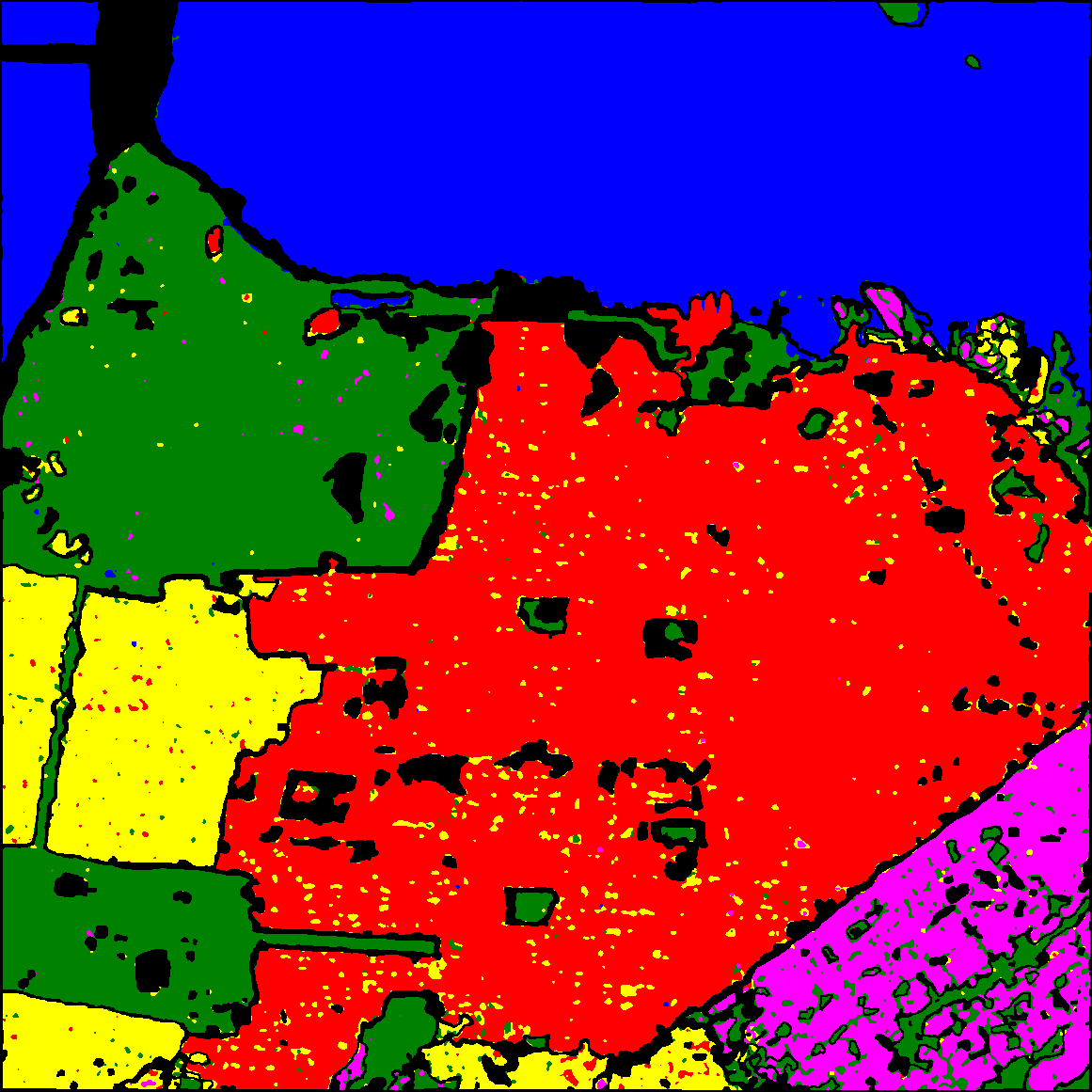}}
\label{WMM}
\subfloat[O-SVM]{\includegraphics[width=0.32\linewidth,height=2.6cm]{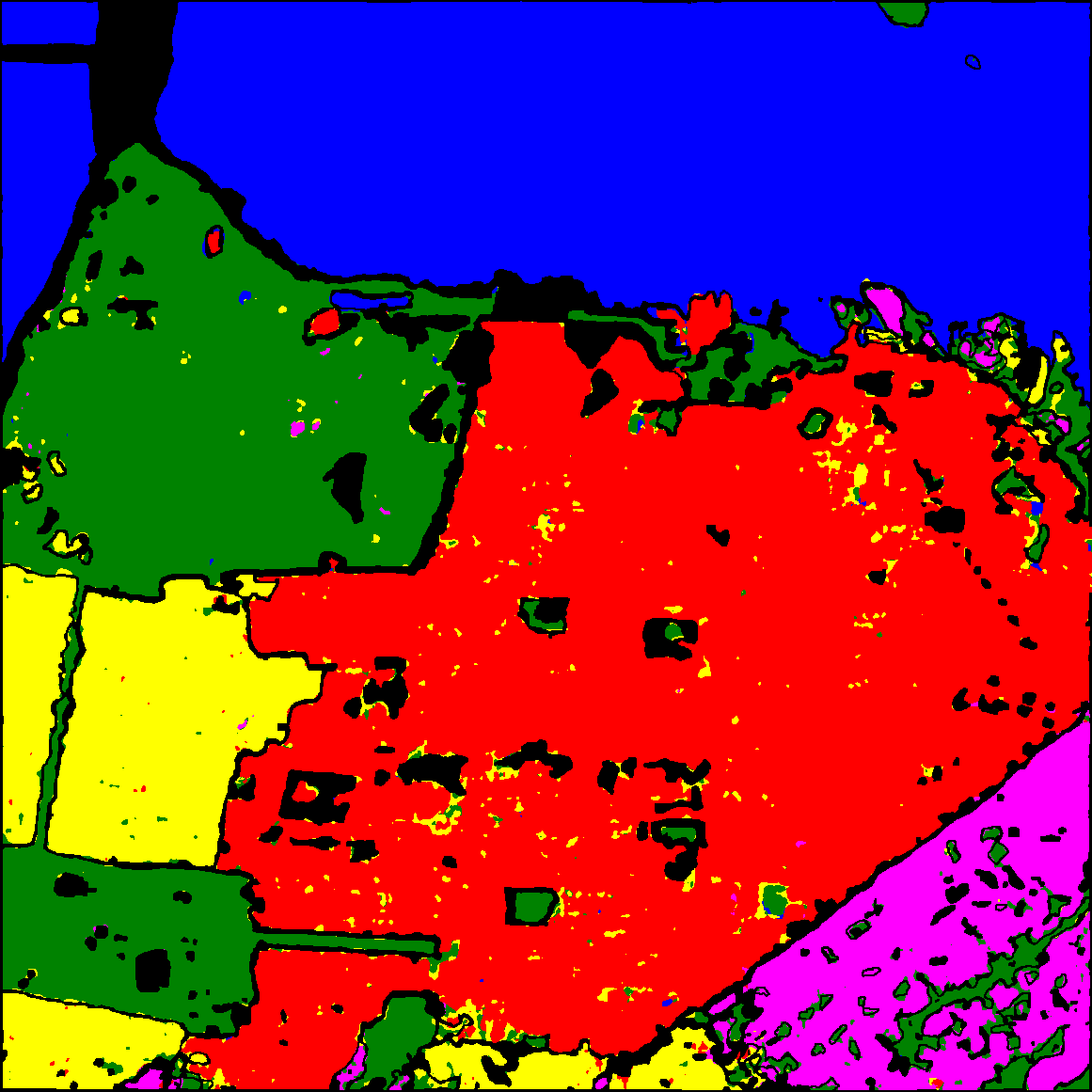}}
\label{O-SVM}
\subfloat[S-SRC]{\includegraphics[width=0.32\linewidth,height=2.6cm]{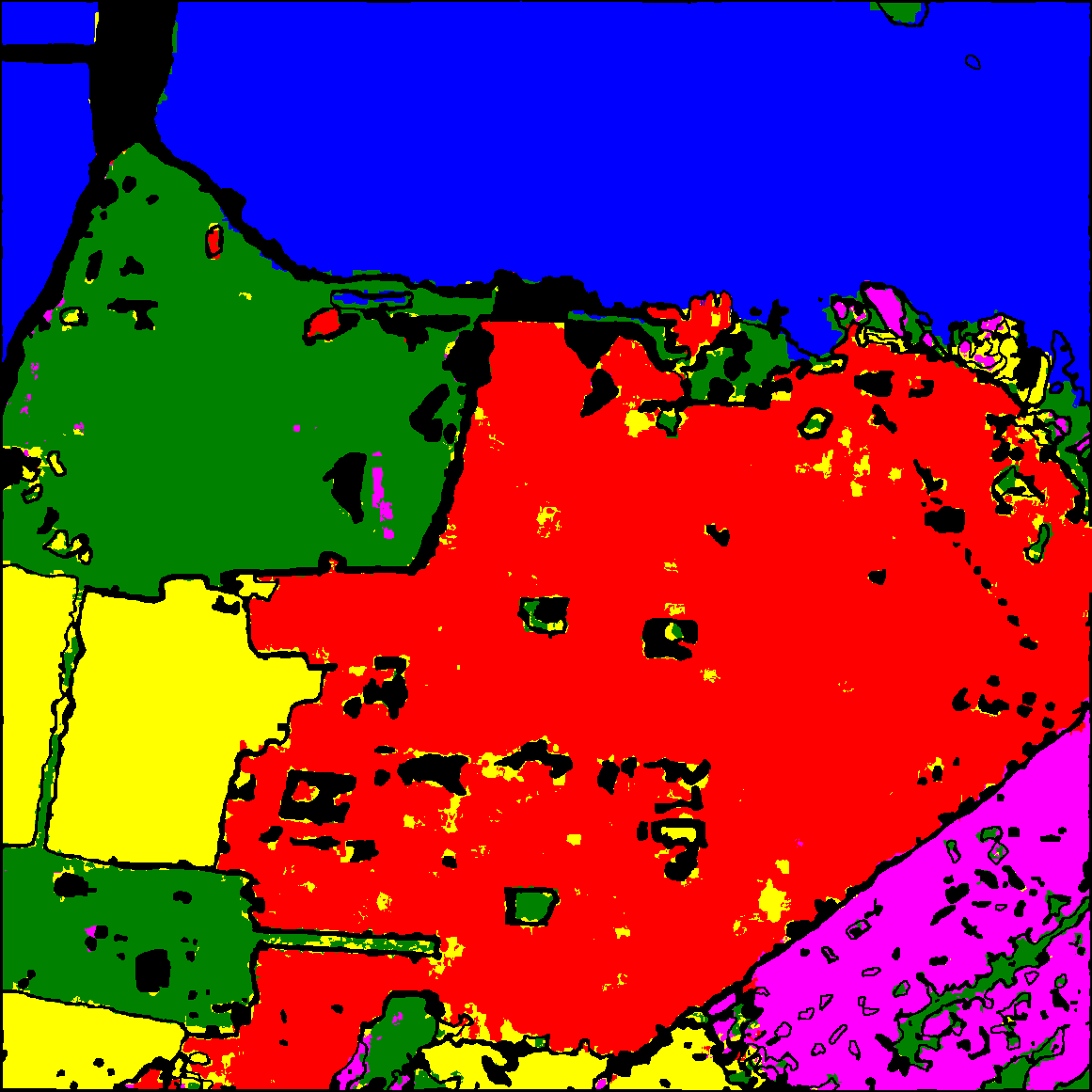}}
\label{S-SRC}
\subfloat[TF-ANN]{\includegraphics[width=0.32\linewidth,height=2.6cm]{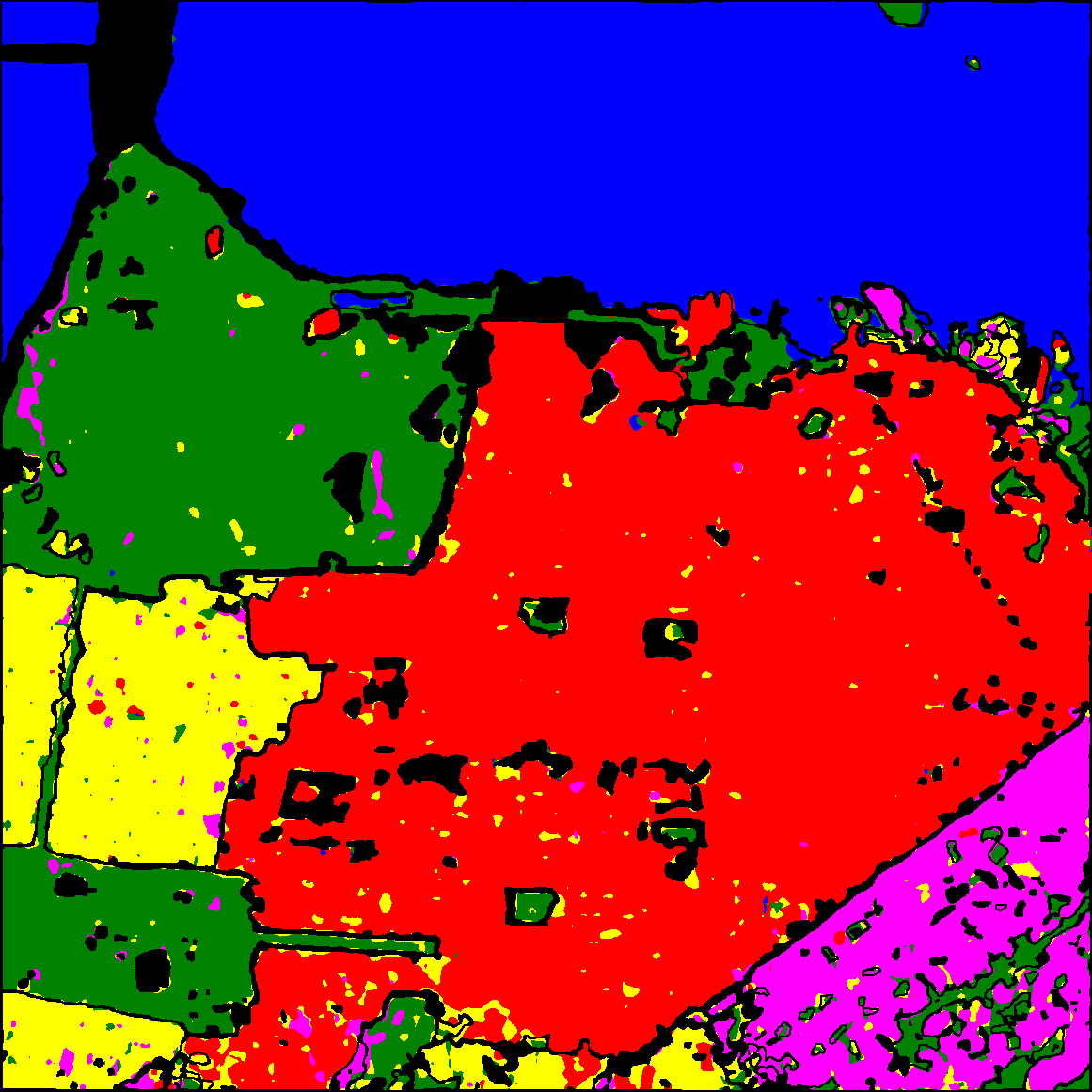}}
\label{TF-ANN}
\subfloat[CRPM-Net]{\includegraphics[width=0.32\linewidth,height=2.6cm]{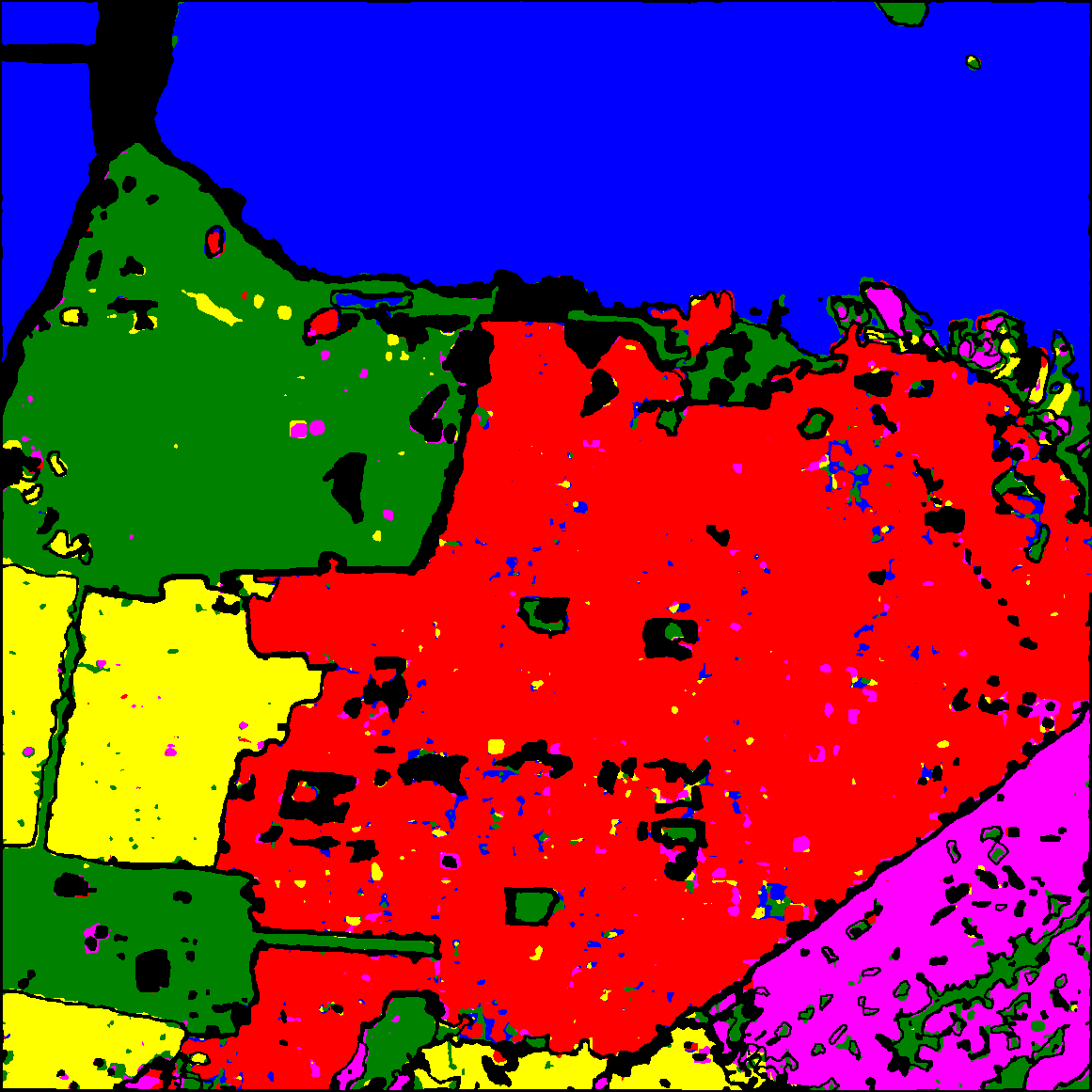}}
\label{CRPM-Net}
\subfloat[Tb-CNN]{\includegraphics[width=0.32\linewidth,height=2.6cm]{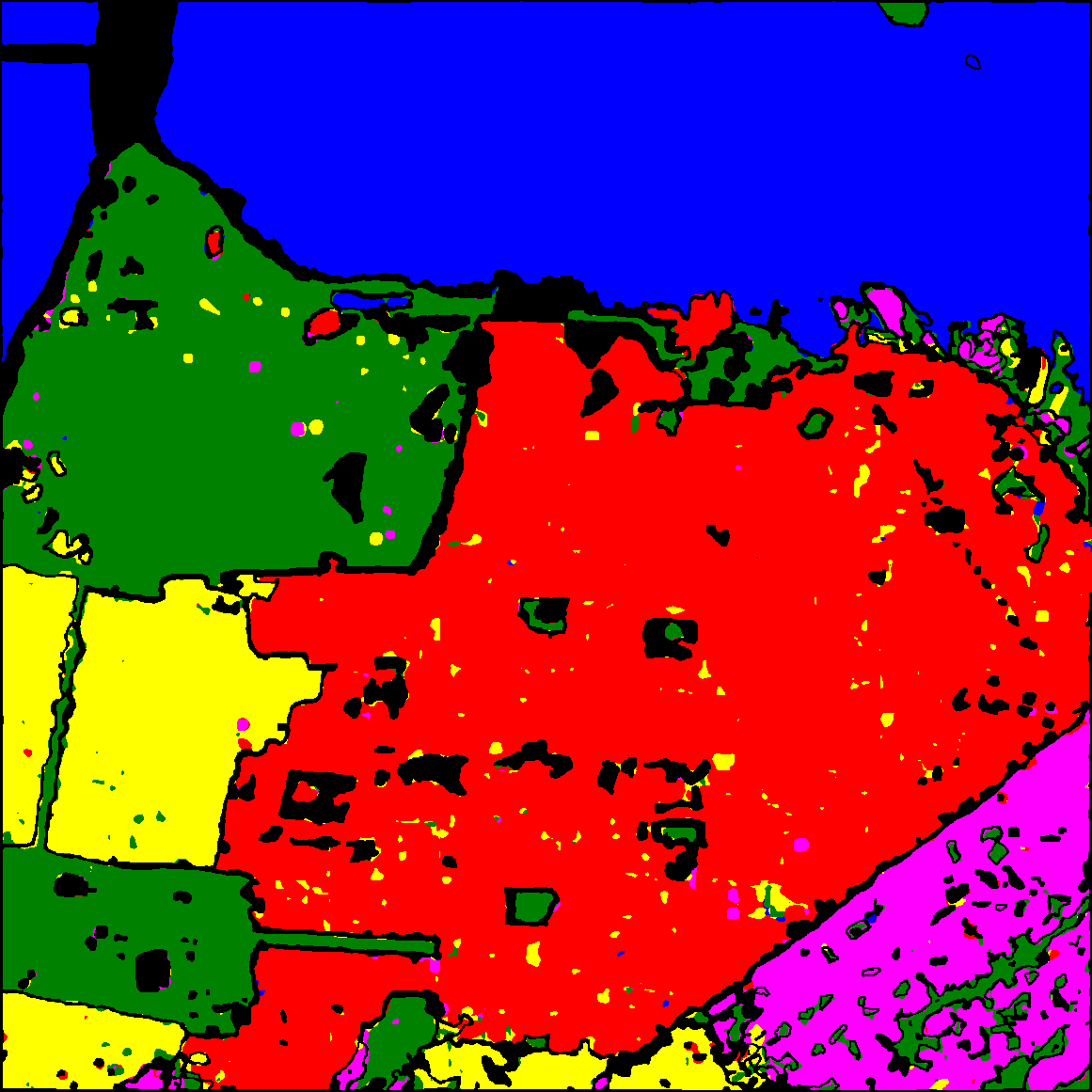}}
\label{Tb-CNN}
\subfloat[DMCNN]{\includegraphics[width=0.32\linewidth,height=2.6cm]{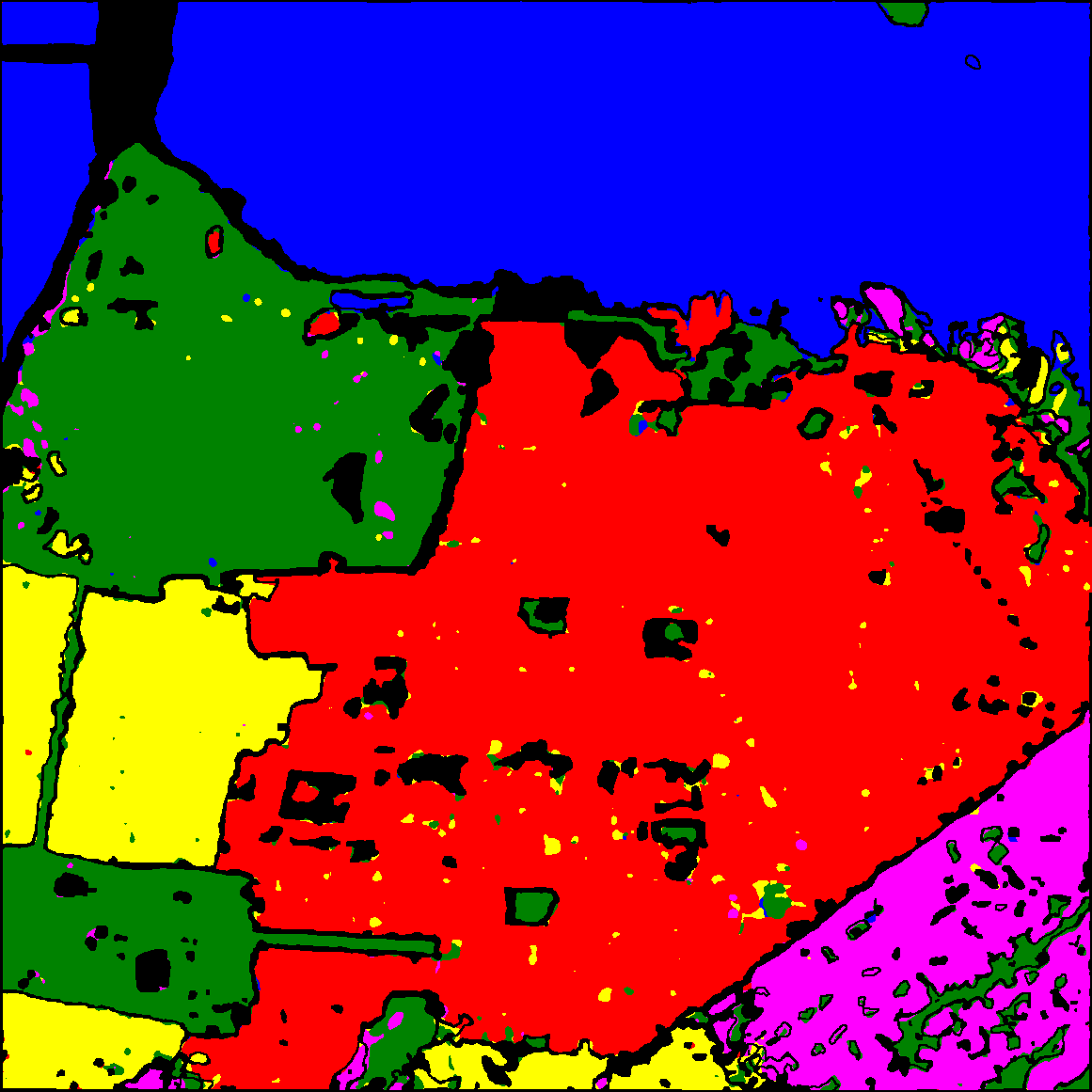}}
\label{DMCNN}
\subfloat[ViT]{\includegraphics[width=0.32\linewidth,height=2.6cm]{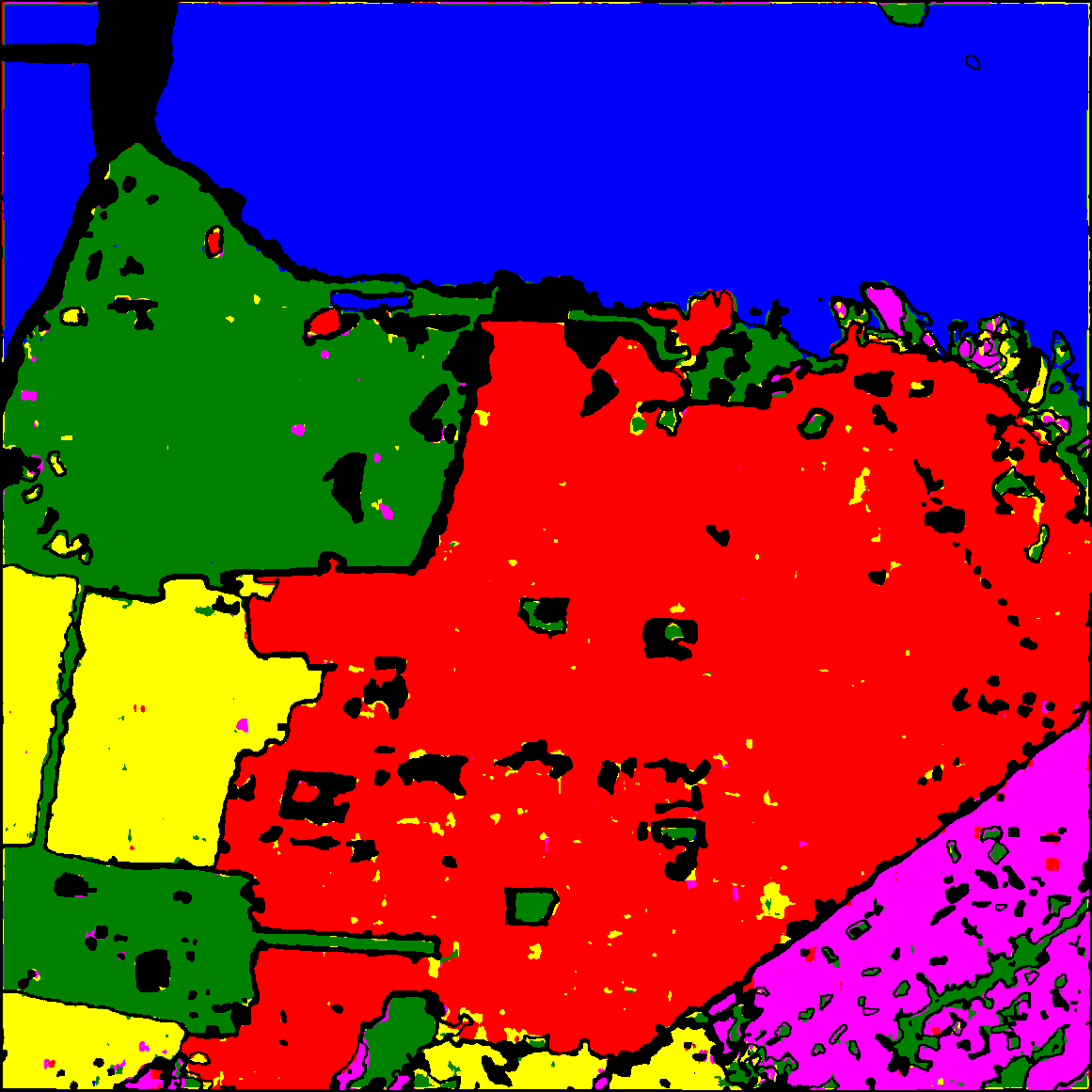}}
\label{ViT}
\subfloat{\includegraphics[width=0.9\linewidth,height=0.4cm]{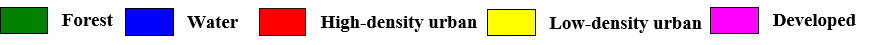}}

\caption{Classification results of different methods on the SanFrancisco dataset.}
\label{figure-SanFran-CL}
\end{figure} 

\subsubsection{Results on the SanFrancisco dataset}
Table \ref{table-SanFran-CL} shows the quantitative comparison on the SanFrancisco dataset, and Fig. \ref{figure-SanFran-CL} provides the corresponding classification maps for visual evaluation. As shown in Table \ref{table-SanFran-CL}, comparing the results of the two single bands, it can be seen that C-band is superior to L-band in classifying Forest and Water. While L-band is better than C-band in identifying different kinds of Urbans. By merging C and L, the classification accuracy of all categories is improved. Meanwhile, the OA, AA, and Kappa obtained by C\&L are all superior to C and L. These suggest that combining multiple frequency bands can exploit the complementarity of bands to improve category discrimination. Compared with the existing methods, the proposed MF-STFnet achieves the best classification performance. It gets over 97\% accuracy for all categories. Particularly, MF-STFnet better distinguishes High-density urban, Low-density urban, and Developed, and the classification accuracy of all three categories exceeds 98\%. This demonstrates that the proposed MF-STFnet can discriminate categories more effectively and maintain a better classification balance among categories.

The classification maps shown in Fig. \ref{figure-SanFran-CL} further verify the effectiveness of the proposed MF-STFnet on the SanFrancisco dataset. For C-band, there exhibits a serious mix between the classification of High-density urban and Low-density urban in Fig. \ref{figure-SanFran-CL}(b) (highlighted by white ovals). While for L-band, many pixels belonging to Forest are misclassified as Water in Fig. \ref{figure-SanFran-CL}(c) (highlighted by white rectangles). However, these above phenomena are greatly improved by merging dual-band information in Fig. \ref{figure-SanFran-CL}(d). As shown in Fig. \ref{figure-SanFran-CL}(e)-(l), the existing methods have a weak ability to accurately distinguish between High-density urban and Low-density urban. However, our proposed MF-STFnet can identify them well. In addition, compared with other methods, MF-STFnet also produces a clearer classification map, which is closer to the ground truth distribution. In general, the experimental results on the SanFrancisco dataset can demonstrate the classification superiority of the proposed MF-STFnet.

\begin{table*}[htp] 
\centering  
\fontsize{7.5}{10}\selectfont  
\caption{Classification Performance of Different Methods on the Woniupan Dataset}  
\label{table-Woniu-SL}  
\begin{tabular}{c|c|c|c|c|c|c|c|c|c|c|c} \hline
\toprule[0.3pt]

\multicolumn{1}{c|}{\multirow{2}{*}{\bf Category}}
& \multicolumn{3}{c|}{\bf MF-STFnet}
& \multicolumn{1}{c|}{\multirow{2}{*}{\bf WMM}}
& \multicolumn{1}{c|}{\multirow{2}{*}{\bf O-SVM}}
& \multicolumn{1}{c|}{\multirow{2}{*}{\bf S-SRC}}
& \multicolumn{1}{c|}{\multirow{2}{*}{\bf TF-ANN}}
& \multicolumn{1}{c|}{\multirow{2}{*}{\bf CRPM-Net}}
& \multicolumn{1}{c|}{\multirow{2}{*}{\bf Tb-CNN}}
& \multicolumn{1}{c|}{\multirow{2}{*}{\bf DMCNN }}
& \multicolumn{1}{c}{\multirow{2}{*}{\bf ViT }} \\ \cline{2-4}

\multicolumn{1}{c|}{}
& \multicolumn{1}{c|}{\bf S}& {\bf L} & {\bf S\&L}
& \multicolumn{1}{c|}{}
& \multicolumn{1}{c}{} \\ \hline 
{Road} &{86.79} &{44.46} &\textbf{{92.13}} &{20.65} &{69.42} &{85.90} &{86.57} &{80.49} &{84.97} &{79.43} &{82.64} \cr 
{Building} &{79.51} &{74.23} &{81.19} &{27.38} &{72.60} &{48.61} &{87.90} &{84.02} &{87.45}  &\textbf{{95.80}} &{88.89} \cr 
{Farmland} &{84.11} &{93.03} &{94.40} &\textbf{{96.89}} &{91.97} &{90.27} &{90.62} &{92.91} &{91.99}  &{94.07}  &{91.40}\cr 
{Forest} &{98.43} &{95.97} &\textbf{{99.64}} &{86.69} &{94.69} &{95.62} &{97.51} &{94.23} &{97.92}  &{98.62} &{97.39} \cr 
{Bareland} &{93.45} &{91.62} &\textbf{{97.73}} &{92.56} &{91.86} &{94.70} &{93.19} &{95.20}  &{96.11} &{94.85}  &{96.27}\cr \hline 
{OA(\%)} &{88.34} &{89.56} &\textbf{{95.41}} &{87.95} &{90.38} &{90.66} &{91.82}&{92.67}  &{93.25} &{93.99}  &{92.81}\cr
{AA(\%)} &{88.46} &{79.86} &\textbf{{93.02}} &{80.70} &{84.11} &{83.02} &{91.16}&{89.37}  &{91.69} &{92.55}  &{91.32}\cr 
{$\kappa$} & {0.8267} &{0.8368} &\textbf{{0.9290}} &{0.8066} &{0.8520} &{0.8571}  &{0.8755} &{0.8869}&{0.8964}  &{0.9074} &{0.8900}\cr \hline

\toprule[0.3pt]
\end{tabular}
\end{table*}

\begin{figure}[htp] 
\centering
\subfloat[Ground Truth]{\includegraphics[width=0.32\linewidth,height=2.6cm]{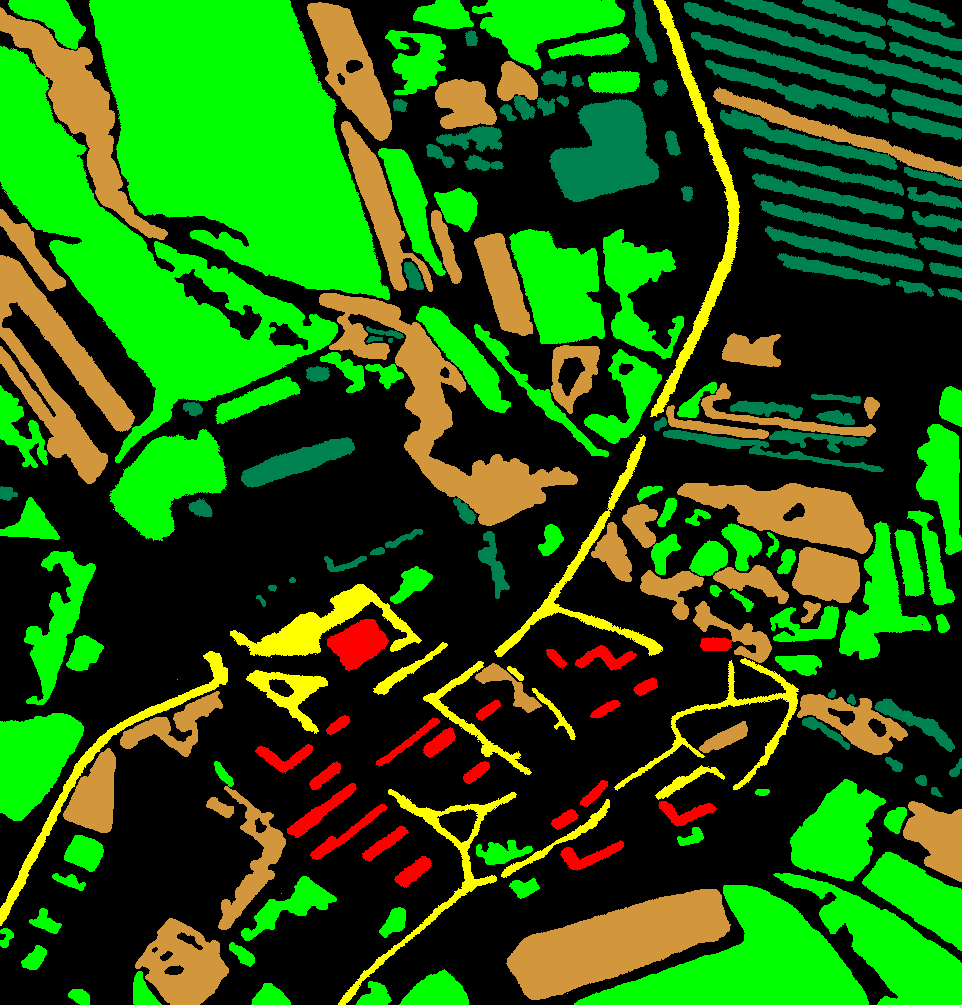}}
\label{Flev-GT}
\subfloat[MF-STFnet(S)]{\includegraphics[width=0.32\linewidth,height=2.6cm]{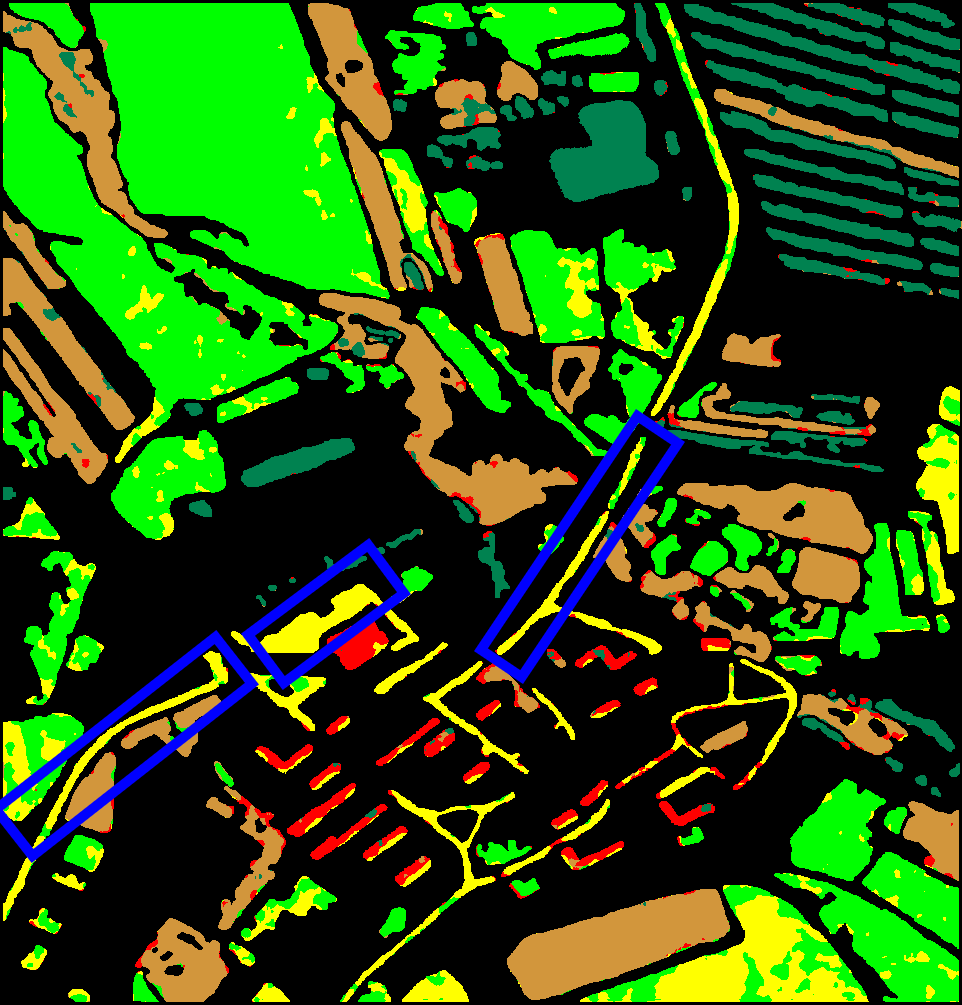}}
\label{STFnet-S}
\subfloat[MF-STFnet(L)]{\includegraphics[width=0.32\linewidth,height=2.6cm]{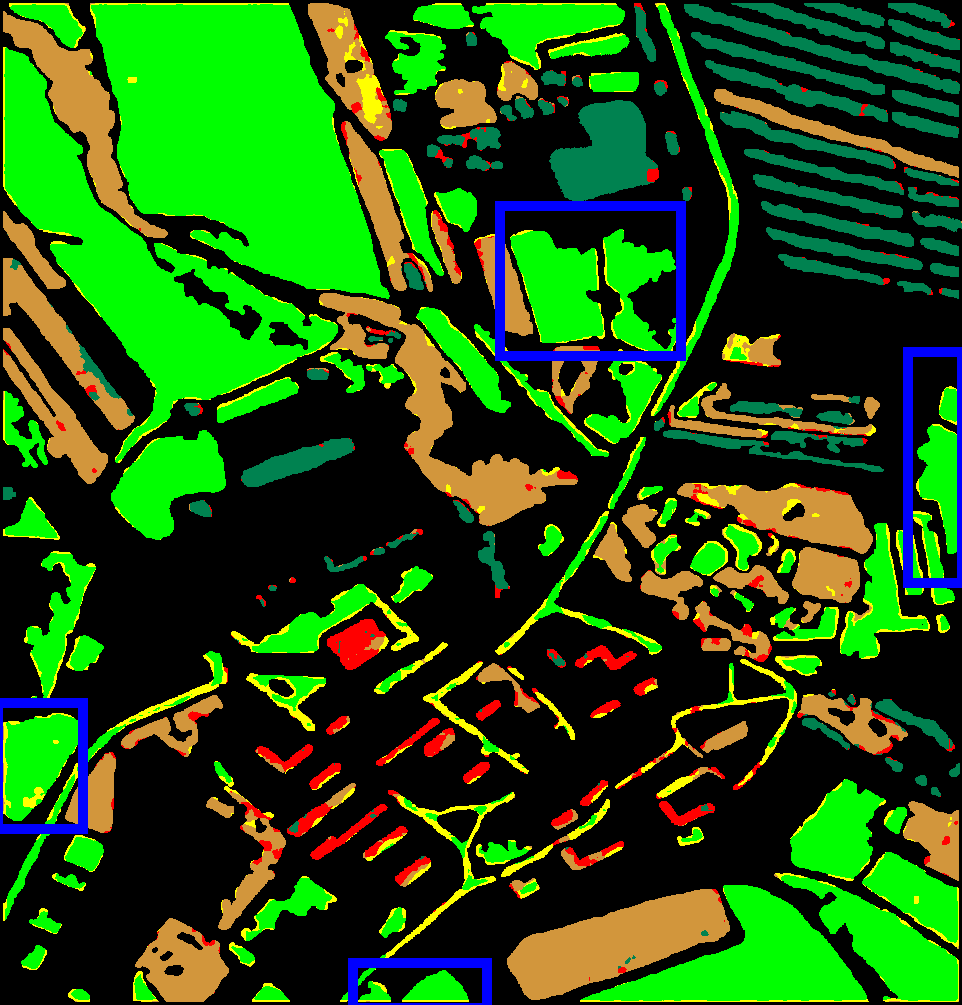}}
\label{STFnet-L}
\subfloat[MF-STFnet(S\&L)]{\includegraphics[width=0.32\linewidth,height=2.6cm]{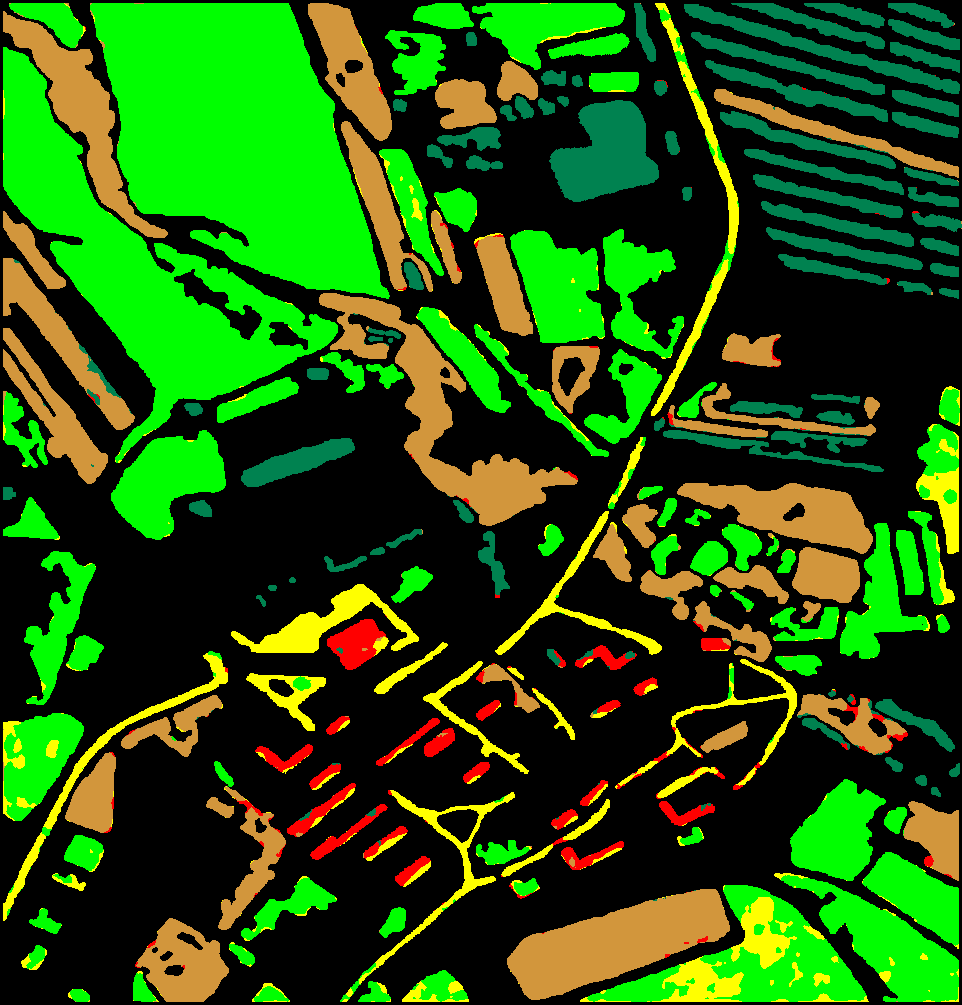}}
\label{STFnet-SL}
\subfloat[WMM]{\includegraphics[width=0.32\linewidth,height=2.6cm]{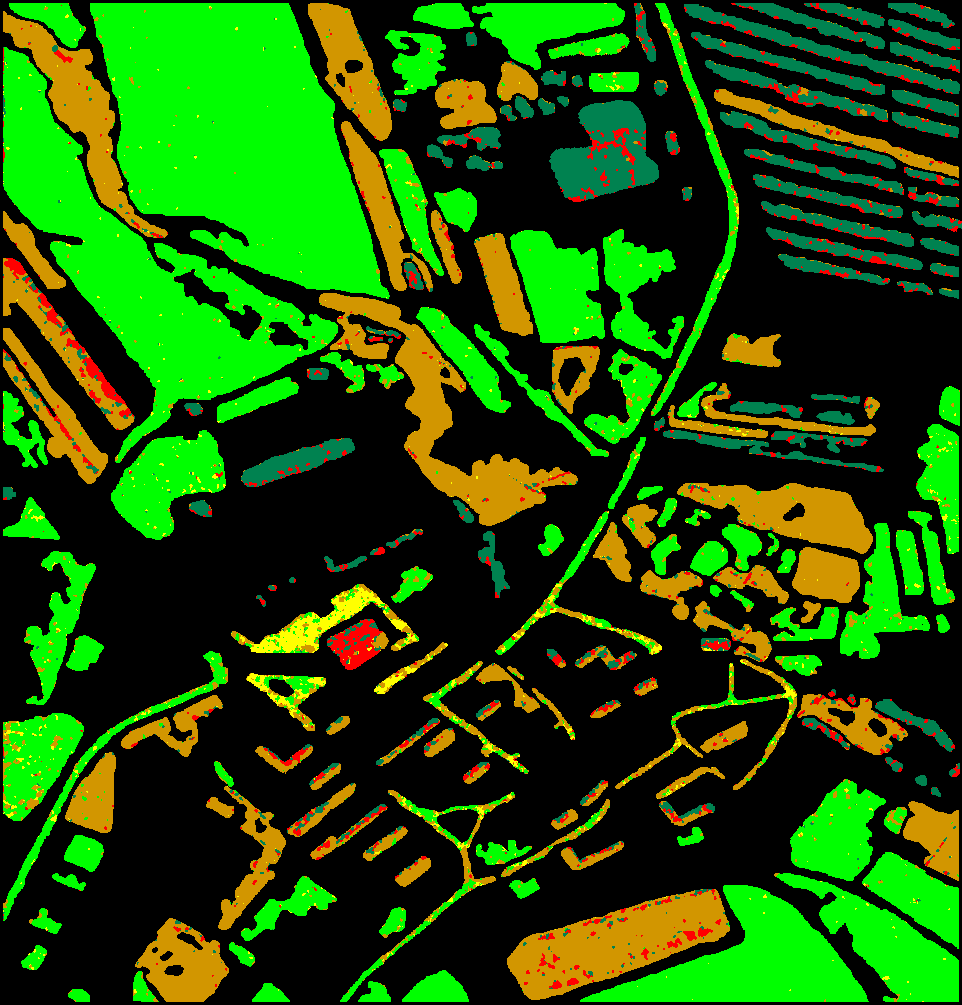}}
\label{WMM}
\subfloat[O-SVM]{\includegraphics[width=0.32\linewidth,height=2.6cm]{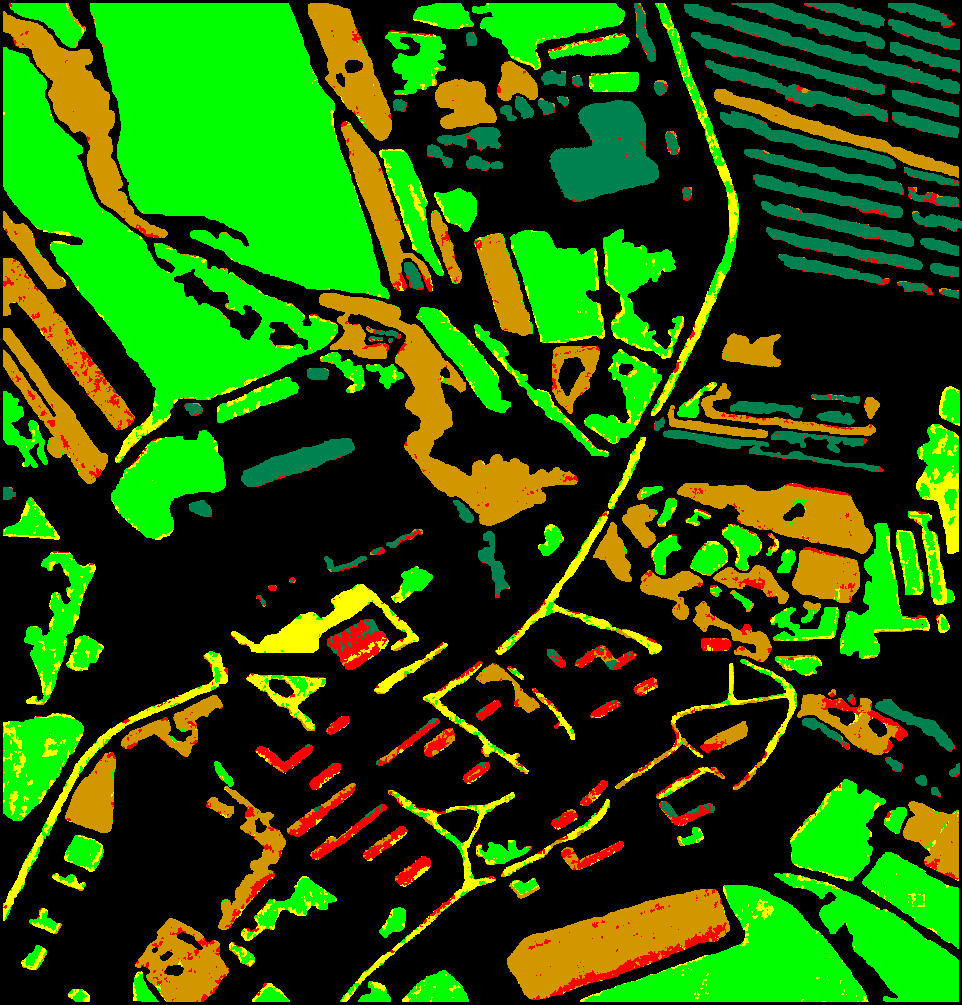}}
\label{O-SVM}
\subfloat[S-SRC]{\includegraphics[width=0.32\linewidth,height=2.6cm]{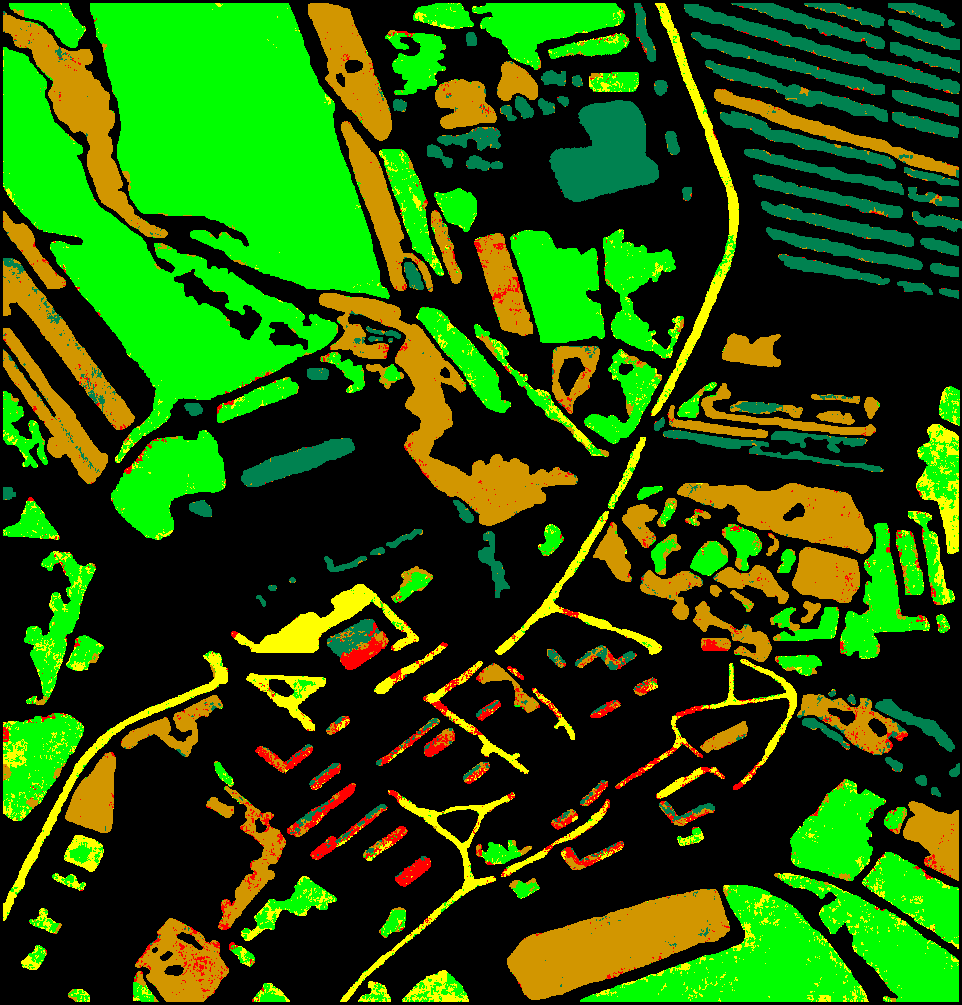}}
\label{S-SRC}
\subfloat[TF-ANN]{\includegraphics[width=0.32\linewidth,height=2.6cm]{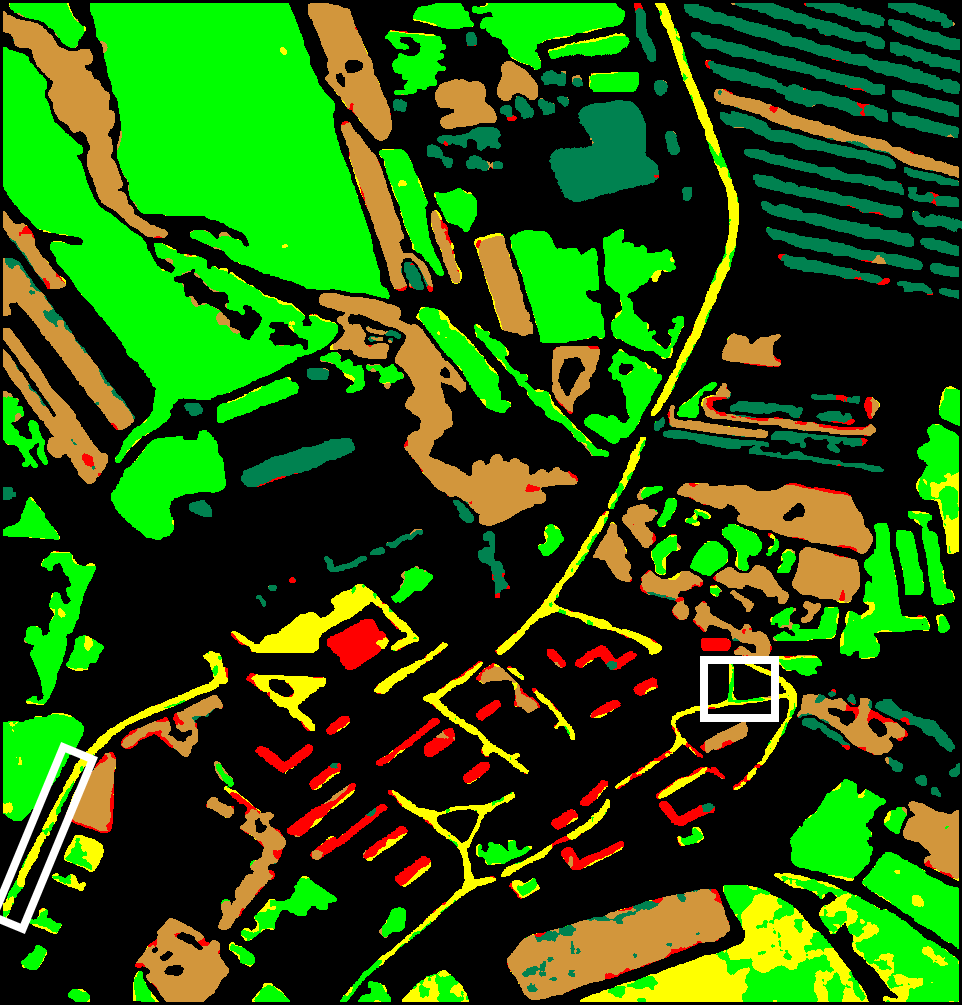}}
\label{TF-ANN}
\subfloat[CRPM-Net]{\includegraphics[width=0.32\linewidth,height=2.6cm]{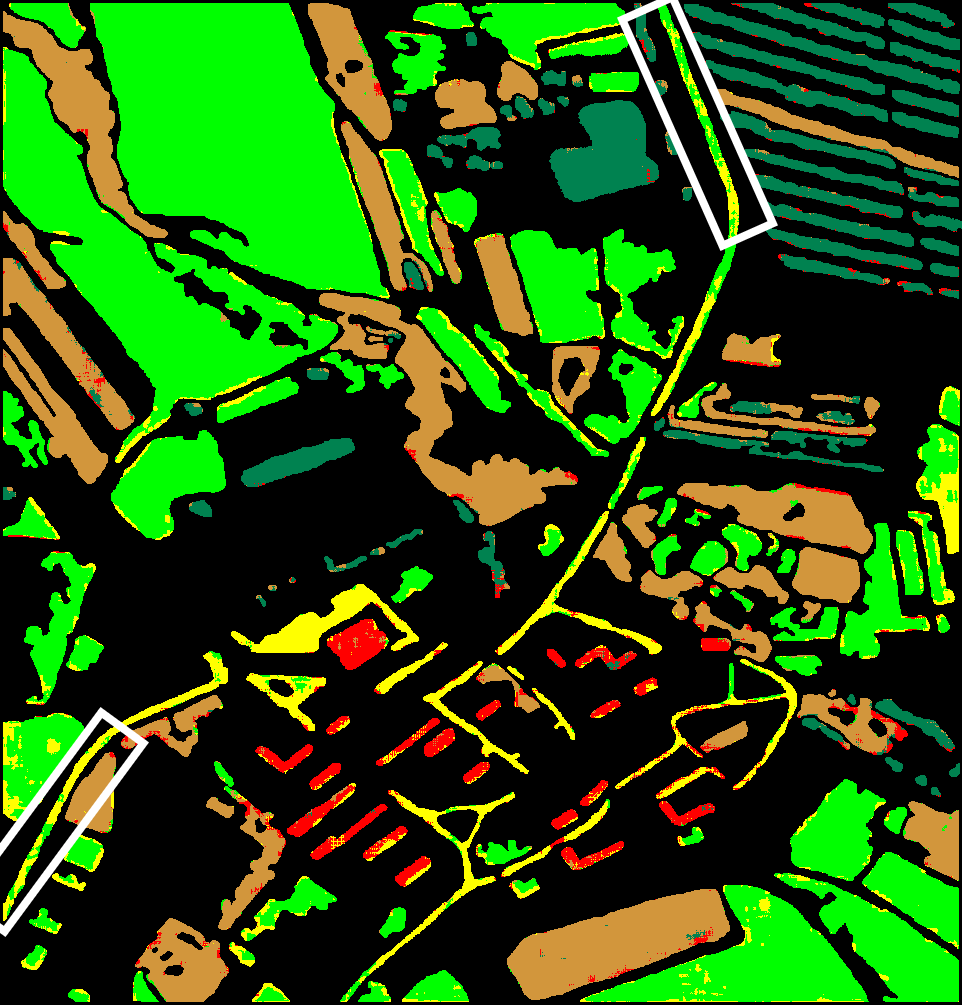}}
\label{CRPM-Net}
\subfloat[Tb-CNN]{\includegraphics[width=0.32\linewidth,height=2.6cm]{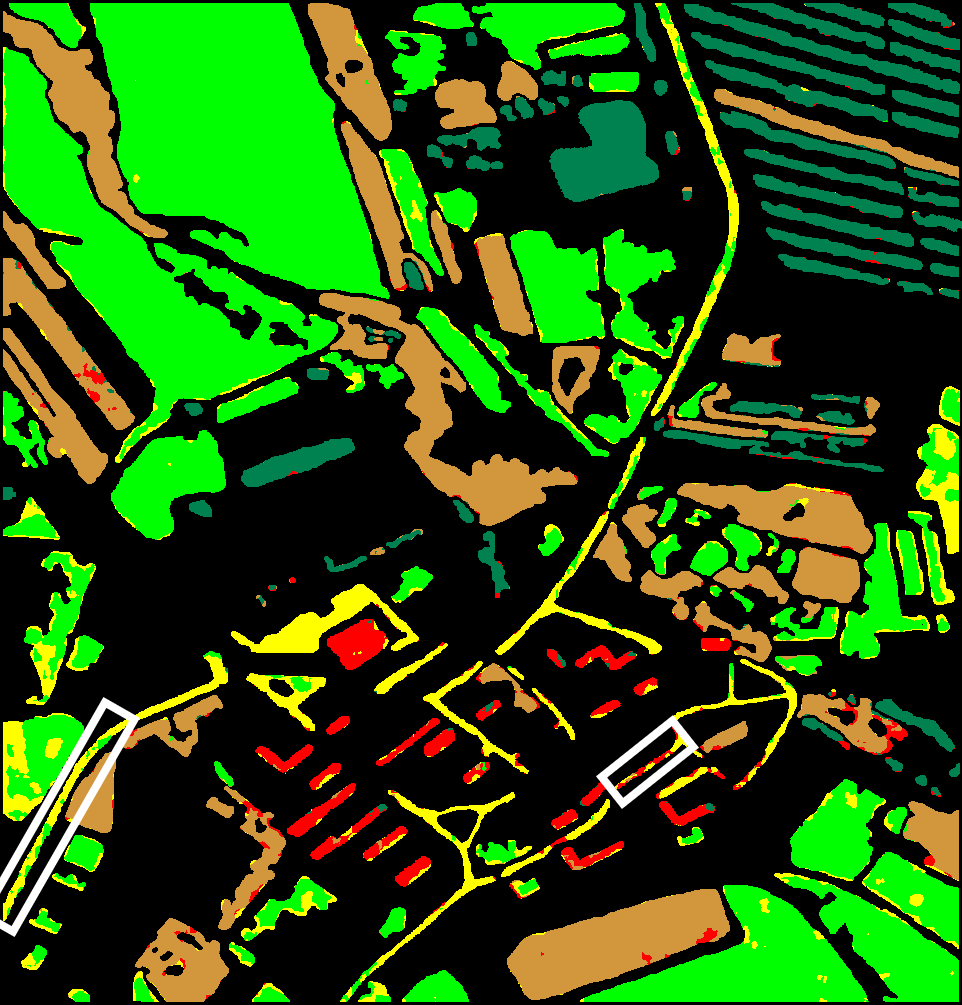}}
\label{Tb-CNN}
\subfloat[DMCNN]{\includegraphics[width=0.32\linewidth,height=2.6cm]{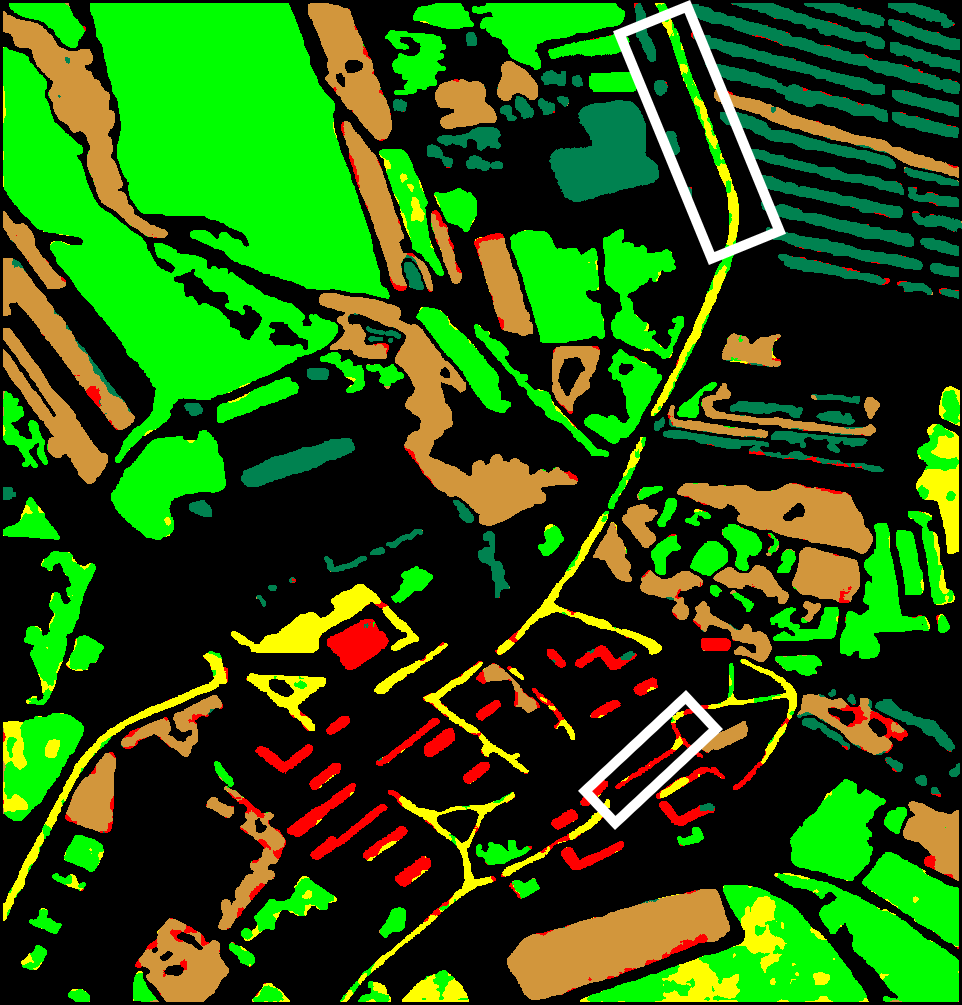}}
\label{DMCNN}
\subfloat[ViT]{\includegraphics[width=0.32\linewidth,height=2.6cm]{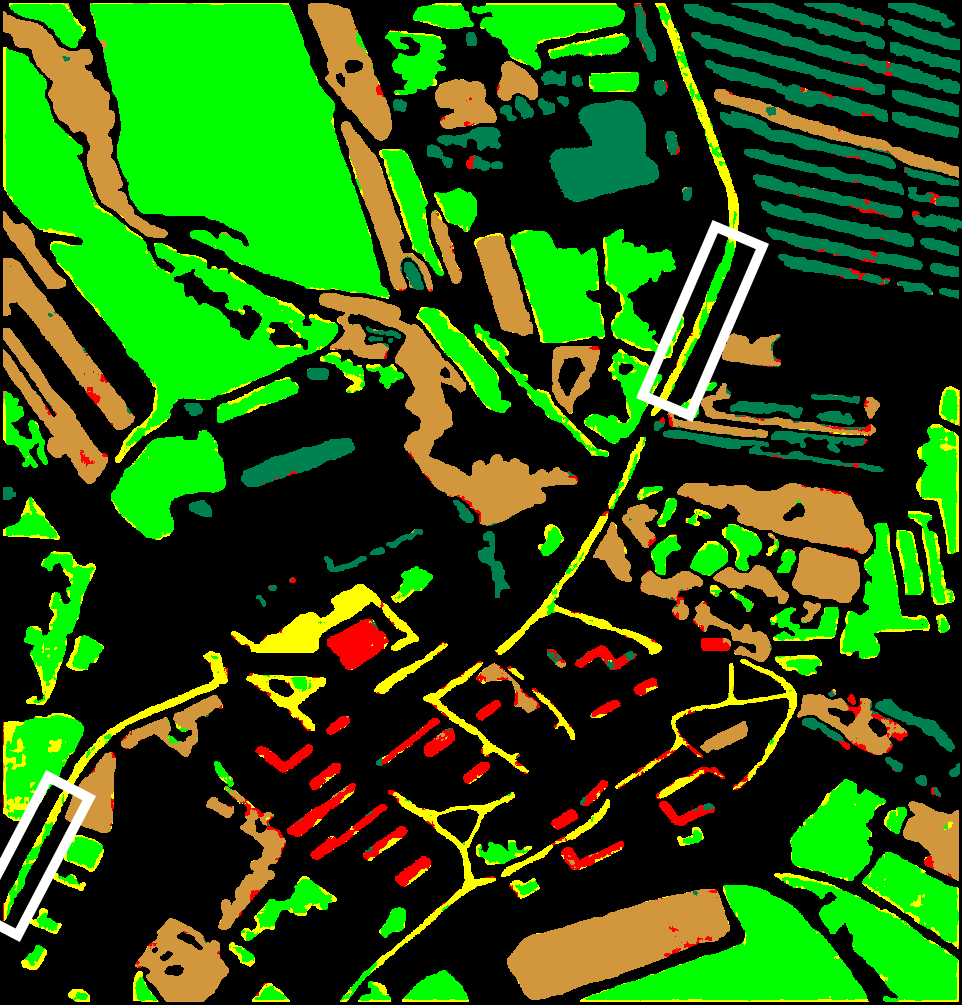}}
\label{ViT}
\subfloat{\includegraphics[width=0.9\linewidth,height=0.4cm]{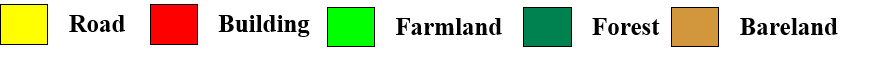}}

\caption{Classification results of different methods on the Woniupan dataset.}
\label{figure-Woniu-SL}
\end{figure} 

\subsubsection{Results on the Woniupan dataset}
For the Woniupan dataset, Table \ref{table-Woniu-SL} and Fig. \ref{figure-Woniu-SL} respectively show the quantitative and qualitative results. As shown in Table \ref{table-Woniu-SL}, comparing S- and L-bands, S-band classifies Road, Building, and Forest better than L-band. While for the classification of Farmland and Bareland, L-band is superior to of S-band. When combining dual bands, the classification accuracy of each category obtained by S\&L is improved. Furthermore, the classification results of S\&L are better than that of S-band and L-band. These results indicate that there is complementarity between the two bands. Compared with traditional methods (WMM, O-SVM, and S-SRC), the DL-based methods (TF-ANN, CRPM-Net, Tb-CNN, DMCNN, ViT, and MF-STFnet) can greatly improve the classification accuracy of Building. The reason may be that the features extracted by DL-based methods have stronger discriminative ability than traditional features and are more conducive to distinguishing similar objects. For all categories except Building, MF-STFnet outperforms other DL-based methods. Notably, the accuracy of Road by MF-STFnet is over 90\%. Besides, MF-STFnet obtains the highest values of OA, AA, and Kappa. All of these can prove that the proposed classification method is effective and can achieve more satisfactory classification performance. 

Visually, comparing Fig. \ref{figure-Woniu-SL}(b) and (c), S-band identifies Road well, while L-band can recognize Farmland well (respectively highlighted by blue rectangles in Fig. \ref{figure-Woniu-SL}(b) and (c)). By observing Fig. \ref{figure-Woniu-SL}(d), it is found that the classification effect of Road and Farmland has a certain complementary improvement compared with single bands. In addition, compared with S- and L-bands, the classification result of Building in Fig. \ref{figure-Woniu-SL}(d) is closer to the ground truth. As shown in Fig. \ref{figure-Woniu-SL}(e)-(g), the non-DL methods have worse classification effects on Building. In contrast, TF-ANN, CRPM-Net, Tb-CNN, DMCNN, and ViT have better visual effects in this category because of more discriminative feature extraction. However, for Road category with narrow structures, they have insufficient recognition ability (highlighted by white rectangles in Fig. \ref{figure-Woniu-SL}(h)-(l)). It can be seen from Fig. \ref{figure-Woniu-SL}(d) that the proposed MF-STFnet can identify Road well and preserve its fine structure better. In addition, the visual result of MF-STFnet is closest to the ground truth.

\begin{table*}[!htbp] 
\centering  
\fontsize{5.5}{9}\selectfont  
\caption{Classification Performance of Different Methods on the Flevoland Dataset} 
\label{table-Flev-CLP}  
\begin{tabular}{c|c|c|c|c|c|c|c|c|c|c|c|c|c|c|c} \hline
\toprule[0.3pt]

\multicolumn{1}{c|}{\multirow{2}{*}{\bf Category}}
& \multicolumn{7}{c|}{\bf MF-STFnet}
& \multicolumn{1}{c|}{\multirow{2}{*}{\bf WMM}}
& \multicolumn{1}{c|}{\multirow{2}{*}{\bf O-SVM}}
& \multicolumn{1}{c|}{\multirow{2}{*}{\bf S-SRC}}
& \multicolumn{1}{c|}{\multirow{2}{*}{\bf TF-ANN}}
& \multicolumn{1}{c|}{\multirow{2}{*}{\bf CRPM-Net}}
& \multicolumn{1}{c|}{\multirow{2}{*}{\bf Tb-CNN}}
& \multicolumn{1}{c|}{\multirow{2}{*}{\bf DMCNN}}
& \multicolumn{1}{c}{\multirow{2}{*}{\bf ViT}} \\ \cline{2-8}

\multicolumn{1}{c|}{}
& \multicolumn{1}{c|}{\bf C}& {\bf L}& {\bf P}& {\bf C\&L}& {\bf C\&P}& {\bf L\&P}& {\bf C\&L\&P}
& \multicolumn{1}{c|}{}
& \multicolumn{1}{c}{} \\ 

\hline 
{Grass} &{74.32} &{72.96} &{62.75} &{80.97} &{86.09} &{77.10} &{85.33} &{42.59} &{82.42} &{80.03} &{83.11} &{82.97} &{82.92} &{78.20} &\textbf{{86.80}} \cr 
{Flax} &{98.89} &{99.79} &{62.08} &\textbf{{99.91}} &{99.49} &{99.68} &{99.81} &{98.11} &{98.79} &{98.99} &{86.66}  &{99.59} &{99.73} &{99.54} &{99.49} \cr 
{Potato} &{95.12} &{94.42} &{88.65} &{97.80} &{97.30} &{96.85} &{98.25} &\textbf{{99.18}} &{97.32} &{97.80} &{98.16}  &{96.64} &{98.65}  &{97.38} &{97.95} \cr 
{Wheat} &{84.30} &{91.70} &{93.37} &{97.18} &{97.78} &{97.11} &\textbf{{98.35}} &{95.28} &{95.19} &{96.60} &{96.89}  &{95.98} &{96.76}  &{98.32} &{96.34} \cr 
{Rapeseed} &{99.84} &{99.47} &{98.62} &{99.75} &{99.76} &{99.84} &{99.90} &\textbf{{99.99}} &{99.59} &{99.91} &{99.77}  &{99.12} &{99.71}  &{99.79} &{99.17} \cr 
{Beet} &{72.73} &{78.96} &{49.04} &{93.60} &{83.44} &{86.21} &{92.35}  &{43.61} &{78.97} &{79.28} &{86.87}  &{83.45} &\textbf{{92.46}}  &{90.30} &{87.40} \cr 
{Barley} &{85.05} &{97.74} &{96.19} &{98.52} &{97.65} &{99.47} &{99.57}  &\textbf{{99.58}} &{97.00} &{98.85} &{98.03}  &{95.03} &{97.92}  &{97.94} &{97.44} \cr 
{Peas} &{67.00} &{67.97} &{67.46} &{82.58} &{72.44} &{72.88} &{73.09}  &{64.48} &{87.24} &\textbf{{93.28}} &{68.35}  &{69.51} &{72.73}  &{71.38} &{71.10} \cr 
{Maize} &{57.10} &{87.57} &{67.69} &{91.11} &{72.44} &{92.52} &\textbf{{97.05}}  &{74.28} &{93.04} &{85.84} &{82.98}  &{84.10} &{91.26}  &{92.73}  &{88.20} \cr 
{Beans} &{97.83} &{54.34} &{19.98} &{84.61} &{96.57} &{70.74} &{97.28}  &\textbf{{98.49}} &{91.34} &{97.09} &{97.12}  &{94.35} &{91.98}  &{92.89} &{92.79} \cr 
{Fruit} &{86.61} &{96.46} &{85.06} &{92.89} &{94.49} &{89.43} &{95.05}  &\textbf{{99.52}} &{97.71} &{98.63} &{94.30}  &{93.85} &{89.57}  &{94.14} &{94.56} \cr 
{Onions} &{59.34} &{58.52} &{28.15} &{96.15} &{78.06} &{45.60} &\textbf{{97.28}}  &{93.66} &{93.52} &{92.50} &{94.92}  &{61.92} &{96.10}  &{82.32} &{96.87} \cr 
{Lucerne} &{70.45} &{99.83} &{90.10} &\textbf{{99.84}} &{97.65} &{99.90} &{99.51}  &{0.12} &{97.13} &{97.76}&{98.66}  &{97.60} &{99.66}  &{99.55} &{98.78} \cr 
{Building} &{65.92} &{73.06} &{73.24} &{76.61} &{73.04} &{75.35} &{76.90}  &{34.20} &{86.24} &{74.25} &\textbf{{87.94}}  &{74.49} &{68.17}  &{73.52} &{75.02} \cr 
{Road} &{62.29} &{72.30} &{61.87} &{81.77} &{80.90} &{84.70} &\textbf{{84.90}}  &{41.81} &{68.36} &{62.76} &{80.38}  &{76.53} &{82.93}  &{85.92} &{85.54} \cr 
\hline 
{OA(\%)} &{83.65} &{89.21} &{81.48} &{94.32} &{93.37} &{92.92} &\textbf{{95.45}}  &{81.06} &{91.33} &{91.37}&{93.34}  &{91.57} &{94.19}  &{94.14} &{93.96} \cr
{AA(\%)} &{78.35} &{82.95} &{69.18} &{91.09} &{88.72} &{85.75} &\textbf{{92.36}}  &{72.32} &{90.92} &{90.24}&{90.28}  &{87.01} &{90.70}  &{90.26} &{91.16} \cr 
{$\kappa$ $\times$ 100} & {81.24} &{87.63} &{78.88} &{93.47} &{92.38} &{91.87} &\textbf{{94.77}}  &{78.09} &{90.04} &{90.09}&{92.34}  &{90.33} &{93.33}  &{93.27} &{93.06} \cr \hline

\toprule[0.3pt]
\end{tabular}
\end{table*}

\begin{figure*}[!htbp] 
\vspace{-0.8cm}
\centering
\subfloat[Ground Truth]{\includegraphics[width=0.22\linewidth,height=3.2cm]{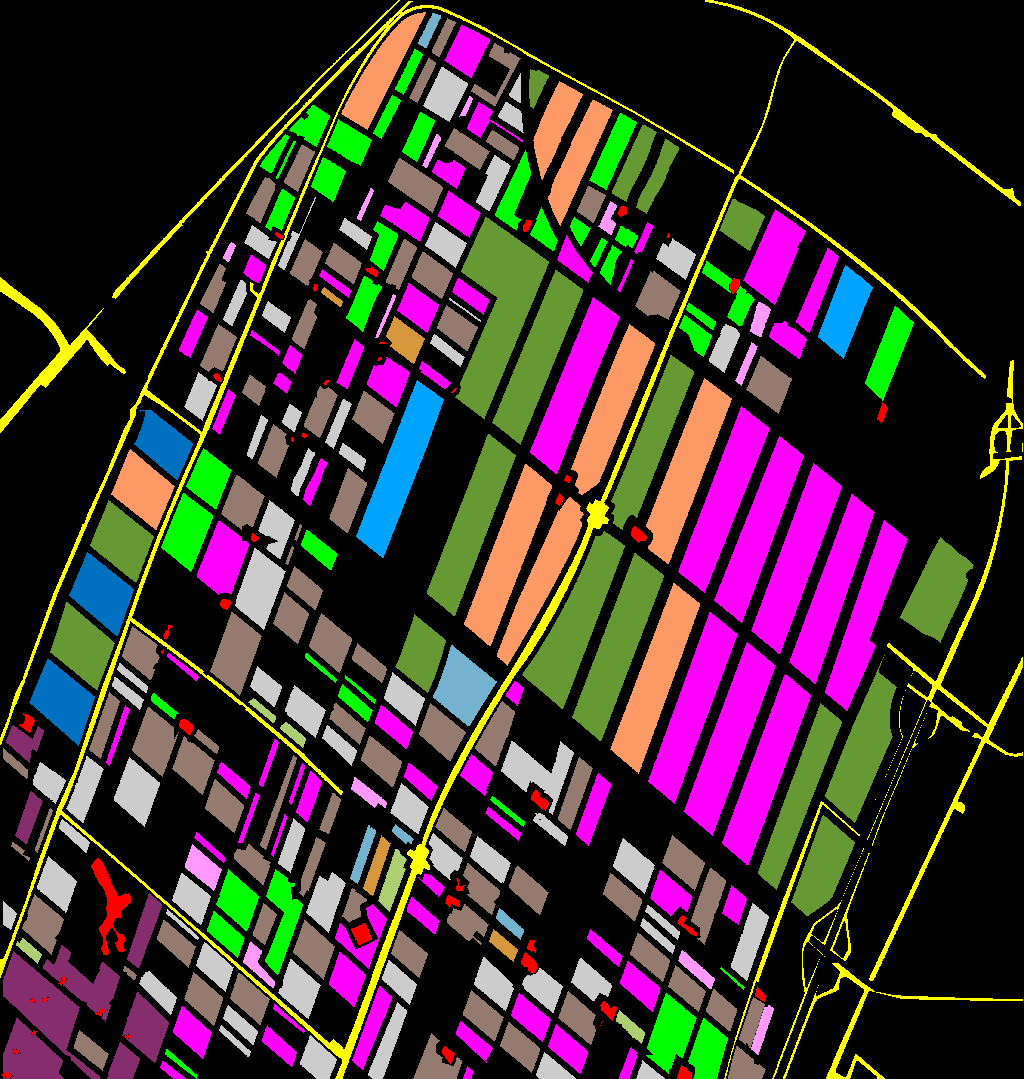}}
\label{Flev-GT}
\subfloat[MF-STFnet(C)]{\includegraphics[width=0.22\linewidth,height=3.2cm]{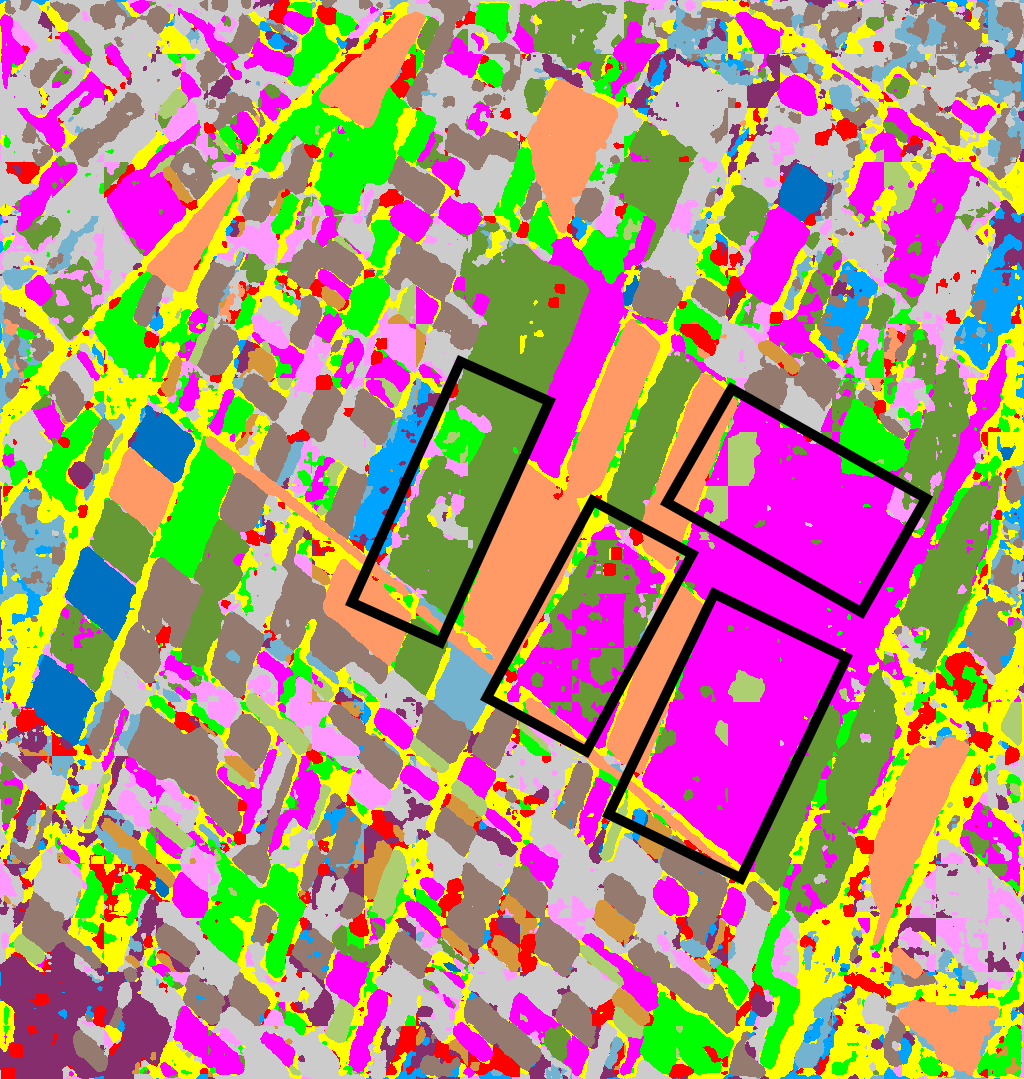}}
\label{STFnet-C}
\subfloat[MF-STFnet(L)]{\includegraphics[width=0.22\linewidth,height=3.2cm]{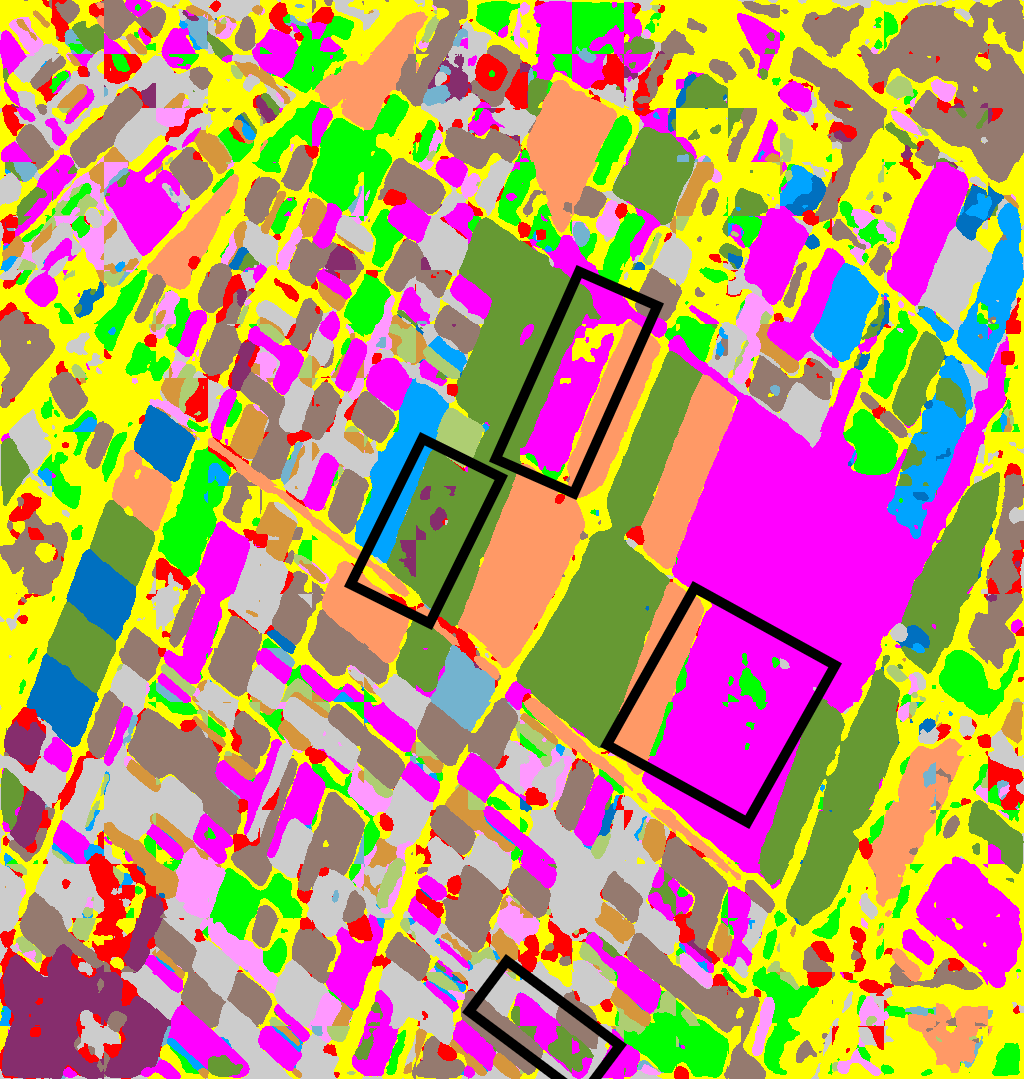}}
\label{STFnet-L}
\subfloat[MF-STFnet(P)]{\includegraphics[width=0.22\linewidth,height=3.2cm]{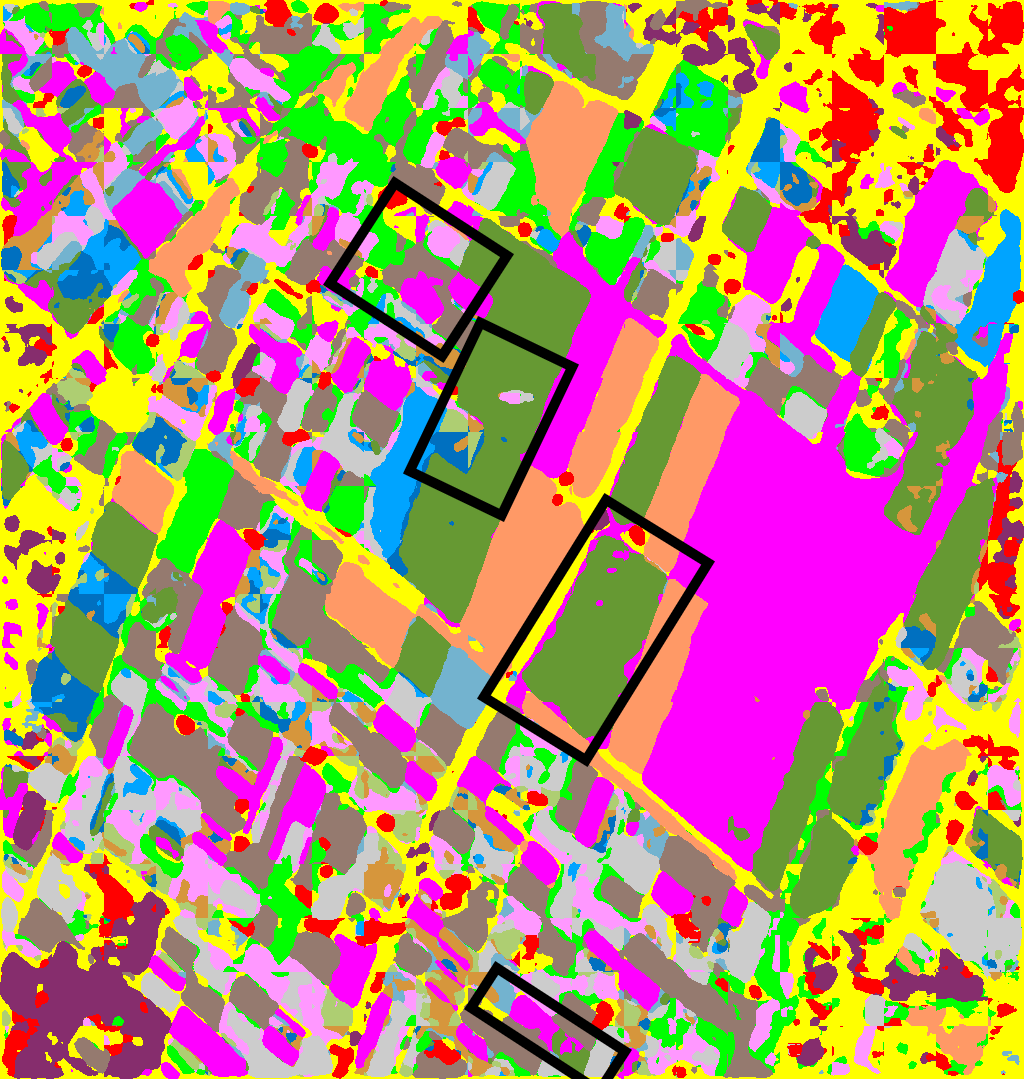}}
\label{STFnet-P}
\subfloat[MF-STFnet(C\&L)]{\includegraphics[width=0.22\linewidth,height=3.2cm]{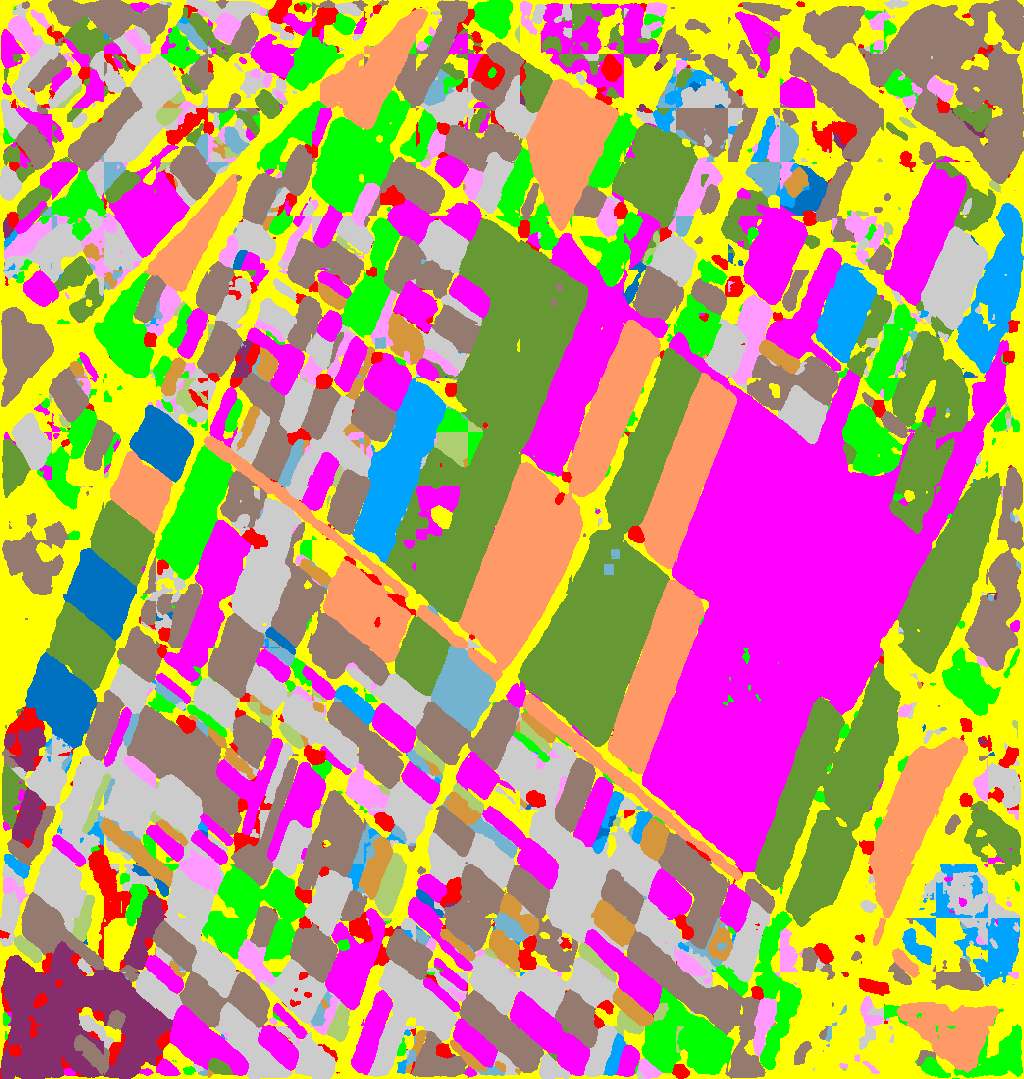}}
\label{STFnet-CL}
\subfloat[MF-STFnet(C\&P)]{\includegraphics[width=0.22\linewidth,height=3.2cm]{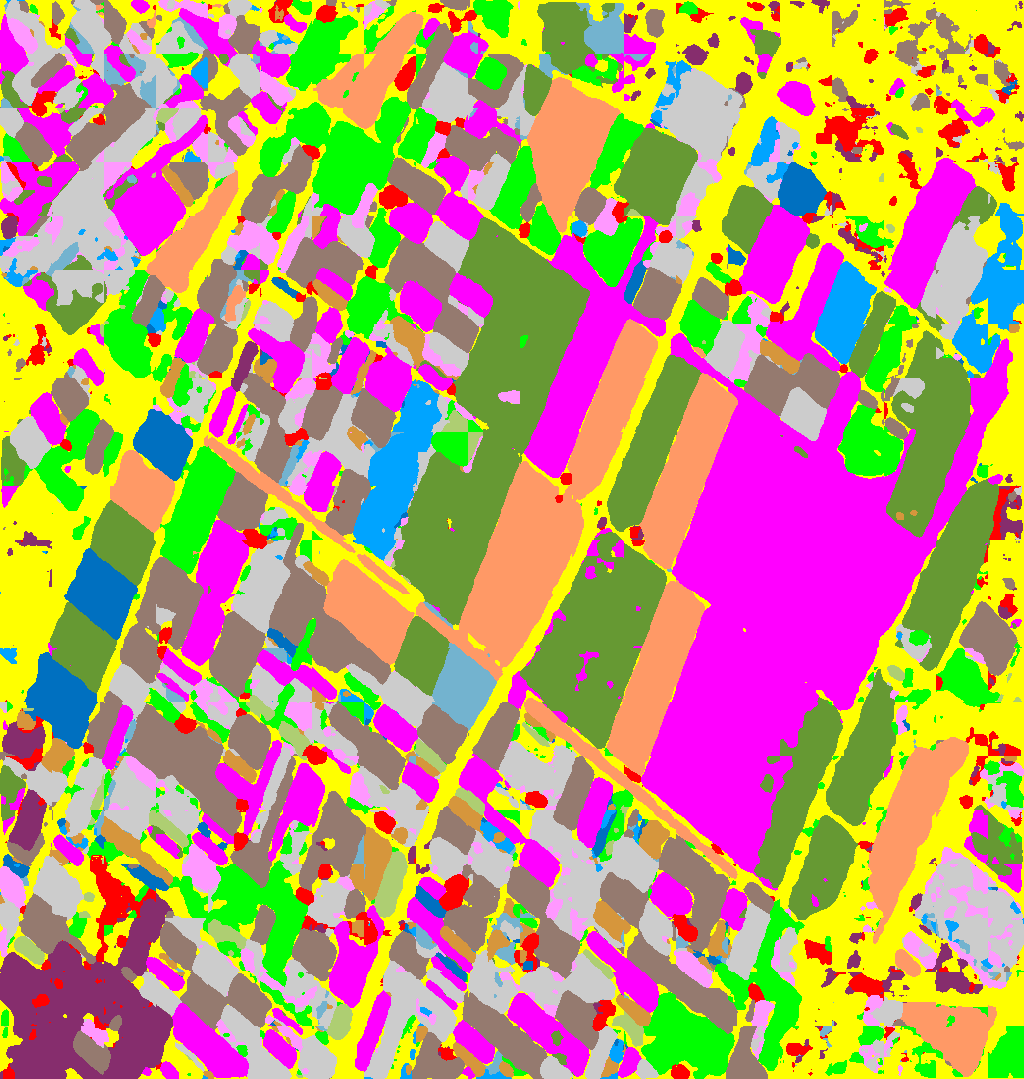}}
\label{STFnet-CP}
\subfloat[MF-STFnet(L\&P)]{\includegraphics[width=0.22\linewidth,height=3.2cm]{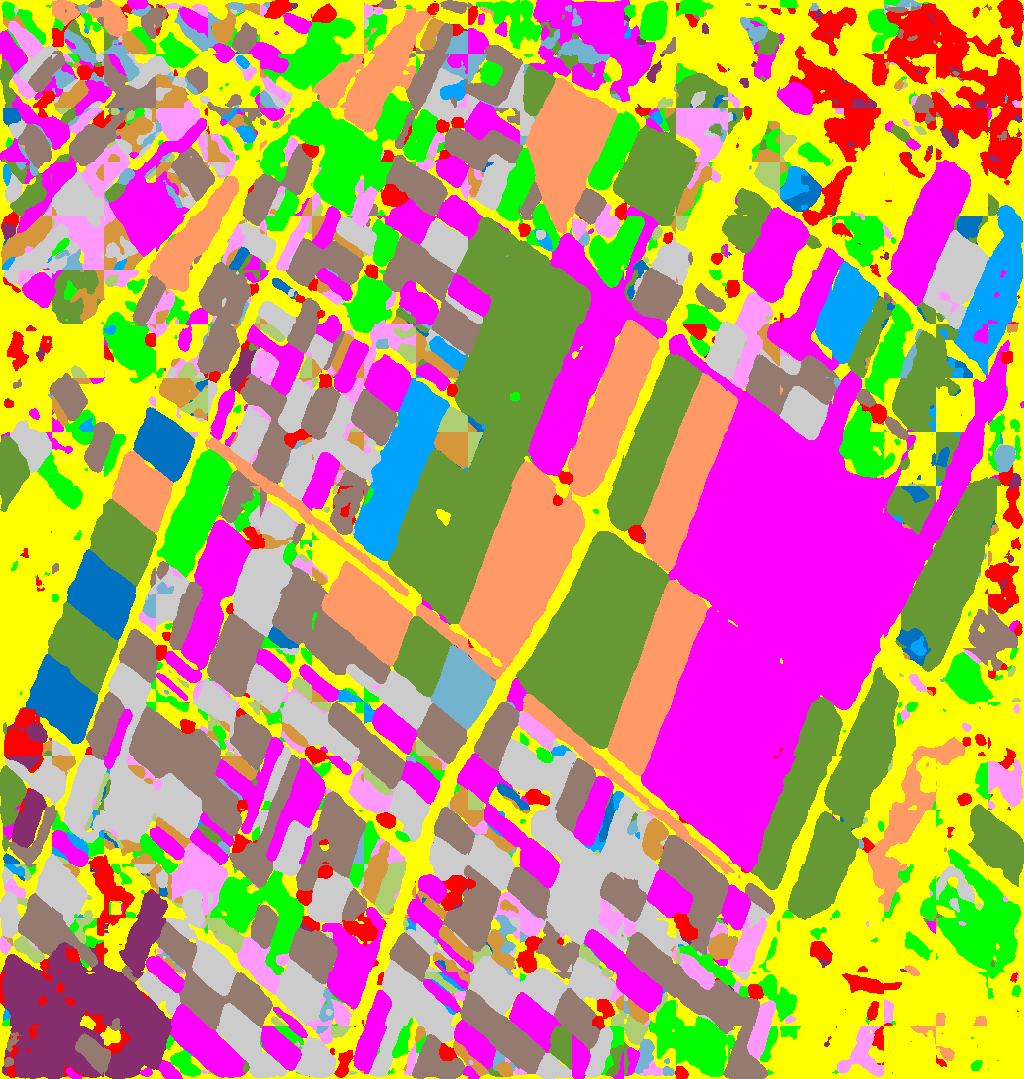}}
\label{STFnet-LP}
\subfloat[MF-STFnet(C\&L\&P)]{\includegraphics[width=0.22\linewidth,height=3.2cm]{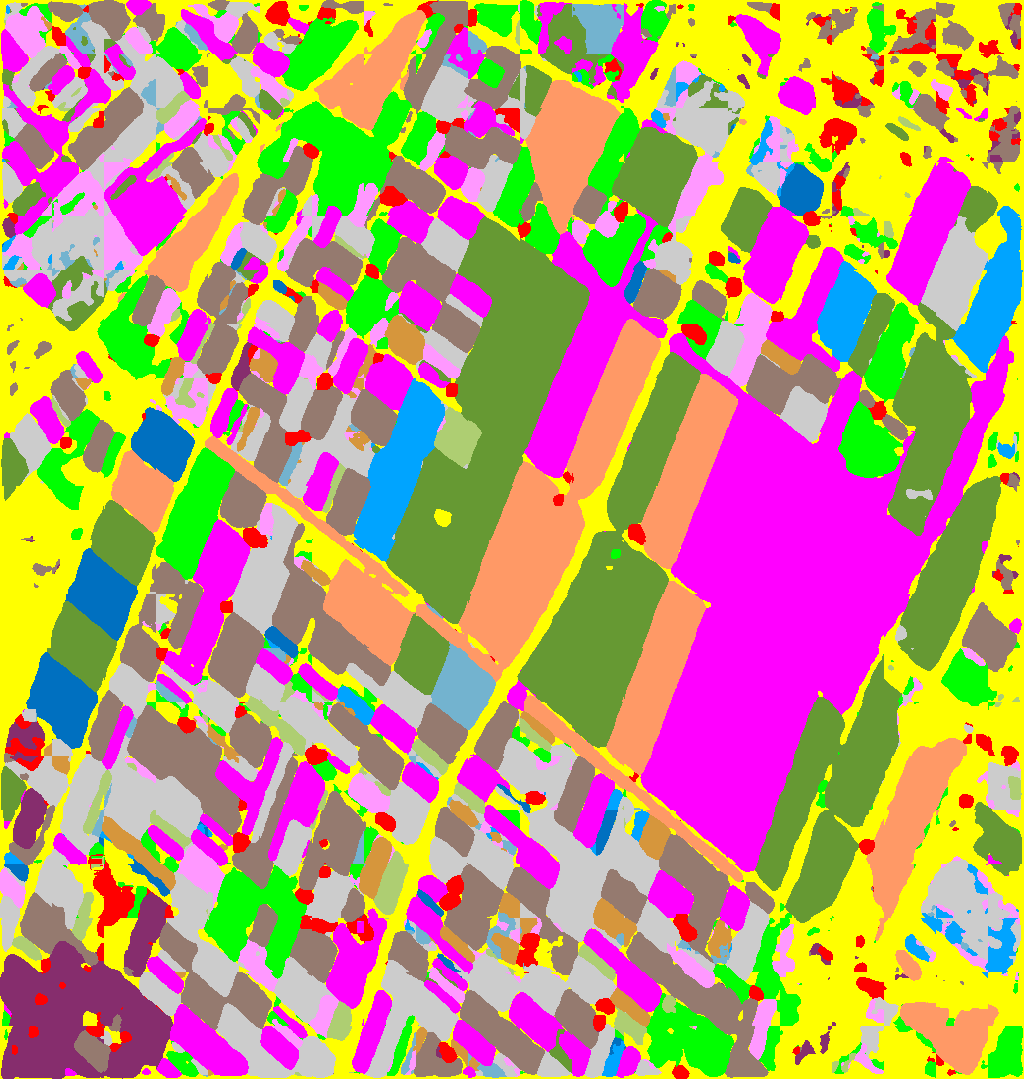}}
\label{STFnet-CLP}
\subfloat[WMM]{\includegraphics[width=0.22\linewidth,height=3.2cm]{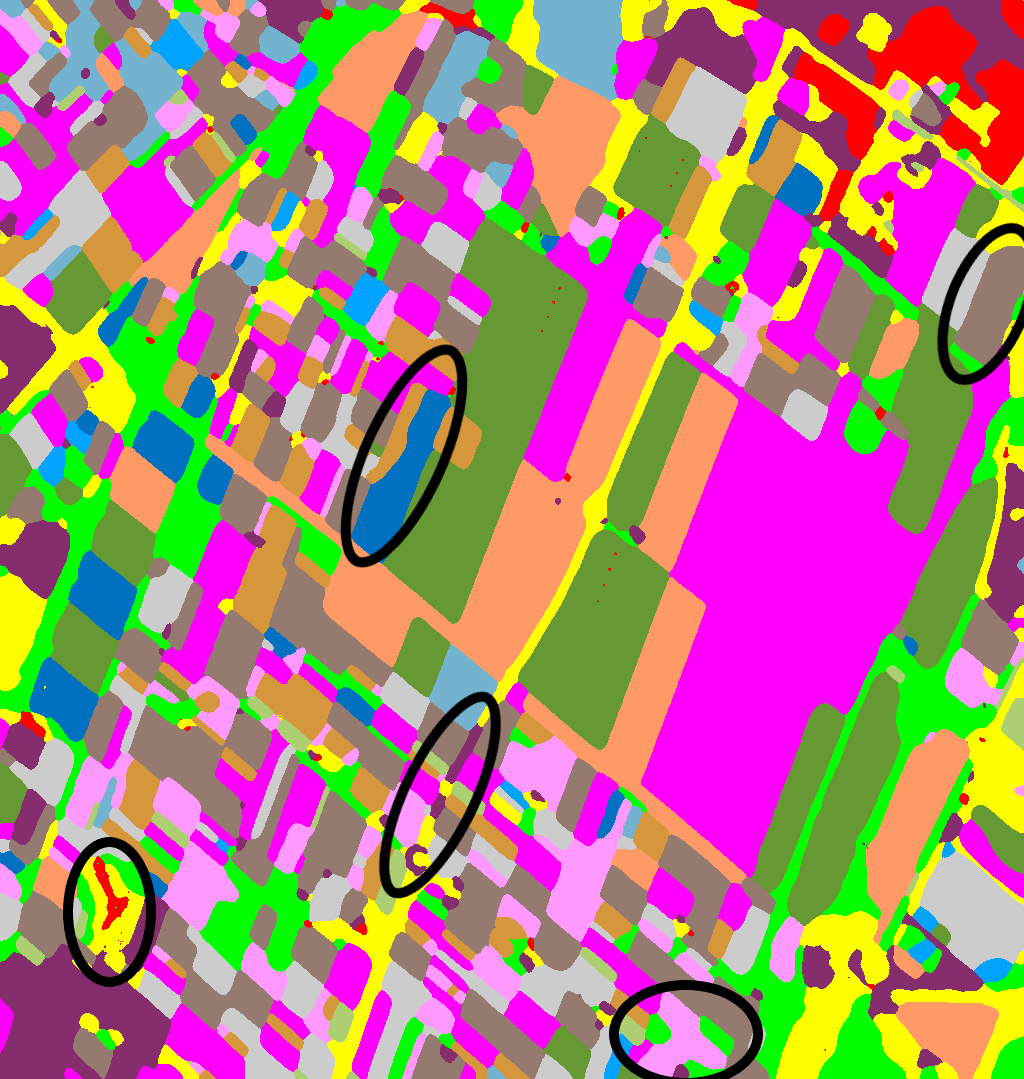}}
\label{WMM}
\subfloat[O-SVM]{\includegraphics[width=0.22\linewidth,height=3.2cm]{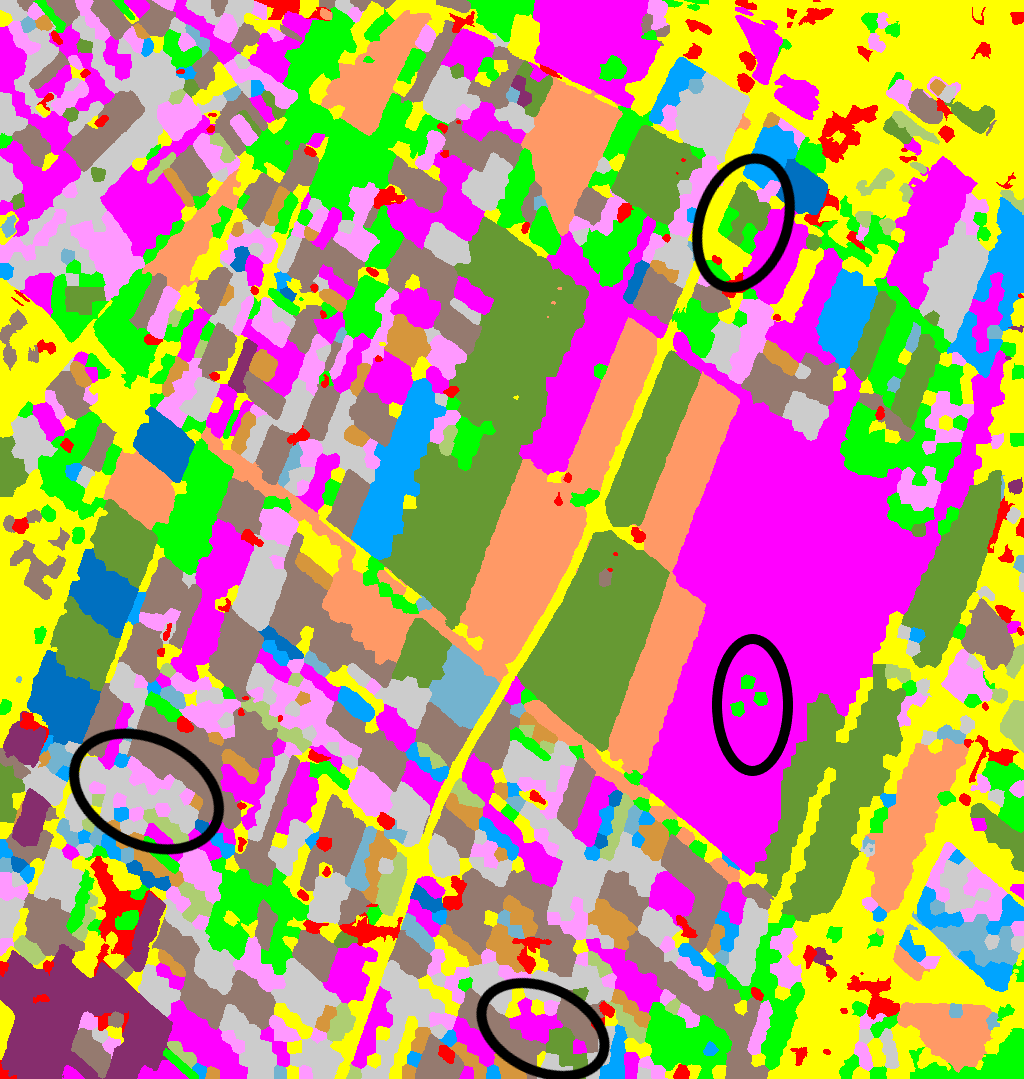}}
\label{O-SVM}
\subfloat[S-SRC]{\includegraphics[width=0.22\linewidth,height=3.2cm]{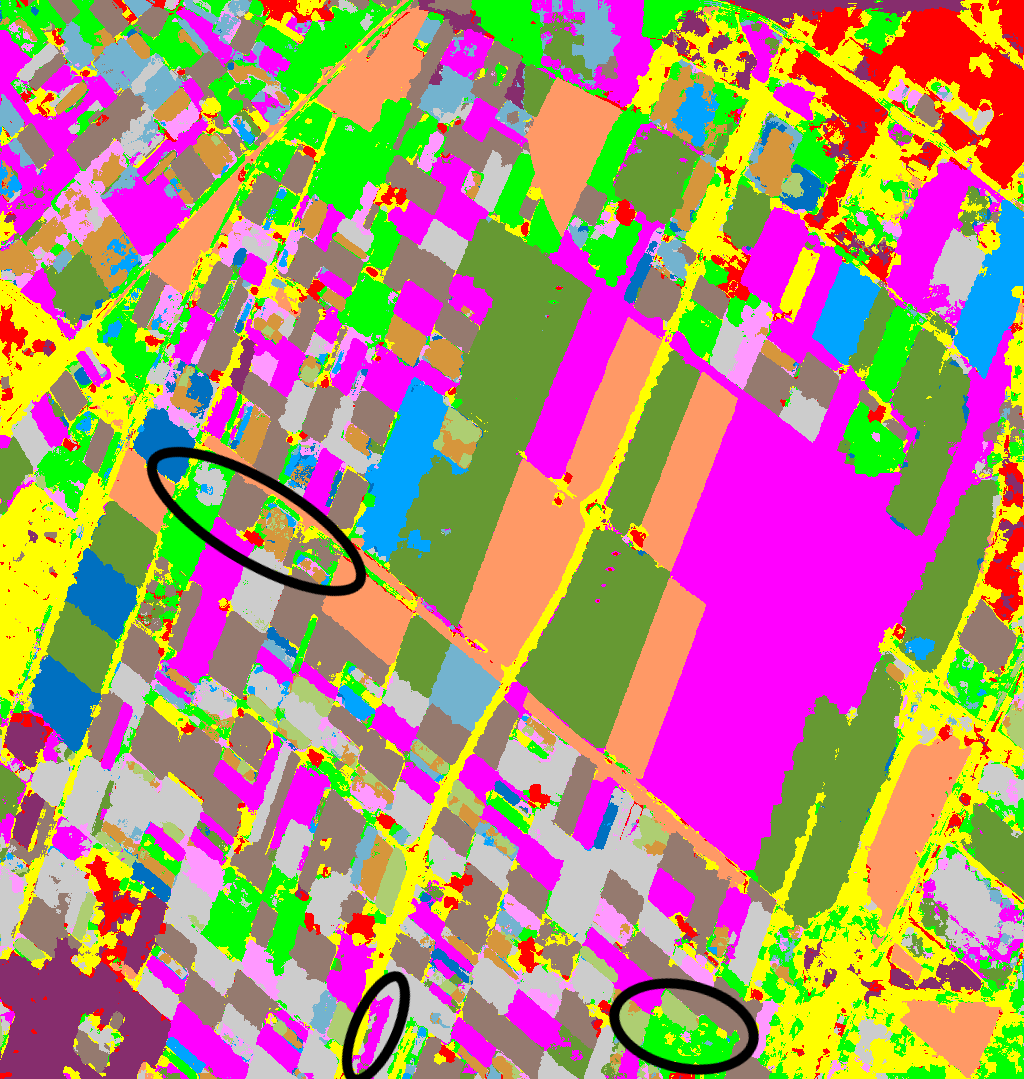}}
\label{S-SRC}
\subfloat[TF-ANN]{\includegraphics[width=0.22\linewidth,height=3.2cm]{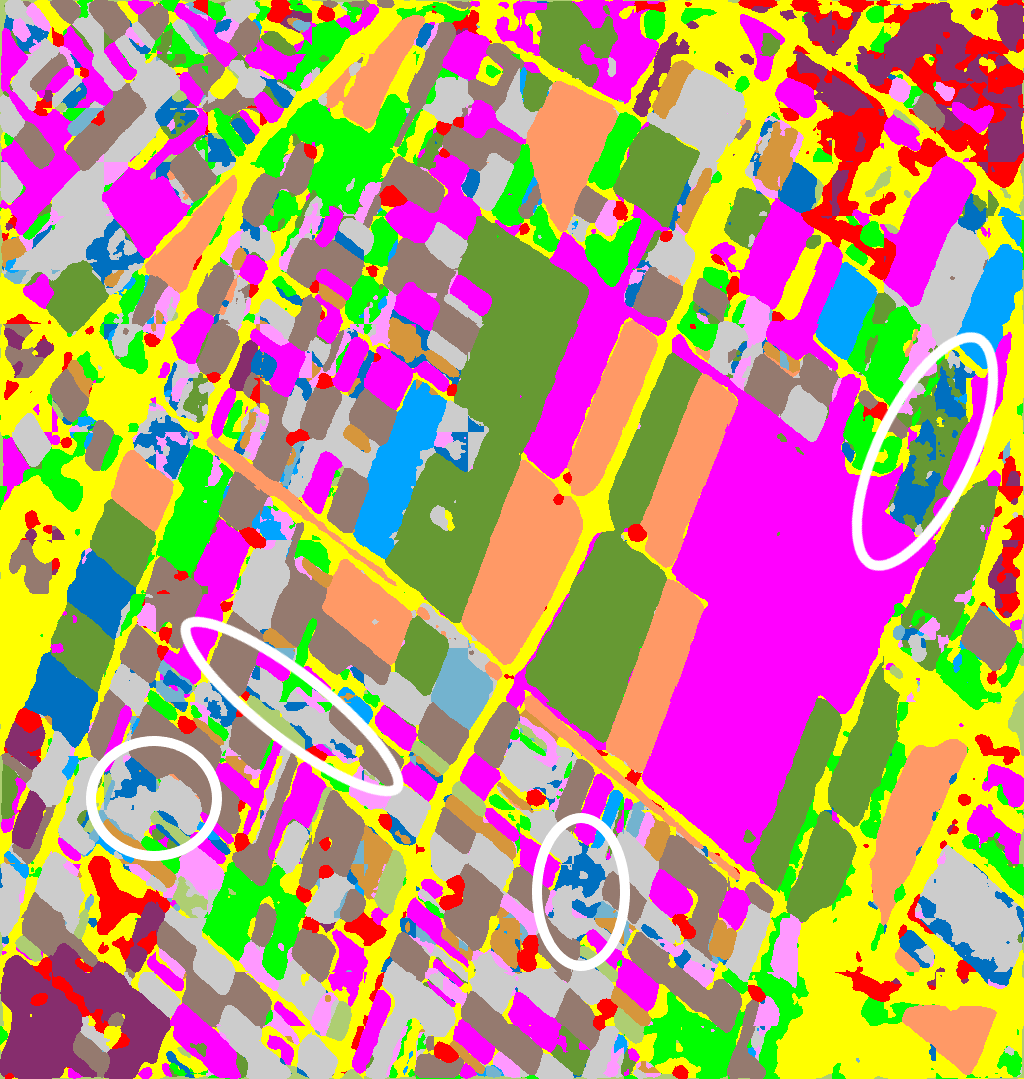}}
\label{TF-ANN}
\subfloat[CRPM-Net]{\includegraphics[width=0.22\linewidth,height=3.2cm]{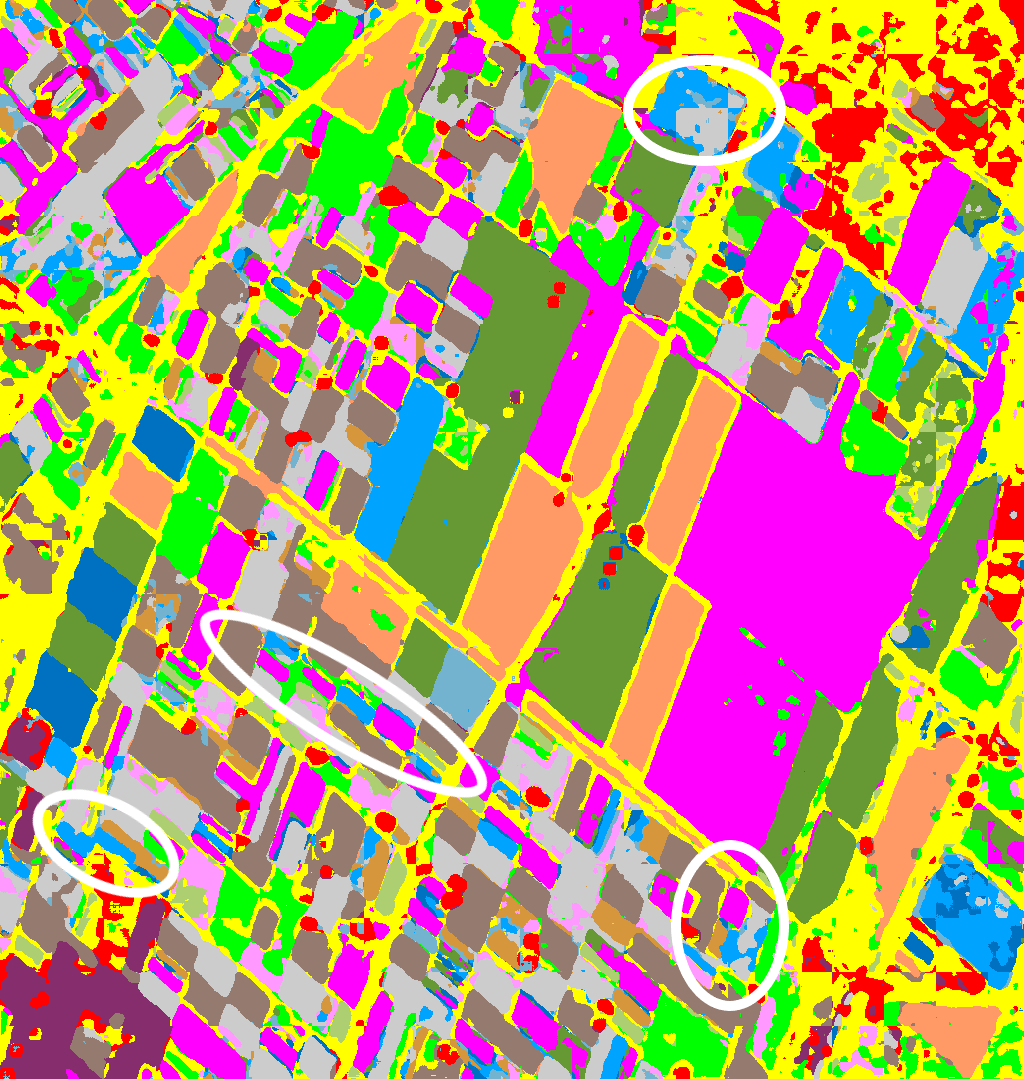}}
\label{CRPM-Net}
\subfloat[Tb-CNN]{\includegraphics[width=0.22\linewidth,height=3.2cm]{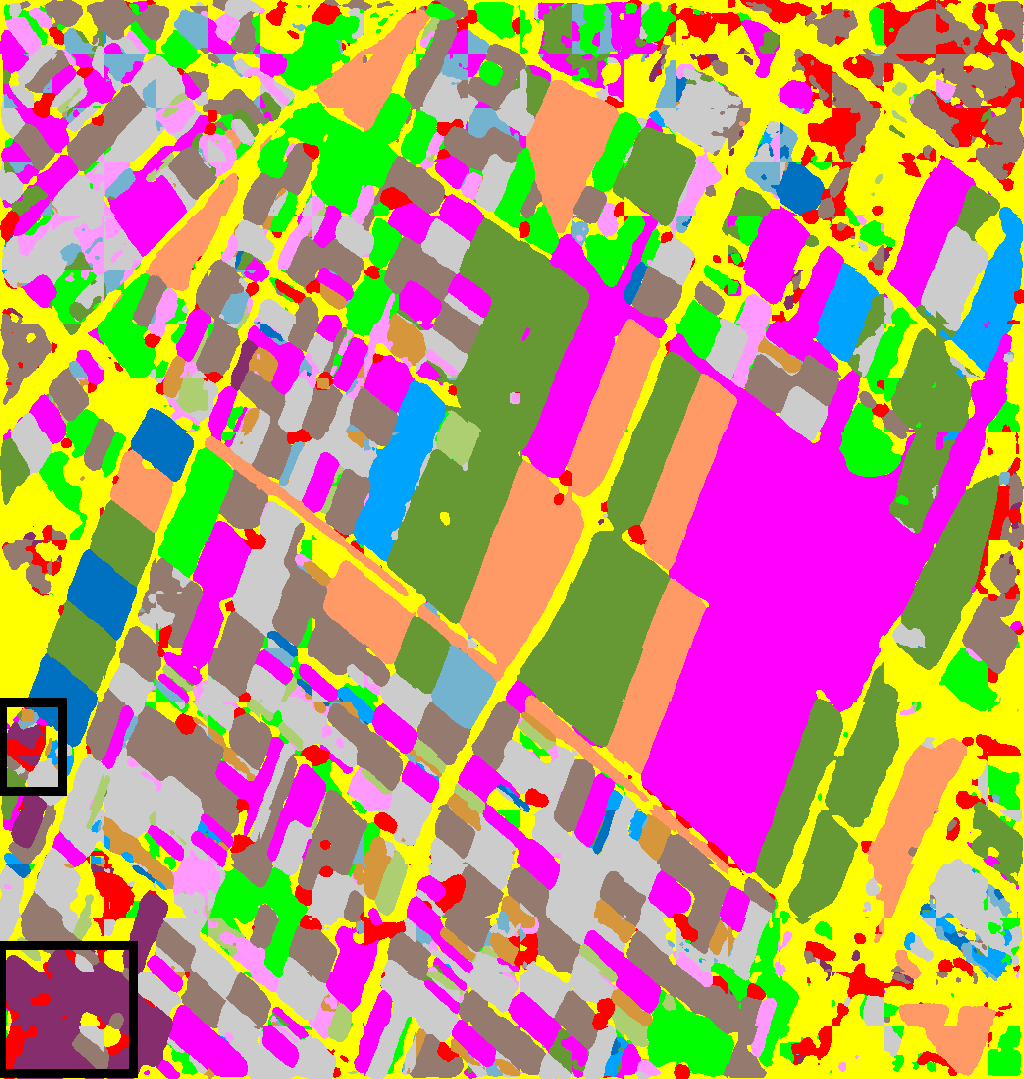}}
\label{Tb-CNN}
\subfloat[DMCNN]{\includegraphics[width=0.22\linewidth,height=3.2cm]{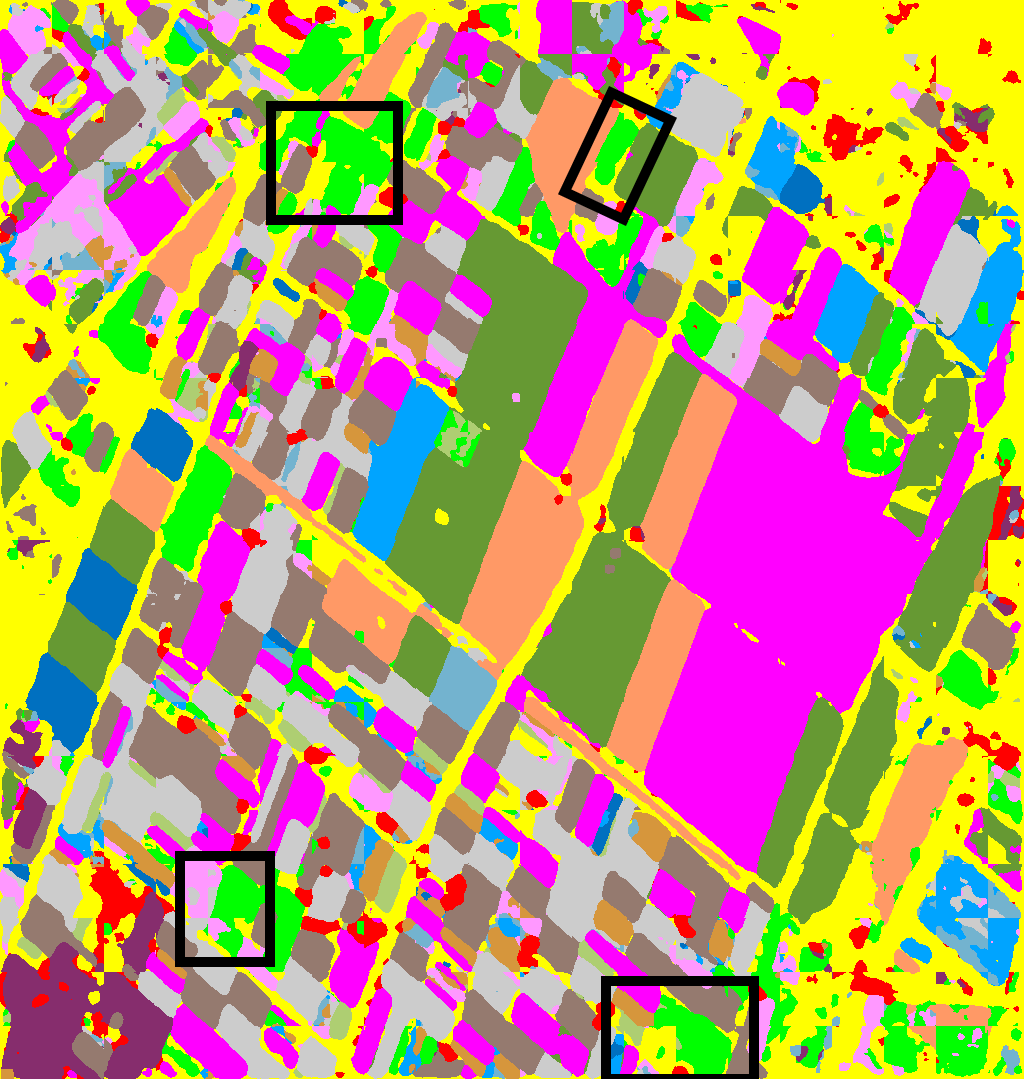}}
\label{DMCNN}
\subfloat[ViT]{\includegraphics[width=0.22\linewidth,height=3.2cm]{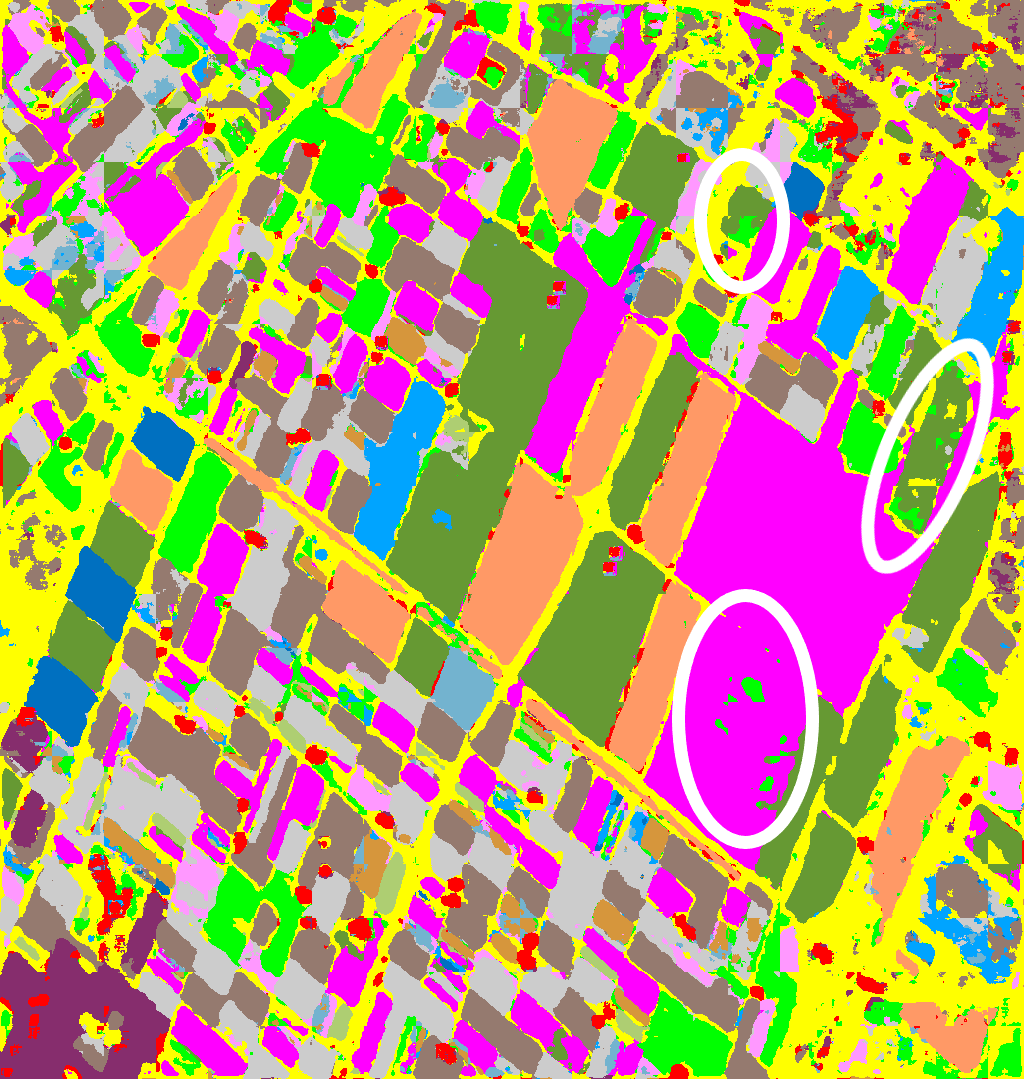}}
\label{ViT}
\subfloat{\includegraphics[width=0.9\linewidth,height=0.4cm]{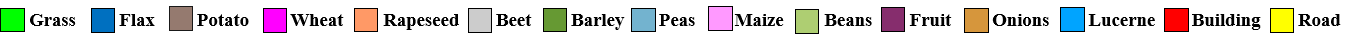}}
\label{color}

\caption{Classification results of different methods on the Flevoland dataset.}
\label{figure-Flev-CLP}
\end{figure*} 

\subsubsection{Results on the Flevoland dataset}
For the Flevoland dataset, Table \ref{table-Flev-CLP} quantitatively demonstrates the performance comparison. Besides, Fig. \ref{figure-Flev-CLP} shows the relevant classification results. As shown in Table \ref{table-Flev-CLP}, L-band achieves higher classification accuracy than C- and P-bands for most categories. C-band is more suitable to identify Grass, Potato, and Beans. P-band is superior to other bands in classifying Wheat. The main reason can be explained by the band difference. As previously analyzed, compared with C- and P-bands, the L-band microwave has a certain penetration ability. It can sense the different responses of the branches under canopies to better distinguish similar crops. Meanwhile, these experimental results reflect that different bands are suitable for identifying different ground objects. The dual-band combinations (C\&L, C\&P, and L\&P) all yield better accuracy than single-band. Notably, the C\&L\&P combination gets the best accuracy than single-band and other combinations. Except for Beans, Fruit, and Lucerne, the accuracy of other categories in the C\&L\&P combination is better than that of single-band. These illustrate that merging multiple bands indeed improves the classification performance under single-band conditions. For the C\&L\&P combination, MF-STFnet exhibits higher OA, AA, and Kappa than existing methods. In addition, compared with other DL-based methods (TF-ANN, CRPM-Net, Tb-CNN, DMCNN, and ViT), MF-STFnet achieves the best classification accuracy in most categories. These results show that our designed network improves feature discrimination, thereby better distinguishing similar categories.

Visually, as illustrated in Fig. \ref{figure-Flev-CLP}(b)-(d), many pixels in single-band classification maps are misclassified. For example, Wheat and Barley are heavily mixed with other categories (highlighted by black rectangles in Fig. \ref{figure-Flev-CLP}(b)-(d)). This phenomenon is alleviated in Fig. \ref{figure-Flev-CLP}(e)-(h) because of the efficient use of multi-band features. Especially, the classification map of C\&L\&P has smoother homogeneous areas and better connectivity. As shown in Fig. \ref{figure-Flev-CLP}(i)-(k), WMM, O-SVM, and S-SRC have good regional category consistency, but there are obvious regional misclassifications (highlighted by black ovals). For TF-ANN, it is clear that many areas are misclassified as Flax in Fig. \ref{figure-Flev-CLP}(l) (highlighted by white ovals). In addition, in Fig. \ref{figure-Flev-CLP}(m), CRPM-Net misclassifies many pixels as Lucerne (highlighted by white ovals). As shown in Fig. \ref{figure-Flev-CLP}(p), the misclassification of Wheat and Barely is more serious (highlighted by white ovals). Compared with TF-ANN, CRPM-Net, and ViT, Tb-CNN and DMCNN have better visual effects and the misclassification points are greatly reduced. Despite this, there are still some unreasonable distributions in Tb-CNN and DMCNN, which are away from the real ones. For example, Building pixels appear in other areas and are mixed with Fruit pixels in Fig. \ref{figure-Flev-CLP}(n) (highlighted by black rectangles), and some pixels belonging to Grass are misclassified in Fig. \ref{figure-Flev-CLP}(o) (highlighted by black rectangles). Compared with existing methods, MF-STFnet obtains the result closer to the ground truth, and takes the lead in terms of boundary position recognition and region label consistency. In summary, according to Table \ref{table-Flev-CLP} and Fig. \ref{figure-Flev-CLP}, it can be concluded that the proposed MF-STFnet outperforms other methods on the Flevoland dataset.

To sum up, the experimental results on three multi-frequency datasets show that merging multi-frequency information can eliminate the classification inaccuracy under single-band, and is beneficial to distinguishing similar categories. In addition, compared with existing methods, the proposed MF-STFnet achieves remarkable performance improvement, which demonstrates our designed modules and network are effective. With the help of CIFEM, as well as jointly considering local and nonlocal spatiality, MF-STFnet can capture and utilize more discriminative information, thereby obtaining more satisfactory performance.

\section*{Conclusion}
This paper proposes MF-STFnet for MF-PolSAR image classification, which aims to fully mine and leverage the complementarity of bands and combine local and nonlocal spatial information to improve classification accuracy. In MF-STFnet, based on the band-specific semantic representation, the proposed CIFEM explicitly models the deep semantic correlation among bands. This realizes the interactive fusion and enhancement of inter-band information, thereby making full use of the complementarity of bands to improve the accuracy of ground object classification. In addition, MF-STFnet adopts the GraphSAGE model to dynamically capture the representation of nonlocal topological relations between samples. In this way, local and nonlocal spatial information are captured simultaneously, which further improves the robustness of classification results. Moreover, based on the total loss, the proposed AWF strategy adaptively updates weights to flexibly fuse inferences from different bands for the final multi-frequency joint classification decision. Experiments on two measured MF-PolSAR datasets show that the proposed modules are effective for improving classification performance. In addition, the proposed MF-STFnet can more effectively capture and utilize the complementarity of bands to eliminate the classification inaccuracy under SF conditions. Furthermore, MF-STFnet can combine local and nonlocal spatial information to obtain more accurate results than other related state-of-art classification methods.

\section*{Acknowledgment}
The authors would like to thank the anonymous reviewers for their constructive comments and suggestions that greatly strengthened this paper.

\ifCLASSOPTIONcaptionsoff
  \newpage
\fi

\raggedend 
\end{document}